\documentclass{UoYCSproject}
\author{Graham Campbell}
\title{Efficient Graph Rewriting}
\date{~\\April 2019\\~\\Revised August 2019}
\supervisor{Dr Detlef Plump}
\BSc

\acknowledgements{
  I should like to thank my supervisor, Detlef Plump, for his support and guidance this year and for giving me the benefit of his invaluable technical understanding of rewriting systems. Our (regularly overrunning!) meetings were of great benefit, and I hope to continue to collaborate after I move to the School of Mathematics, Statistics and Physics at Newcastle.
}

\definecolor{gp2green}{RGB}{69, 191, 156}
\definecolor{gp2blue}{RGB}{153, 187, 255}
\definecolor{gp2red}{RGB}{236, 107, 116}
\definecolor{gp2pink}{RGB}{239, 161, 193}
\definecolor{gp2grey}{RGB}{196, 192, 200}
\definecolor{performanceBlue}{RGB}{0, 136, 255}
\definecolor{performanceYellow}{RGB}{252, 199, 17}

\input{proof.tex}

\begin{document}
\setcounter{tocdepth}{2}
\pagenumbering{roman}
\maketitle

\cleardoublepage
\phantomsection
\addcontentsline{toc}{chapter}{\listfigurename}
\listoffigures

\renewcommand{\listtheoremname}{List of Theorems}
\cleardoublepage
\phantomsection
\addcontentsline{toc}{chapter}{\listtheoremname}
\listoftheorems[ignoreall,show=theorem]

\begin{summary}
\vspace{-0.4em}
\section*{Introduction, Motivation and Goals}

\vspace{-0.1em}
Graph transformation is the rule-based modification of graphs, and is a discipline dating back to the 1970s, with the \enquote{algebraic approach} invented at the Technical University of Berlin by Ehrig, Pfender, and Schneider \parencite{Ehrig-Pfender-Schneider73} \parencite{Ehrig79}. It is a comprehensive framework in which the local transformation of structures can be modelled and studied in a uniform manner \parencite{Corradini-Montanari-Rossi-Ehrig-Heckel-Lowe97} \parencite{Ehrig97} \parencite{Ehrig06}. Applications in Computer Science are wide-reaching including compiler construction, software engineering, natural language processing, modelling of concurrent systems, and logical and functional programming \parencite{CorradiniRossiParisiPresicce91} \parencite{CorradiniEhrigKreowskiRozenberg02} \parencite{EhrigEngelsParisi-PresicceRozenberg04}. There are a number of GT languages and tools \parencite{Schurr-Winter-Zundorf99} \parencite{Runge-Ermel-Taentzer11} \parencite{Agrawal-Karsai-Neema-Shi-Vizhanyo06} \parencite{Ghamarian-Mol-Rensink-Zambon-Zimakova12} \parencite{Jakumeit-Buchwald-Kroll10} \parencite{Plump11}.

The declarative nature of graph rewriting rules comes at a cost. In general, to match the left-hand graph of a fixed rule within a host graph requires polynomial time. To improve matching performance, D\"orr \parencite{Doerr95} proposed to equip rules and host graphs with distinguished \enquote{root} nodes, and to match roots in rules with roots in host graphs. This concept has been
implemented by Bak and Plump in GP\,2, allowing programs to rival the performance of traditional implementations in languages such as C \parencite{Bak-Plump12}.

Graph transformation with root nodes and relabelling is not yet well understood. With only relabelling, Habel and Plump have been able to recover many, but not all, of the standard results \parencite{Habel-Plump02} \parencite{HabelPlump12}. Moreover, Bak and Plump's model suffers from the problem that derivations are not necessarily invertible. This motivates us to develop a new model of rooted graph transformation with relabelling which does not suffer this problem. If we have termination and invertibility, then we have an algorithm for testing graph language membership, and if we have confluence (and constant time matching), then we have an efficient algorithm too \parencite{DoddsPlump06} \parencite{Plump10}.

Testing for \enquote{confluence} is not possible in general \parencite{Plump93}, however we can sometimes use \enquote{critical pair} analysis to show confluence. Confluence remains poorly understood, and while there are techniques for classifying \enquote{conflicts} \parencite{EhrigLambersOrejas08} \parencite{Hristakiev18}, it is rarely possible to actually show confluence. Moreover, in general, confluence is stronger than required for language efficient membership testing, motivating a weaker definition of confluence.

Our method will be to use mathematical definitions and proofs, as is usual in theoretical computer science. We aim to:

\vspace{-0.6em}
\begin{minipage}{\textwidth}
\begin{enumerate}[itemsep=-0.7ex,topsep=-0.7ex]
\item Outline rooted DPO graph transformation with relabelling;
\item Repair the problem of lack of invertibility in rooted GT systems;
\item Develop a new example of linear time graph algorithm;
\item Develop new results for confluence analysis of GT systems.
\vspace{-0.8em}
\end{enumerate}
\end{minipage}


\section*{Outline, Results and Evaluation}

We regard this project as a success, having achieved our four original goals. Each of our goals have been addressed by the first four chapters, respectively. We started by reviewing the current state of graph transformation, with a particular focus on the \enquote{injective DPO} approach with relabelling and graph programming languages, establishing issues with the current approach to rooted graph transformation due to its \enquote{pointed} implementation. We also briefly reviewed DPO-based graph programming languages.

We address the lack of invertibility of rooted derivations by defining rootedness using a partial function onto a two-point set rather than pointing graphs with root nodes. We have shown rule application corresponds to \enquote{NDPOs}, how Dodds' complexity theory \parencite{Dodds08} applies in our system, and briefly discussed the equivalence of and refinement of GT systems. Developing a fully-fledged theory of correctness and refinement for (rooted) GT systems remains future work, as does establishing if the Local Church-Rosser and Parallelism theorems hold \parencite{EhrigGolasHabelLambersOrejas2014} \parencite{HabelPlump12}. Applications of our model to efficient graph class recognition are exciting due to the invertibility of derivations.

We have shown a new result that the graph class of trees can be recognised by a rooted GT system in linear time, given an input graph of bounded degree. Moreover, we have given empirical evidence by implementing the algorithm in GP\,2 and collecting timing results. We have submitted our program and results for publication \parencite{Campbell-Courtehoute-Plump19}. Overcoming the restriction of host graphs to be of bounded degree remains open research, as well as showing further case studies and applications.

We have defined a new notion of \enquote{confluence modulo garbage} and \enquote{non-garbage critical pairs}, and shown that it is sufficient to require strong joinability of only the non-garbage critical pairs to establish confluence modulo garbage. We have applied this theory to Extended Flow Diagrams \parencite{Farrow-Kennedy-Zucconi76} and the encoding of partially labelled (rooted) GT systems as standard GT systems, performing non-garbage critical pair analysis on the encoded system. Further exploring the relationship between confluence modulo garbage and weak garbage separation remains open work, as does improving the analysis of (non-garbage) critical pairs to allow us to decide confluence in more cases than currently possible via pair analysis.


\section*{Ethical Considerations}

This project is of a theoretical nature. As such, no human participants were required, and no confidential data has been collected. Moreover, there are no anticipated ethical implications of this work or its applications.

\end{summary}

\chapter{Theoretical Background} \label{chapter:theoryintro}

Before reading the main text, the reader should first skim read Appendix \ref{appendix:notation} in order to set up notation and definitions.

In this chapter, we will review the rewriting of totally labelled graphs with relabelling, and Bak and Plump's modifications adding \enquote{root} nodes \parencite{Bak-Plump12}. We will see how (rooted) graph transformation systems are instances of abstract reduction systems, and will look at graph programming languages.


\section{Graphs and Morphisms} \label{sec:maingraphs}

There are various definitions of a \enquote{graph}. In particular, we are interested in graphs where edges are directed and parallel edges are permitted.

\begin{definition} \label{def:graph}
We can formally define a \textbf{concrete graph} as:
\begin{align*}
G = (V, E, s: E \to V, t: E \to V)
\end{align*}
where \(V\) is a \textbf{finite} set of \textbf{vertices}, \(E\) is a \textbf{finite} set of \textbf{edges}. We call \(s: E \to V\) the \textbf{source} function, and \(t: E \to V\) the \textbf{target} function.
\end{definition}

\begin{definition} \label{def:graphsize}
If \(G\) is a \textbf{concrete graph}, then \(\abs{G} = \abs{V_G} + \abs{E_G}\).
\end{definition}

\begin{figure}[H]
\centering
\noindent
\begin{tikzpicture}[every node/.style={align=center}]
    \node (a) at (0.8,0.0)    [draw, circle, thick, fill=black, scale=0.3] {\,};
    \node (b) at (0.0,-0.0)   [draw, circle, thick, fill=black, scale=0.3] {\,};
    \node (c) at (-0.8,-0.0)  [draw, circle, thick, fill=black, scale=0.3] {\,};

    \node (A) at (0.8,-0.18)  {\tiny{3}};
    \node (B) at (-0.8,-0.18) {\tiny{1}};
    \node (C) at (0.0,-0.18)  {\tiny{2}};

    \draw (b) edge[->,thick, bend left] (c)
          (c) edge[->,thick, bend left] (b)
          (a) edge[->,thick] (b)
          (a) edge[->,thick,in=35,out=-35,loop] (a);
\end{tikzpicture}
\vspace{-0.5em}
\caption{Example Concrete Graph}
\label{fig:eg1}
\end{figure}
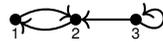

\begin{example}
Consider the concrete graph \(G = (\{1, 2, 3\}, \{a, b, c, d\}, s, t)\) where \(s = \{(a, 1), (b, 2), (c, 3), (d, 3)\}\), \(t = \{(a, 2), (b, 1), (c, 1), (d, 3)\}\) (treating functions as sets). Its graphical representation is given in Figure \ref{fig:eg1}. Note that the numbers are not \enquote{labels}, but \enquote{node ids}.
\end{example}

\vspace{-0.6em}
\vspace{-0.2em}

\begin{definition}\label{def:morph}
Given two concrete graphs \(G\) and \(H\), a \textbf{graph morphism} \(g: G \to H\) is a pair of maps \(g = (g_V: V_G \to V_H, g_E: E_G \to E_H)\) such that sources and targets are preserved. That is, \(\forall e \in E_G,\) \(g_V(s_G(e))\) \(= s_H(g_E(e))\) and \(g_V(t_G(e)) = t_H(g_E(e))\). Equivalently, both of the squares in Figure \ref{fig:comsqrsmorph} commute.
\end{definition}

\vspace{-0.8em}
\begin{figure}[H]
\centering
\noindent
\begin{equation*}
\begin{tikzcd}
  E_G \arrow[r, "s_G"] \arrow[d, "g_E"]
    & V_G \arrow[d, "g_V"]
    & E_G \arrow[r, "t_G"] \arrow[d, "g_E"]
    & V_G \arrow[d, "g_V"] \\
  E_H \arrow[r, "s_H"]
    & V_H
    & E_H \arrow[r, "t_H"]
    & V_H
\end{tikzcd}
\end{equation*}
\vspace{-1.0em}
\caption{Graph Morphism Commuting Diagrams}
\label{fig:comsqrsmorph}
\end{figure}

\newpage

\begin{definition}
A graph morphism \(g: G \to H\) is \textbf{injective}/\textbf{surjective} iff both \(g_V\) and \(g_E\) are injective/surjective as functions. We say \(g\) is an \textbf{isomorphism} iff it is both injective and surjective.
\end{definition}

\begin{example}
The identity morphism \((id_V, id_E)\) is an isomorphism between any graph and itself.
\end{example}

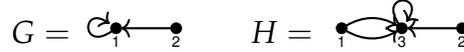
\begin{figure}[H]
\centering
\noindent
\begin{tikzpicture}[every node/.style={align=center}]
    \node (a) at (-0.2,-0.05)  {$G = $};
    \node (b) at (0.8,0.0)    [draw, circle, thick, fill=black, scale=0.3] {\,};
    \node (c) at (1.6,0.0)    [draw, circle, thick, fill=black, scale=0.3] {\,};

    \node (B) at (0.8,-0.18)  {\tiny{1}};
    \node (C) at (1.6,-0.18)  {\tiny{2}};

    \node (d) at (3.0,-0.05) {$H = $};
    \node (e) at (3.8,0.0)    [draw, circle, thick, fill=black, scale=0.3] {\,};
    \node (f) at (5.4,0.0)    [draw, circle, thick, fill=black, scale=0.3] {\,};
    \node (g) at (4.6,0.0)    [draw, circle, thick, fill=black, scale=0.3] {\,};

    \node (E) at (3.8,-0.18)  {\tiny{1}};
    \node (F) at (5.4,-0.18)  {\tiny{2}};
    \node (G) at (4.6,-0.18)  {\tiny{3}};

    \draw (b) edge[->,thick,in=145,out=215,loop] (b)
          (c) edge[->,thick] (b)
          (e) edge[->,thick, bend left] (g)
          (e) edge[->,thick, bend right] (g)
          (f) edge[->,thick] (g)
          (g) edge[->,thick,in=55,out=125,loop] (g);
\end{tikzpicture}
\vspace{-0.6em}
\caption{Example Concrete Graphs}
\label{fig:eg2}
\end{figure}

\begin{example} \label{eg:graphmorpheg1}
Consider the graphs in Figure \ref{fig:eg2}. There are four morphisms \(G \to H\), three of which are injective, none of which are surjective. There are actually also four morphisms \(H \to G\), three of which are surjective.
\end{example}

\vspace{-1.2em}
\vspace{-0.2em}

\begin{definition}
We say that graphs \(G, H\) are \textbf{isomorphic} iff there exists a \textbf{graph isomorphism} \(g: G \to H\), and we write \(G \cong H\). This naturally gives rise to \textbf{equivalence classes} \([G]\), called \textbf{abstract graphs}.
\end{definition}

\begin{proposition}
The \textbf{quotient} (Definition \ref{dfn:quotient}) of the \textbf{collection} of all \textbf{concrete graphs} with \(\cong\) is the \textbf{countable set} of all \textbf{abstract graphs}.
\end{proposition}


\section{Graph Transformation} \label{sec:gtintro}

There are various approaches to graph transformation, most notably the \enquote{edge replacement} \parencite{Drewes-Kreowski-Habel97}, \enquote{node replacement} \parencite{Engelfriet-Rozenberg97}, and \enquote{algebraic} approaches \parencite{Corradini-Montanari-Rossi-Ehrig-Heckel-Lowe97} \parencite{Ehrig97}. The two major approaches to algebraic graph transformation are the so called \enquote{double pushout} (DPO) approach, and the \enquote{single pushout} (SPO) approach. Because the DPO approach operates in a structure-preserving manner (rule application in SPO is without an interface graph, so there are no dangling condition checks), this approach is more widely used than the SPO \parencite[p.9-14]{Ehrig06} \parencite{Ehrig97}. For this reason, we will focus only on the DPO approach with injective matching. Moreover, DPO graph grammars can generate every recursively enumerable set of graphs \parencite{Uesu78}.

Given an unlabelled graph (Definition \ref{def:graph}), there are two common approaches to augmenting it with data: \textbf{typed graphs} and \textbf{totally labelled graphs}. We choose to work with the \textbf{labelled approach} (Section \ref{sec:graphpl}) because it is easy to understand and reason about, has a \textbf{relabelling} theory (Section \ref{section:relabelling}), and a \enquote{rooted} modification (Section \ref{section:rooted}). Details of the \textbf{typed approach} can be found in Section \ref{section:typed}. Note that typed (attributed) (hyper)graphs have a rich theory \parencite{LoweKorffWagner93} \parencite{CorradiniMontanariRossi96} \parencite{BertholdFischerKoch00} \parencite{HeckelKusterMalteTaentzer02} \parencite{EhrigPrangeTaentzer04} \parencite{Plump10}.

A review of graph transformation of labelled graphs using the DPO approach with injective matching can be found in Appendix \ref{appendix:transformation}. We will also cover the definitions and results for our new type of system in Chapter \ref{chapter:newtheory}, so we will not repeat ourselves in this chapter by giving all of the detail again. Additionally, an example system and grammar can be found in Chapter \ref{chapter:treerecognision}.


\newpage
\section{Adding Relabelling} \label{section:relabelling}

The origin of partially labelled graphs is from the desire to have \enquote{relabelling}. If the interface \(K\) is totally labelled, then any node which has context (incident edges) cannot be deleted, and so we must preserve its label to avoid breaking uniqueness of rule application. We can get around this problem with partial labelling of interface graphs, and thus with modest modifications to the theory for totally labelled graphs we allow rules to \enquote{relabel} nodes. We shall be using this foundation going forward. All the relevant definitions and theorems are in Appendix \ref{appendix:transformation}.

We are in fact using a restricted version of the theory presented by Habel and Plump \parencite{Habel-Plump02}, the restriction being that we allow the interface \(K\) to be partially labelled, but require \(L\), \(R\) and \(G\) to be totally labelled, ensuring that given a totally labelled input graph \(G\), the result graph \(H\) is also totally labelled. Thus, derivations are defined only on totally labelled graphs, but allow us to relabel nodes.

\begin{example}
Consider the following totally labelled \enquote{rule}, over the label alphabet \((\{1, 2\}, \{\Square\})\) where \(x, y\) are to be determined:

\vspace{0.2em}
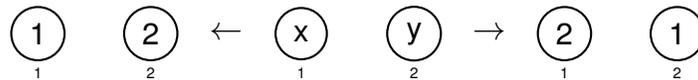
\begin{figure}[H]
\centering
\noindent
\begin{tikzpicture}[every node/.style={inner sep=0pt, text width=6.5mm, align=center}]
    \node (a) at (0.0,0) [draw, circle, thick] {1};
    \node (b) at (1.5,0) [draw, circle, thick] {2};

    \node (c) at (2.5,0) {$\leftarrow$};

    \node (d) at (3.5,0) [draw, circle, thick] {x};
    \node (e) at (5.0,0) [draw, circle, thick] {y};

    \node (f) at (6.0,0) {$\rightarrow$};

    \node (g) at (7.0,0) [draw, circle, thick] {2};
    \node (h) at (8.5,0) [draw, circle, thick] {1};

    \node (A) at (0.0,-.52) {\tiny{1}};
    \node (B) at (1.5,-.52) {\tiny{2}};
    \node (D) at (3.5,-.52) {\tiny{1}};
    \node (E) at (5.0,-.52) {\tiny{2}};
    \node (G) at (7.0,-.52) {\tiny{1}};
    \node (H) at (8.5,-.52) {\tiny{2}};
\end{tikzpicture}
\vspace{-0.3em}
\caption{Relabelling Non-Example}
\end{figure}
\vspace{-0.2em}

\vspace{-0.4em}
We want to swap the labels without deleting the nodes, because they may have context. There is no value we can choose for \(x\) or \(y\) such that the conditions to be a totally labelled graph morphism are satisfied. Now consider the setting where we allow the interface graph to have a partial node label map. We could simply not label the interface nodes, and then we have exactly what we want.
\end{example}


\section{Rooted Graph Transformation} \label{section:rooted}

Rooted graph transformation first appeared when D\"orr \parencite{Doerr95} proposed to equip rules and host graphs with distinguished (root) nodes, and to match roots in rules with roots in host graphs. More recently, Bak and Plump \parencite{Bak-Plump12} \parencite{Bak15} have used rooted graph transformation in conjunction with the theory of partially labelled graph transformation in GP\,2.

The motivation for root nodes is to improve the complexity of finding a match of the left-hand graph \(L\) of a rule within a host graph \(G\). In general, linear time graph algorithms may, instead, take polynomial time when expressed as graph transformation systems \parencite{GeissBatzGrundHackSzalkowski06} \parencite{DoddsPlump06} \parencite{Bak-Plump12} \parencite{Campbell-Romo-Plump18}. An excellent account of this is available in Part II of Dodds' Thesis \parencite{Dodds08}.

We can define rooted graphs in a pointed style, just as for typed graphs. An account of the theoretical modifications is provided in Section \ref{section:rootedgt}, using Bak's approach \parencite{Bak-Plump12}. Note that Dodds \parencite{DoddsPlump06} \parencite{Dodds08} previously implemented root nodes via an augmentation of the label alphabet, however Bak's approach makes for more concise theory, and has been implemented in GP\,2.

We can formalise the problem of applying a rule:

\begin{definition}[Graph Matching Problem (GMP)] \label{dfn:gmp}
Given a graph \(G\) and a rule \(r = \langle L \leftarrow K \rightarrow R \rangle\), find the set of injective graph morphisms \(L \to G\).
\end{definition}

\begin{definition}[Rule Application Problem (RAP)] \label{dfn:rap}
Given a graph \(G\), a rule \(r = \langle L \leftarrow K \rightarrow R \rangle\), and an injective match \(g: L \to G\), find the result graph \(H\). That is, does it satisfy the \enquote{dangling condition}, and if so, construct \(H\).
\end{definition}

\begin{proposition} \label{prop:gmptime}
The GMP requires time \(O(\abs{G}^{\abs{L}})\) time given the assumptions in Figure \ref{fig:perfass}. Moreover, given a match, one can decide if it is applicable in \(O(|r|)\) time. That is, the RAP requires \(O(|r|)\) time. \parencite{Dodds08}
\end{proposition}

\vspace{-0.6em}
To improve matching performance, one can add root nodes to rules and match roots in rules with roots in host graphs, meaning we need only consider subgraphs of bounded size for matching, vastly improving the time complexity. That is, given a graph \(G\) of bounded degree containing a bounded number of root nodes, and a rule \(r\) of bounded size with \(L\) containing a single root node, then the time complexity of GMP reduces to constant time  \parencite{Dodds08}.

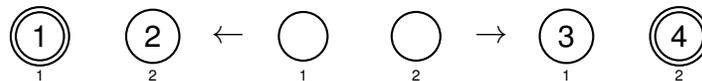
\begin{figure}[H]
\centering
\noindent
\begin{tikzpicture}[every node/.style={inner sep=0pt, text width=6.5mm, align=center}]
    \node (a) at (0.0,0) [draw, circle, thick, double, double distance=0.4mm] {1};
    \node (b) at (1.5,0) [draw, circle, thick] {2};

    \node (c) at (2.5,0) {$\leftarrow$};

    \node (d) at (3.5,0) [draw, circle, thick] {\,};
    \node (e) at (5.0,0) [draw, circle, thick] {\,};

    \node (f) at (6.0,0) {$\rightarrow$};

    \node (g) at (7.0,0) [draw, circle, thick] {3};
    \node (h) at (8.5,0) [draw, circle, thick, double, double distance=0.4mm] {4};

    \node (A) at (0.0,-.52) {\tiny{1}};
    \node (B) at (1.5,-.52) {\tiny{2}};
    \node (D) at (3.5,-.52) {\tiny{1}};
    \node (E) at (5.0,-.52) {\tiny{2}};
    \node (G) at (7.0,-.52) {\tiny{1}};
    \node (H) at (8.5,-.52) {\tiny{2}};
\end{tikzpicture}
\vspace{-0.2em}
\caption{Example Rooted Rule}
\label{fig:egrootedrule}
\end{figure}

\begin{example}
Figure \ref{fig:egrootedrule} \enquote{moves the root node} and also \enquote{relabels} the nodes in the host graph. A \enquote{fast} rooted implementation of the 2-colouring problem is available at \parencite{Bak-Plump12}, showcasing root nodes in GP\,2.
\end{example}

\vspace{0em}
\vspace{0.2em}
\vspace{-0.2em}
\vspace{-0.4em}

We will revisit time complexity in Section \ref{section:complexitythms}, showing that if rules are of a certain type, then derivations take only constant time, allowing us to use only derivation length as a measure of time complexity, as in standard complexity analysis theory for (non-deterministic) Turing Machines, first considered by Hartmanis and Stearns in 1965 \parencite{HartmanisStearns65}.


\section{Abstract Reduction Systems} \label{section:arsgt}

\textbf{Abstract reduction systems} (or simply \textbf{reduction systems} or \textbf{ARS}) are a much more general setting than \textbf{graph transformation systems} (\textbf{GT systems} or \textbf{GTS}), and model the step-wise transformation of objects (see Appendix \ref{appendix:ars}). These systems were studied for the first time by Newman in the early 40s \parencite{Newman42}. Turing Machines and GT systems clearly fit into this model of reduction. Moreover, the formal semantics of programming languages is often defined in terms of a step-wise computation relation.

\begin{example}
\((\mathbb{N}, >)\) is a \textbf{terminating} (Definition \ref{dfn:ars3}), \textbf{finitely branching} (Definition \ref{dfn:branching}), \textbf{confluent} (Definition \ref{dfn:ars3}) ARS (Definition \ref{dfn:ars0}).
\end{example}

\begin{example}
\((\mathbb{Z}, >)\) by comparison is not \textbf{terminating} or \textbf{finitely branching}, but it is \textbf{confluent}!
\end{example}

\begin{definition} \label{dfn:gl}
Let \(\mathcal{L}\) be some fixed label alphabet (Definition \ref{dfn:labelalphabet}). We let \(\mathcal{G}(\mathcal{L})\) be the \textbf{collection} of all totally labelled \textbf{abstract graphs}, and \(\widehat{\mathcal{G}}(\mathcal{L})\) be the \textbf{collection} of all totally labelled, totally rooted \textbf{abstract graphs}.
\end{definition}

\begin{proposition} \label{prop:graphuniverse}
Given some \(\mathcal{L}\), \(\mathcal{G}(\mathcal{L})\) and \(\widehat{\mathcal{G}}(\mathcal{L})\) are \textbf{countable sets}.
\end{proposition}

\begin{definition}
Let \(T = (\mathcal{L}, \mathcal{R})\) be a (rooted) \textbf{GTS}. Then \((\mathcal{G}(\mathcal{L}), \rightarrow_{\mathcal{R}})\) is the induced \textbf{ARS} defined by \(\forall [G], [H] \in \mathcal{G}(\mathcal{L}), [G] \rightarrow_{\mathcal{R}} [H]\) iff \(G \Rightarrow_{\mathcal{R}} H\).
\end{definition}

\begin{lemma} \label{lem:gtprops}
Consider the \textbf{ARS} \((\mathcal{G}(\mathcal{L}), \rightarrow)\) induced by a (rooted) \textbf{GTS}. Then \(\rightarrow\) is a \textbf{binary relation} (Definition \ref{def:binrel}) on \(\mathcal{G}(\mathcal{L})\). Moreover, it is \textbf{finitely branching} (Definition \ref{dfn:branching}) and \textbf{decidable} (Definition \ref{dfn:decidable}).
\end{lemma}

\vspace{0.4em}
\begin{proof}
By Proposition \ref{prop:graphuniverse}, \(\mathcal{G}(\mathcal{L})\) is a countable set, and so \(\rightarrow\) is a countable set (by Theorem \ref{thm:setprod}), and is well-defined since derivations are unique up to isomorphism (Theorem \ref{thm:uniquederivations}). Finally, we have only finitely many rules, and for each rule, there can only exist finitely many matches \(L \to G\), so there can only ever be finitely many result graphs \(H\) (up to isomorphism) \(G \Rightarrow_{\mathcal{R}} H\) for any given \(G\).
\end{proof}
\vspace{-0.2em}

\begin{theorem}[Property Undecidability] \label{thm:undecidablegtsrealdeal}
Consider the \textbf{ARS} \((\mathcal{G}(\mathcal{L}), \rightarrow)\) induced by a (rooted) \textbf{GTS}. Then testing if \(\rightarrow\) is \textbf{terminating}, \textbf{acyclic}, or (\textbf{locally}) \textbf{confluent} is \textbf{undecidable} in general.
\end{theorem}

\vspace{0.4em}
\begin{proof}
Testing for acyclicity or termination was shown to be undecidable in general by Plump in 1998 \parencite{Plump98}. Undecidability of (local) confluence checking was shown by Plump in 1993 \parencite{Plump93}, even for terminating GT systems \parencite{Plump05}.
\end{proof}
\vspace{-0.2em}


\section{Graph Programming Languages}

GT systems naturally lend themselves to expressing computation by considering the normal forms of the input graph.

\begin{example} \label{eg:gtsemfunc}
Given a GT system \(T = (\mathcal{L}, \mathcal{R})\), consider the \textbf{state space} \(\Sigma = \mathcal{G}(\mathcal{L}) \cup \{\bot\}\) and the induced ARS \((\mathcal{G}(\mathcal{L}), \rightarrow_{\mathcal{R}})\). We may define the \textbf{semantic function} \(f_T: \mathcal{G}(\mathcal{L}) \to \mathcal{P}(\Sigma)\) by \(f_T([G]) = \{[H] \mid [H]\) is a normal form of \([G]\) with respect to \(\rightarrow_{\mathcal{R}}\} \cup \{\bot \mid\) there is an infinite reduction sequence starting from \([G]\}\) and \(f_T(\bot) = \{\bot\}\).
\end{example}

There are a number of GT languages and tools, such as
AGG \parencite{Runge-Ermel-Taentzer11},
GMTE \parencite{Hannachi13},
Dactl \parencite{Glauert-Kennaway-Sleep91},
GP\,2 \parencite{Plump11},
GReAT \parencite{Agrawal-Karsai-Neema-Shi-Vizhanyo06},
GROOVE \parencite{Ghamarian-Mol-Rensink-Zambon-Zimakova12},
GrGen.Net \parencite{Jakumeit-Buchwald-Kroll10},
Henshin \parencite{Arendt-Biermann-Jurack-Krause-Taentzer10},
PROGRES \parencite{Schurr-Winter-Zundorf99},
and PORGY \parencite{Fernandez-Kirchner-Mackie-Pinaud14}.
Habel and Plump \parencite{Habel-Plump01} show that such languages can be \enquote{computationally complete}:

\begin{proposition}
To be \textbf{computationally complete}, the three constructs:
\begin{enumerate}[itemsep=-0.6ex,topsep=-0.6ex]
\item Nondeterministic application of a rule from a set of rules (\(\mathcal{R}\));
\item Sequential composition (\(P1;\,P2\));
\item Iteration in the form that rules are applied as long as possible \(P\!\!\downarrow\).
\end{enumerate}
\noindent
are not only \textbf{sufficient}, but \textbf{necessary} (using DPO-based rule application).
\end{proposition}

\begin{example}
The semantics of some program \(P\) is a binary relation \(\rightarrow_P\) on some set of abstract (rooted) graphs \(\mathcal{G}\), inductively defined as follows:
\begin{enumerate}[itemsep=-0.6ex,topsep=-0.6ex]
\item \(\rightarrow_{\mathcal{R}} \defeq \rightarrow\) \: (where \(\rightarrow\) is the induced ARS relation on \(\mathcal{R})\).
\item \(\rightarrow_{P1;\,P2} \defeq \rightarrow_{P2} \circ \rightarrow_{P1}\).
\item \(\rightarrow_{P\!\downarrow} \defeq \{([G], [H]) \mid [G] \rightarrow_P^* [H] \text{ and } [H] \text{ is in normal form}\footnotemark\}\).\footnotetext{\([H]\) is in normal form iff it is not reducible using \(\rightarrow_P\)}
\end{enumerate}
\vspace{-1.6em}
\end{example}

\begin{remark}
While GT systems can \enquote{simulate} any Turing Machine, this does not make them \enquote{computationally complete} in the strong sense that any computable function on arbitrary graphs can be programmed.
\end{remark}

\vspace{-0.6em}
GP\,2 is an experimental rule-based language for problem solving in the domain of graphs, developed at York, the successor of GP \parencite{Plump09} \parencite{Plump11}. GP\,2 is of interest because it has been designed to support formal reasoning on programs \parencite{Plump16}, with a semantics defined in terms of partially labelled graphs, using the injective DPO approach with relabelling \parencite{Habel-Muller-Plump01} \parencite{Habel-Plump02}. Poskitt and Plump have set up the foundations for verification of GP\,2 programs \parencite{Poskitt-Plump12} \parencite{Poskitt-Plump13} \parencite{Poskitt13} \parencite{Poskitt-Plump14} using a Hoare-Style \parencite{Hoare69} system (actually for GP \parencite{ManningPlump08} \parencite{Plump09}), Hristakiev and Plump have developed static analysis for confluence checking \parencite{HristakievPlump17} \parencite{Hristakiev18}, and Bak and Plump have extended the language, adding root nodes \parencite{Bak-Plump12} \parencite{Bak15}. Plump has shown computational completeness \parencite{Plump17}.

GP\,2 uses a model of \enquote{rule schemata} with \enquote{application conditions}, rather than \enquote{rules} as we have seen up until now. The label alphabet used for both nodes and edges is \((\mathbb{Z} \cup {Char}^*)^* \times \mathcal{B}\). Roughly speaking, rule application works by finding an injective \enquote{premorphism} by ignoring labels, and then checking if there is an assignment of values such that after evaluating the label expressions, the morphism is label-preserving. The application condition is then checked, then rule application continues. \parencite{Plump11}

The formal semantics of GP\,2 is given in the style of Plotkin's structural operational semantics \parencite{Plotkin04}. Inference rules inductively define a small-step transition relation \(\rightarrow\) on configurations. The inference rules and definition of the semantic function \(\llbracket . \rrbracket: ComSeq \to \mathcal{G} \to \mathcal{P}(\mathcal{G} \cup \{fail, \bot\})\) were first defined in \parencite{Plump11}. Up-to-date versions can be found in \parencite{Bak15}.

\chapter{A New Theory} \label{chapter:newtheory}

\begin{figure}[H]
\centering
\noindent
\scalebox{.9}{\begin{tikzpicture}[every node/.style={inner sep=0pt, text width=6.5mm, align=center}]
    \node (a) at (0.0,0.0) [draw, circle, thick] {\,};
    \node (b) at (1.0,0.0) {$\leftarrow$};
    \node (c) at (2.0,0.0) [draw, circle, thick] {\,};
    \node (d) at (3.0,0.0) {$\rightarrow$};
    \node (e) at (4.0,0.0) [draw, circle, thick, double, double distance=0.4mm] {\,};

    \node (f) at (0.0,-1.0) {$\big\downarrow$};
    \node (g) at (1.0,-1.0) {NPO};
    \node (h) at (2.0,-1.0) {$\big\downarrow$};
    \node (i) at (3.0,-1.0) {PO};
    \node (j) at (4.0,-1.0) {$\big\downarrow$};

    \node (k) at (0.0,-2.0) [draw, circle, thick, double, double distance=0.4mm] {\,};
    \node (l) at (1.0,-2.0) {$\leftarrow$};
    \node (m) at (2.0,-2.0) [draw, circle, thick, double, double distance=0.4mm] {\,};
    \node (n) at (3.0,-2.0) {$\rightarrow$};
    \node (o) at (4.0,-2.0) [draw, circle, thick, double, double distance=0.4mm] {\,};
\end{tikzpicture}}
\vspace{-0.4em}
\caption{Example Rooted Derivation}
\label{fig:egrd}
\end{figure}
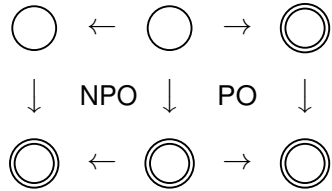

Graph transformation with relabelling as described in Sections \ref{sec:gtintro}, \ref{section:relabelling}, \ref{section:gt1} and \ref{section:gt2} has desirable properties. It was shown by Habel and Plump in 2002 \parencite{Habel-Plump02} that derivations are natural double pushouts (Theorem \ref{theorem:uniquederivations}) and thus are invertible. Unfortunately, Bak and Plump's modifications to add root nodes (Sections \ref{section:rooted} and \ref{section:rootedgt}) mean that derivations no longer exhibit these properties. That is, only the right square of a derivation in a rooted GT system need be a natural pushout (Figure \ref{fig:egrd}). This asymmetry is unfortunate, because derivations are no longer invertible.

\vspace{0.2em}
\vspace{-0.2em}

\vspace{-0.2em}
We propose an alternative theory for rooted graph transformation with relabelling, with some more desirable properties. Critically, we \textbf{restore invertibility} of derivations (Corollary \ref{cor:invertiblederivations}), and remove some undesirable matching cases (Lemma \ref{lem:rootedmatching}), allowing us to prove a handy root node invariance result (Corollary \ref{cor:invariantroots}).


\section{Graphs and Morphisms}

Fix some common label alphabet (Definition \ref{dfn:labelalphabet}) \(\mathcal{L} = (\mathcal{L}_V, \mathcal{L}_E)\). In this section we define our new notions of graphs and morphisms.

\begin{definition}
A \textbf{graph} over \(\mathcal{L}\) is a tuple \(G = (V, E, s, t, l, m, p)\) where:
\begin{enumerate}[itemsep=-0.6ex,topsep=-0.6ex]
\item \(V\) is a \textbf{finite} set of \textbf{vertices};
\item \(E\) is a \textbf{finite} set of \textbf{edges};
\item \(s: E \to V\) is a \textbf{total} source function;
\item \(t: E \to V\) is a \textbf{total} target function;
\item \(l: V \to \mathcal{L}_V\) is a \textbf{partial} function, labelling the vertices;
\item \(m: E \to \mathcal{L}_E\) is a \textbf{total} function, labelling the edges;
\item \(p: V \to \mathbb{Z}_2\)\footnote{\(\mathbb{Z}_2\) is the quotient \(\mathbb{Z}/2\mathbb{Z} = \{0, 1\}\).} is a \textbf{partial} function, determining vertex rootedness.
\end{enumerate}
\end{definition}

\begin{definition}
A graph \(G\) is \textbf{totally labelled} iff \(l_G\) is total, and \textbf{totally rooted} if \(p_G\) is total. If \(G\) is both, then we call it a \textbf{TLRG}.
\end{definition}

\begin{remark}
A totally rooted graph need not have every node a root node, only \(p_G\) must be total. \(0\) denotes unrooted, and \(1\) rooted. When we draw graphs, we shall denote the absence of rootedness with \textbf{diagonal stripes}. If a node has a \textbf{double border}, it is rooted, otherwise, it is unrooted.
\end{remark}

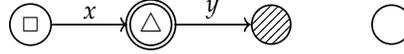
\begin{figure}[H]
\centering
\noindent
\scalebox{0.8}{\begin{tikzpicture}[every node/.style={inner sep=0pt, text width=6.5mm, align=center}]
    \node (a) at (0,0) [draw, circle, thick] {\(\Square\)};
    \node (b) at (2,0) [draw, circle, thick, double, double distance=0.4mm] {\(\triangle\)};
    \node (c) at (4,0) [draw, circle, thick, pattern=north east lines] {\,};
    \node (d) at (6,0) [draw, circle, thick] {\,};

    \draw (a) edge[->,thick] node[above, yshift=2.5pt] {\(x\)} (b)
          (b) edge[->,thick] node[above, yshift=2.5pt] {\(y\)} (c);
\end{tikzpicture}}
\vspace{-0.4em}
\caption{Example Graph}
\label{fig:eg-graph}
\end{figure}

\begin{example}
Let \(\mathcal{L} = (\{\Square, \triangle\}, \{x, y\})\). Then \(G = (V, E, s, t, l, m, p)\) is a graph over \(\mathcal{L}\) where \(V = \{1, 2, 3, 4\}\), \(E = \{1, 2\}\), \(s = \{(1, 1), (2, 2)\}\), \(t = \{(1, 2), (2, 3)\}\), \(l = \{(1, \Square), (2, \triangle)\}\), \(m = \{(1, x), (2, y)\}\), and \(p = \{(1, 0), (2, 1), (4, 0)\}\). Its graphical representation is shown in Figure \ref{fig:eg-graph}. \(G\) is neither totally rooted nor totally labelled, since node \(3\) has both undefined rootedness and no label, and node \(4\) also has no label.
\end{example}

\vspace{-0.6em}
\vspace{-0.4em}

\begin{definition}
A \textbf{graph morphism} between graphs \(G\) and \(H\) is a pair of functions \(g = (g_V: V_G \to V_H, g_E: E_G \to E_H)\) such that sources, targets, labels, and rootedness are preserved. That is:
\begin{enumerate}[itemsep=-0.6ex,topsep=-0.6ex]
\item \(\forall e \in E_G, \, g_V(s_G(e))= s_H(g_E(e))\);                 \hspace{2.185cm} [Sources]
\item \(\forall e \in E_G, \, g_V(t_G(e)) = t_H(g_E(e))\);                \hspace{2.230cm} [Targets]
\item \(\forall e \in E_G, \, m_G(e) = m_H(g_E(e))\);                     \hspace{2.709cm} [Edge Labels]
\item \(\forall v \in l_G^{-1}(\mathcal{L}_V), \, l_G(v) = l_H(g_V(v))\); \hspace{1.958cm} [Node Labels]
\item \(\forall v \in p_G^{-1}(\mathbb{Z}_2), \, p_G(v) = p_H(g_V(v))\).  \hspace{1.615cm} [Rootedness]
\end{enumerate}
\end{definition}

\begin{remark}
If \(G\) and \(H\) are \textbf{TLRG}s, then this is equivalent to the following diagram commuting (for \(s_G, s_H\) and \(t_G, t_H\) separately):

\vspace{-0.9em}
\vspace{0.2em}
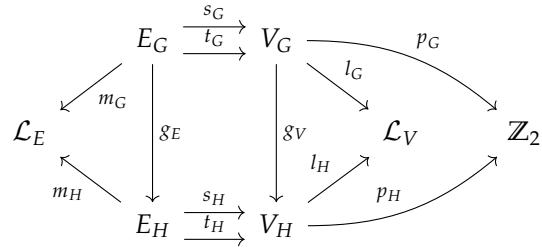
\begin{figure}[H]
\centering
\noindent
\begin{tikzpicture}
\node[scale=.85] (a) at (0,0){
\begin{tikzcd}
  \text{}
    & E_G \arrow[r, shift left=1ex, "s_G"] \arrow[r, shift right=1ex, "t_G"] \arrow[dd, "g_E"] \arrow[ld, "m_G"]
    & V_G \arrow[dd, "g_V"] \arrow[rd, "{l_G}"] \arrow[rrd, bend left=20, "{p_G}"]
    & \text{}
    & \text{} \\
  \mathcal{L}_E 
    & \text{}
    & \text{}
    & \mathcal{L}_V
    & \mathbb{Z}_2 \\
  \text{}
    & E_H \arrow[r, shift left=1ex, "s_H"] \arrow[r, shift right=1ex, "t_H"] \arrow[lu, "m_H"]
    & V_H \arrow[ru, "{l_H}"] \arrow[rru, bend right=20, "{p_H}"]
    & \text{}
    & \text{}
\end{tikzcd}
};
\end{tikzpicture}
\vspace{-0.8em}
\caption{Graph Morphism Commuting Diagrams}
\vspace{0.2em}
\end{figure}
\vspace{-0.2em}
\end{remark}

\begin{proposition}
Our new notion of graphs and morphisms is a \textbf{locally small category} (Definition \ref{def:cat}), just like the previous notions.
\end{proposition}

\begin{definition}
A graph morphism \(g: G \to H\) is \textbf{injective}/\textbf{surjective} iff the underlying functions \(g_V\), \(g_E\) are injective/surjective. We say that \(g\) is an \textbf{isomorphism} iff it is \textbf{injective} and \textbf{surjective}, and \(g^{-1}: H \to G\) is a graph morphism.
\end{definition}

\begin{definition}
We say \(H\) is a \textbf{subgraph} of \(G\) iff there exists an \textbf{inclusion morphism} \(H \hookrightarrow G\). This happens iff \(V_H \subseteq V_G\), \(E_H \subseteq E_G\), \(s_H = \restr{s_G}{E_H}\), \(t_H = \restr{t_G}{E_H}\), \(m_H = \restr{m_G}{E_H}\), \(l_H \subseteq l_G\), \(p_H \subseteq p_G\) (treating functions as sets).
\end{definition}

\begin{definition}
We say that graphs \(G, H\) are \textbf{isomorphic} iff there exists a \textbf{graph isomorphism} \(g: G \to H\). This gives \textbf{equivalence classes} \([G]\) over \(\mathcal{L}\). We denote by \(\mathcal{G}^{\varoplus}(\mathcal{L})\) the collection of totally labelled, totally rooted \textbf{abstract graphs} over some fixed \(\mathcal{L}\).
\end{definition}

\begin{proposition}
\(\mathcal{G}^{\varoplus}(\mathcal{L})\) is a \textbf{countable set}.
\end{proposition}

\begin{definition}
If \(G\) is a \textbf{graph}, then \(\abs{G} = \abs{V_G} + \abs{E_G}\).
\end{definition}


\section{Rules and Derivations}

Fixing some common \(\mathcal{L} = (\mathcal{L}_V, \mathcal{L}_E)\), we define rules and derivations.

\begin{definition}
A \textbf{rule} \(r = \langle L \leftarrow K \rightarrow R \rangle\) consists of left/right \textbf{TLRGs} \(L\), \(R\), the interface \textbf{graph} \(K\), and \textbf{inclusions} \(K \hookrightarrow L\) and \(K \hookrightarrow R\).
\end{definition}

\begin{example}
See Figure \ref{fig:tree2}.
\end{example}

\begin{definition}
We define the \textbf{inverse rule} to be \(r^{-1} = \langle R \leftarrow K \rightarrow L \rangle\).
\end{definition}

\begin{definition}
If \(r = \langle L \leftarrow K \rightarrow R \rangle\) is a \textbf{rule}, then \(\abs{r} = max \{\abs{L},\abs{R}\}\).
\end{definition}

\begin{definition}
Given a \textbf{rule} \(r = \langle L \leftarrow K \rightarrow R \rangle\) and a \textbf{TLRG} \(G\), we say that an \textbf{injective} morphism \(g: L \hookrightarrow G\) satisfies the \textbf{dangling condition} iff no edge in \(G \setminus g(L)\) is incident to a node in \(g(L \setminus K)\).
\end{definition}

\begin{samepage}
\begin{definition} \label{dfn:ruleapplication}
To \textbf{apply} a rule \(r = \langle L \leftarrow K \rightarrow R \rangle\) to some \textbf{TLRG} \(G\), find an \textbf{injective} graph morphism \(g: L \hookrightarrow G\) satisfying the \textbf{dangling condition}, then:

\begin{enumerate}[itemsep=-0.8ex,topsep=-0.8ex]
\item Delete \(g(L \setminus K)\) from \(G\). For each unlabelled node \(v\) in \(K\), make \(g_V(v)\) unlabelled, and for each node \(v\) in \(K\) with undefined rootedness, make \(g_V(v)\) have undefined rootedness, giving \textbf{intermediate graph} \(D\).
\item Add disjointly \(R \setminus K\) to D, keeping their labels and rootedness. For each unlabelled node \(v\) in \(K\), label \(g_V(v)\) with \(l_R(v)\), and for each node with undefined rootedness \(v\) in \(K\), make \(g_V(v)\) have rootedness \(p_R(v)\), giving the \textbf{result graph} \(H\).
\end{enumerate}

\noindent
If the \textbf{dangling condition} fails, then the rule is not applicable using the \textbf{match} \(g\). We can exhaustively check all matches to determine applicability.
\end{definition}

\begin{definition}
We write \(G \Rightarrow_{r,g} M\) for a successful application of \(r\) to \(G\) using match \(g\), obtaining result \(M \cong H\). We call this a \textbf{direct derivation}. We may omit \(g\) when it is not relevant, writing simply \(G \Rightarrow_{r} M\).
\end{definition}

\begin{definition}
For a given set of rules \(\mathcal{R}\), we write \(G \Rightarrow_{\mathcal{R}} H\) iff \(H\) is \textbf{directly derived} from \(G\) using any of the rules from \(\mathcal{R}\).
\end{definition}

\begin{definition}
We write \(G \Rightarrow_{\mathcal{R}}^{+} H\) iff \(H\) is \textbf{derived} from \(G\) in one or more \textbf{direct derivations}, and \(G \Rightarrow_{\mathcal{R}}^{*} H\) iff \(G \cong H\) or \(G \Rightarrow_{\mathcal{R}}^{+} H\).
\end{definition}
\vspace{-0.2em}
\end{samepage}


\section{Foundational Theorems}

We will show that gluing and deletions correspond to natural pushouts and natural pushout complements, respectively. Thus, derivations are invertible.

\vspace{-0.8em}
\begin{figure}[H]
\centering
\noindent
\begin{tikzpicture}
\node[scale=1.0] (a) at (0,0){
\begin{tikzcd}
  L \arrow[d, ""] \arrow[dr, phantom, "(1)"]
  & K \arrow[l, ""] \arrow[d, ""] \arrow[r, ""] \arrow[dr, phantom, "(2)"]
  & R \arrow[d, ""] \\
  G 
    & D \arrow[l, ""] \arrow[r, ""]
    & H
\end{tikzcd}
};
\end{tikzpicture}
\vspace{-0.8em}
\caption{Commuting Squares}
\label{fig:comsqrs}
\end{figure}
\vspace{-0.2em}

\vspace{-0.4em}
\begin{lemma} \label{lem:pullbacksexist}
Given \textbf{graph morphisms} \(g: L \to G\) and \(c: D \to G\), there exist a \textbf{graph} \(K\) and \textbf{graph morphisms} \(b: K \to L\), \(d: K \to D\) such that the resulting square is a \textbf{pullback} (Definition \ref{dfn:pullback}).
\end{lemma}

\vspace{0.4em}
\begin{proof}
The constructions are exactly as in Lemma 1 of \parencite{Habel-Plump02}, with the rootedness function defined analogously to the node labelling function. Trivial modifications to the proof give the result.
\end{proof}
\vspace{-0.2em}

\begin{lemma} \label{lem:pushoutsexist}
Let \(b : K \to R\), \(d: K \to D\) be \textbf{graph morphisms} such that \(d\) is \textbf{injective}, \(\forall v \in V_R, \abs{l_R(\{v\}) \cup l_D(d_V(b_V^{-1}(\{v\})))} \leq 1\), and \(\forall v \in V_R, \abs{p_R(\{v\}) \cup p_D(d_V(b_V^{-1}(\{v\})))} \leq 1\). Then, there exist a \textbf{graph} \(H\) and \textbf{graph morphisms} \(h: R \to H\), \(c: D \to H\) such that the resulting square is a \textbf{pushout} (Definition \ref{dfn:pushout}).
\end{lemma}

\vspace{0.4em}
\begin{proof}
The constructions are exactly as in Lemma 2 of \parencite{Habel-Plump02}, with the rootedness function defined analogously to the node labelling function.
\end{proof}
\vspace{-0.2em}

\begin{lemma}
Given two \textbf{graph morphisms} \(b: K \to L\) and \(d: K \to D\) such that \(b\) is \textbf{injective} and \(L\) is a \textbf{TLRG}, then the pushout (1) is \textbf{natural} (Definition \ref{dfn:npo}) iff \(l_D(d_V(V_K \setminus l_K^{-1}(\mathcal{L}_V))) = \emptyset = p_D(d_V(V_K \setminus p_K^{-1}(\mathbb{Z}_2)))\).
\end{lemma}

\vspace{0.4em}
\begin{proof}
Let square (1) in Figure \ref{fig:comsqrs} be a natural pushout with graph morphisms \(g: L \to G\) and \(c: D \to G\). Once again, we can proceed as in Lemma 3 of \parencite{Habel-Plump02} with the obvious modifications. Similar for the other direction.
\end{proof}
\vspace{-0.2em}

\begin{lemma}
Let \(g: L \to G\) be an \textbf{injective graph morphism} and \(K \to L\) an \textbf{inclusion morphism}. Then, there exist a \textbf{graph} \(D\) and \textbf{morphisms} \(K \to D\) and \(D \to G\) such that the square (1) is a \textbf{natural pushout} iff \(g\) satisfies the \textbf{dangling condition}. Moreover, in this case, \(D\) is unique up to isomorphism.
\end{lemma}

\vspace{0.4em}
\begin{proof}
Proceed as in Lemma 4 of \parencite{Habel-Plump02} with the obvious modifications.
\end{proof}
\vspace{-0.2em}

\begin{theorem}[Derivation Uniqueness] \label{thm:derivationtheorem}
Given a rule \(\langle L \leftarrow K \rightarrow R \rangle\) and an \textbf{injective graph morphism} \(g: L \to G\), then there exists a \textbf{natural DPO} diagram as above iff \(g\) satisfies the \textbf{dangling condition}. In this case, \(D\) and \(H\) are unique up to isomorphism. This exactly corresponds to Definition \ref{dfn:ruleapplication}. Moreover, if \(G \Rightarrow_r H\), then \(G\) is a \textbf{TLRG} iff \(H\) is a \textbf{TLRG}.
\end{theorem}

\newpage

\vspace{0.4em}
\begin{proof}
Proceed as in Theorem 1 of \parencite{Habel-Plump02} with the obvious modifications. Totality of labelling is given by Theorem 2 of \parencite{Habel-Plump02}, and totality of rootedness is given by replacing all occurrences of the labelling function with the rootedness function in the proof.
\end{proof}
\vspace{-0.2em}

\begin{corollary} \label{cor:invertiblederivations}
Derivations are \textbf{invertible}. That is \(G \Rightarrow_{r} H\) iff \(H \Rightarrow_{r^{-1}} G\).
\end{corollary}

\vspace{0.4em}
\begin{proof}
By the last theorem, \(G \Rightarrow_{r} H\) means we have a match \(g: L \to G\), and a comatch \(h: R \to H\), and so by symmetry, we have the result.
\end{proof}
\vspace{-0.2em}

\vspace{-0.6em}
This symmetry is unique to this new approach to rooted graph transformation. In Bak's approach (Appendix \ref{section:rootedgt}), derivations are not, in general, invertible (Figure \ref{fig:egrd}). In Bak's system, the intermediate graph \(D\) must not have a root if we want to invert the derivation. Finally, we can now show our root node invariance result:

\begin{lemma} \label{lem:rootedmatching}
Let \(G\) be a \textbf{TLRG}, and \(r = \langle L \leftarrow K \rightarrow R \rangle\) a rule. Then root nodes in \(L\) can only be matched against root nodes in \(G\), and similarly for non-root nodes.
\end{lemma}

\vspace{0.4em}
\begin{proof}
Immediate from the definitions.
\end{proof}
\vspace{-0.2em}

\vspace{-0.6em}
By comparison, in Bak's system, non-root nodes could be matched against root nodes.

\begin{corollary} \label{cor:invariantroots}
Let \(G\) be a \textbf{TLRG}, and \(r = \langle L \leftarrow K \rightarrow R \rangle\) a rule such that \(L\) and \(R\) both contain \(k\) root nodes, for some fixed \(k \in \mathbb{N}\). Then any \textbf{TLRG} \(H\) derived from \(G\) using \(r\) contains \(n\) root nodes iff \(G\) contains \(n\) root nodes.
\end{corollary}

\vspace{0.4em}
\begin{proof}
By Lemma \ref{lem:rootedmatching} (non-)roots in \(L\) can only be identified with (non-)roots in \(G\), and by symmetry the same for \(R\) in \(H\). By Theorem \ref{thm:derivationtheorem}, NDPO existence corresponds to Definition \ref{dfn:ruleapplication}, so, \(\abs{p_G^{-1}(\{1\})} = \abs{p_H^{-1}(\{1\})}\).
\end{proof}
\vspace{-0.2em}
\vspace{-1.6em}


\section{Equivalence of Rules} \label{sec:ruleequiv}

We now consider equivalence of rules, starting by formalising what it means to say that two rules are isomorphic, and then we will show that we can find a normal form for rules, unique up to isomorphism.

\begin{definition} \label{dfn:ruleiso}
Given \textbf{rules} \(r_1 = \langle L_1 \leftarrow K_1 \rightarrow R_1 \rangle\), \(r_2 = \langle L_2 \leftarrow K_2 \rightarrow R_2 \rangle\). We call \(r_1\) and \(r_2\) \textbf{isomorphic} iff there exists isomorphisms \(f: L_1\) \(\to L_2\), \(g: R_1 \to R_2\) such that \(\restr{f}{K_1} = \restr{g}{K_1}\) and \(f(K_1) = K_2\). Write \(r_1 \cong r_2\).
\end{definition}

\begin{proposition}
The above notion of \textbf{rule isomorphism} is an \textbf{equivalence}, and gives rise to \textbf{abstract rules} \([r]\).
\end{proposition}

\begin{definition}
Given a \textbf{rule} \(r = \langle L \leftarrow K \rightarrow R \rangle\), define its \textbf{normal form} \(r\!\!\downarrow  = \langle L \leftarrow K'\) \(\rightarrow R \rangle\) where \(K' = (V_K, \emptyset, \emptyset, \emptyset, \emptyset, \emptyset, \emptyset)\). We say two rules \(r_1\), \(r_2\) are \textbf{normalisation equivalent} iff \(r_1\!\!\downarrow \cong r_2\!\!\downarrow\). We write \(r_1 \simeq r_2\).
\end{definition}

\begin{proposition}
Clearly, this gives us a \textbf{coarser} notion of \textbf{equivalence} for rules than the notion of \textbf{isomorphism}.
\end{proposition}

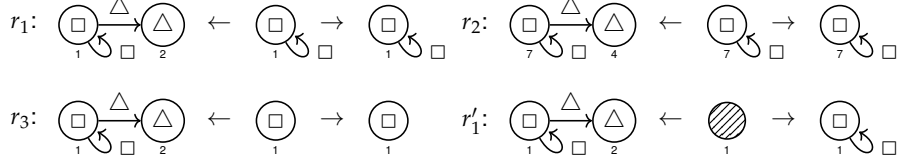
\begin{figure}[H]
\centering
\noindent
\scalebox{.75}{\begin{tikzpicture}[every node/.style={inner sep=0pt, text width=6.5mm, align=center}]
    \node (a) at (0.0,0) {$r_1$:};
    \node (b) at (1.0,0) [draw, circle, thick] {$\Square$};
    \node (c) at (2.5,0) [draw, circle, thick] {$\triangle$};
    \node (d) at (3.5,0) {$\leftarrow$};
    \node (e) at (4.5,0) [draw, circle, thick] {$\Square$};
    \node (f) at (5.5,0) {$\rightarrow$};
    \node (g) at (6.5,0) [draw, circle, thick] {$\Square$};

    \node (B) at (1.0,-.52) {\tiny{1}};
    \node (C) at (2.5,-.52) {\tiny{2}};
    \node (E) at (4.5,-.52) {\tiny{1}};
    \node (G) at (6.5,-.52) {\tiny{1}};

    \draw (b) edge[->,thick] node[above, yshift=2.5pt] {$\triangle$} (c)
          (b) edge[->,in=-30,out=-60,loop,thick] node[right, yshift=1.5pt] {$\Square$} (b)
          (e) edge[->,in=-30,out=-60,loop,thick] node[right, yshift=1.5pt] {$\Square$} (e)
          (g) edge[->,in=-30,out=-60,loop,thick] node[right, yshift=1.5pt] {$\Square$} (g);

    \node (h) at (8.0,0)  {$r_2$:};
    \node (i) at (9.0,0)  [draw, circle, thick] {$\Square$};
    \node (j) at (10.5,0) [draw, circle, thick] {$\triangle$};
    \node (k) at (11.5,0) {$\leftarrow$};
    \node (l) at (12.5,0) [draw, circle, thick] {$\Square$};
    \node (m) at (13.5,0) {$\rightarrow$};
    \node (n) at (14.5,0) [draw, circle, thick] {$\Square$};

    \node (I) at (9.0,-.52)  {\tiny{7}};
    \node (J) at (10.5,-.52) {\tiny{4}};
    \node (L) at (12.5,-.52) {\tiny{7}};
    \node (N) at (14.5,-.52) {\tiny{7}};

    \draw (i) edge[->,thick] node[above, yshift=2.5pt] {$\triangle$} (j)
          (i) edge[->,in=-30,out=-60,loop,thick] node[right, yshift=1.5pt] {$\Square$} (i)
          (l) edge[->,in=-30,out=-60,loop,thick] node[right, yshift=1.5pt] {$\Square$} (l)
          (n) edge[->,in=-30,out=-60,loop,thick] node[right, yshift=1.5pt] {$\Square$} (n);

    \node (a) at (0.0,-1.7) {$r_3$:};
    \node (b) at (1.0,-1.7) [draw, circle, thick] {$\Square$};
    \node (c) at (2.5,-1.7) [draw, circle, thick] {$\triangle$};
    \node (d) at (3.5,-1.7) {$\leftarrow$};
    \node (e) at (4.5,-1.7) [draw, circle, thick] {$\Square$};
    \node (f) at (5.5,-1.7) {$\rightarrow$};
    \node (i) at (6.5,-1.7) [draw, circle, thick] {$\Square$};

    \node (B) at (1.0,-2.22) {\tiny{1}};
    \node (C) at (2.5,-2.22) {\tiny{2}};
    \node (E) at (4.5,-2.22) {\tiny{1}};
    \node (I) at (6.5,-2.22) {\tiny{1}};

    \draw (b) edge[->,thick] node[above, yshift=2.5pt] {$\triangle$} (c)
          (b) edge[->,in=-30,out=-60,loop,thick] node[right, yshift=1.5pt] {$\Square$} (b);

    \node (j) at (8.0,-1.7)  {$r_1'$:};
    \node (k) at (9.0,-1.7)  [draw, circle, thick] {$\Square$};
    \node (l) at (10.5,-1.7) [draw, circle, thick] {$\triangle$};
    \node (m) at (11.5,-1.7) {$\leftarrow$};
    \node (n) at (12.5,-1.7) [draw, circle, thick, pattern=north east lines] {\,};
    \node (o) at (13.5,-1.7) {$\rightarrow$};
    \node (p) at (14.5,-1.7) [draw, circle, thick] {$\Square$};

    \node (K) at (9.0,-2.22)  {\tiny{1}};
    \node (L) at (10.5,-2.22) {\tiny{2}};
    \node (N) at (12.5,-2.22) {\tiny{1}};
    \node (P) at (14.5,-2.22) {\tiny{1}};

    \draw (k) edge[->,thick] node[above, yshift=2.5pt] {$\triangle$} (l)
          (k) edge[->,in=-30,out=-60,loop,thick] node[right, yshift=1.5pt] {$\Square$} (k)
          (p) edge[->,in=-30,out=-60,loop,thick] node[right, yshift=1.5pt] {$\Square$} (p);
\end{tikzpicture}}
\vspace{-0.7em}
\caption{Example (Non-)Isomorphic Rules}
\label{fig:ruleiso}
\end{figure}

\begin{example}
Consider the rules over \((\{\Square, \triangle\}, \{\Square, \triangle\})\) as given in Figure \ref{fig:ruleiso}. Clearly \(r_1\) and \(r_2\) are isomorphic, but \(r_3\) is not isomorphic to either. Rule \(r_1\) has normal form \(r_1'\).
\end{example}

\vspace{-1.2em}
\vspace{-0.6em}

\begin{theorem}[Well Behaved Derivations] \label{thm:wellbehavedderivations}
Given a \textbf{rule} \(r = \langle L \leftarrow K \rightarrow R \rangle\) and its normal form \(r\!\!\downarrow = \langle L \leftarrow K' \rightarrow R \rangle\), then for all TLRGs \(G\), \(H\), \(G \Rightarrow_{r} H\) iff \(G \Rightarrow_{r\!\downarrow} H\).
\end{theorem}

\vspace{-1.8em}
\begin{figure}[H]
\centering
\noindent
\begin{tikzpicture}[baseline= (a).base]
\node[scale=.7] (a) at (0,0){
\begin{tikzcd}[sep={1.2em,between origins}]
               &  &  & K' \arrow[rrrr] \arrow[llldd] \arrow[rdd] \arrow[dddd] &                                           &  &  & R \arrow[dddd] \\
               &  &  &                                                        &                                           &  &  &                \\
L \arrow[dddd] &  &  &                                                        & K \arrow[dddd] \arrow[llll] \arrow[rrruu] &  &  &                \\
               &  &  &                                                        &                                           &  &  &                \\
               &  &  & D' \arrow[rdd] \arrow[llldd] \arrow[rrrr]              &                                           &  &  & H              \\
               &  &  &                                                        &                                           &  &  &                \\
G              &  &  &                                                        & D \arrow[llll] \arrow[rrruu]              &  &  &               
\end{tikzcd}
};
\end{tikzpicture}
\vspace{-0.75em}
\caption{Derivations Diagram}
\end{figure}
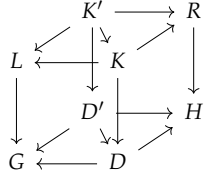
\vspace{-0.4em}

\vspace{0.4em}
\begin{proof}
Consider some fixed graph \(G\). The set of injective morphisms \(g: L \to G\) satisfying the dangling condition must be identical for both rules since \(L\) is the same and so is \(V_K\). Then, by the explicit construction of \(H\) given by Definition \ref{dfn:ruleapplication}, \(G \Rightarrow_{r,g} H\) iff \(G \Rightarrow_{r\!\downarrow,g} H\).
\end{proof}
\vspace{-0.2em}

\begin{remark}
Normal forms for rules is not actually a new observation, and is the foundation of rule schemata in GP\,2 \parencite{Plump11}. Moreover, maximising the number of edges in the interface of rules leads to a reduction of the number of critical pairs (Section \ref{section:critpairs}) of a GT system \parencite{HristakievPlump17}.
\end{remark}


\section{Transformation Systems}

We can now define graph transformation systems using our new definitions of graphs and rules. Next, we will look at equivalence and complexity.

\begin{definition}
A \textbf{graph transformation system} \(T = (\mathcal{L}, \mathcal{R})\), consists of a label alphabet \(\mathcal{L} = (\mathcal{L}_V, \mathcal{L}_E)\), and a \textbf{finite} set \(\mathcal{R}\) of rules over \(\mathcal{L}\).
\end{definition}

\begin{definition}
Given a \textbf{graph transformation system} \(T = (\mathcal{L}, \mathcal{R})\), we define the inverse system \(T^{-1} = (\mathcal{L}, \mathcal{R}^{-1})\) where \(\mathcal{R}^{-1} = \{r^{-1} \mid r \in \mathcal{R}\}\).
\end{definition}

\begin{definition}
Given a \textbf{graph transformation system} \(T = (\mathcal{L}, \mathcal{R})\), a subalphabet of \textbf{non-terminals} \(\mathcal{N}\), and a \textbf{start graph} \(S\) over \(\mathcal{L}\), then a \textbf{graph grammar} is the system \(\pmb{G} = (\mathcal{L}, \mathcal{N}, \mathcal{R}, S)\).
\end{definition}

\begin{definition}
Given a \textbf{graph grammar} \(\pmb{G}\) as defined above, we say that a graph \(G\) is \textbf{terminally labelled} iff \(l(V) \cap \mathcal{N}_V = \emptyset\) and \(m(E) \cap \mathcal{N}_E = \emptyset\). Thus, we can define the \textbf{graph language} generated by \(\pmb{G}\):
\begin{align*}
\pmb{L}(\pmb{G}) = \{[G] \mid S \Rightarrow_{\mathcal{R}}^{*} G, G \text{ terminally labelled}\}
\end{align*}
\end{definition}

\vspace{-0.8em}
\begin{theorem}[Membership Test]
Given a \textbf{grammar} \(\pmb{G} = (\mathcal{L}, \mathcal{N}, \mathcal{R}, S)\), \([G] \in \pmb{L}(\pmb{G})\) iff \(G \Rightarrow_{\mathcal{R}^{-1}}^* S\) and \(G\) is terminally labelled.
\end{theorem}

\begin{lemma} \label{lem:gtars}
A GT system \(T = (\mathcal{L}, \mathcal{R})\) induces a \textbf{decidable}, \textbf{finitely branching} ARS \((\mathcal{G}^{\varoplus}(\mathcal{L}), \rightarrow)\) where \([G] \rightarrow [H]\) iff \(G \Rightarrow_{\mathcal{R}} H\), just like in Section \ref{section:arsgt}. 
\end{lemma}

\begin{remark}
This does not, in general, imply that \(\rightarrow^*\) is decidable. We say that \(T\) is (\textbf{locally}) \textbf{confluent} (\textbf{terminating}) iff its \textbf{induced ARS} is.
\end{remark}


\section{Equivalence of GT Systems} \label{sec:gtequiv}

Building on the work from Section \ref{sec:ruleequiv}, we can ask when two graph transformations are equivalent, or rather, when they are distinct. We will give various notions of equivalence, and show there is a hierarchy of inclusion, as each notion is more and more general than the last.

\begin{definition}
Two \textbf{GT systems} \(T_1\), \(T_2\) over a common alphabet are:
\begin{enumerate}[itemsep=-0.6ex,topsep=-0.6ex]
\item \textbf{Isomorphic} (\(T_1 \cong T_2\)) iff \(\mathcal{R}_1 /\!\!\cong \,\,= \mathcal{R}_2 /\!\!\cong\); \footnote{This is a quotient (Definition \ref{dfn:quotient}) by rule isomorphism (Definition \ref{dfn:ruleiso}).}
\item \textbf{Normalisation equivalent} (\(T_1 \simeq_N T_2\)) iff \(\mathcal{R}_1 /\!\!\simeq \,\,= \mathcal{R}_2 /\!\!\simeq\);
\item \textbf{Step-wise equivalent} (\(T_1 \simeq_S T_2\)) iff the induced \textbf{ARSs}\footnote{Induced ARSs are as defined in Lemma \ref{lem:gtars}.} are identical;
\item \textbf{Semantically equivalent} (\(T_1 \simeq_F T_2\)) iff the \textbf{semantic functions} (modify Example \ref{eg:gtsemfunc} in the obvious way) are identical.
\end{enumerate}
\end{definition}

\begin{proposition}
Each of the above notions are equivalences.
\end{proposition}

\begin{proposition}
This notion of isomorphism gives rise to \textbf{abstract graph transformation systems} \([T]\) over some fixed label alphabet \(\mathcal{L}\). Let \(\mathcal{T}(\mathcal{L})\) denote the collection of all such classes. Then, \(\mathcal{T}(\mathcal{L})\) is a \textbf{countable set}.
\end{proposition}

\begin{remark}
Clearly \textbf{isomorphism} and \textbf{normalisation equivalence} are well behaved. That is, it is \textbf{decidable} to check if two GT systems share the same class. The same is not true of \textbf{semantic equivalence}.
\end{remark}

\begin{theorem}[GT System Equivalence] \label{thm:gtequiv}
GT system \textbf{isomorphism} is finer than \textbf{normalisation equivalence} is finer than \textbf{step-wise equivalence} is finer than \textbf{semantic equivalence}. Moreover, the inclusion is strict, in general. That is, \(T_1 \cong T_2 \Rightarrow T_1 \simeq_N T_2 \Rightarrow T_1 \simeq_S T_2 \Rightarrow T_1 \simeq_F T_2\).
\end{theorem}

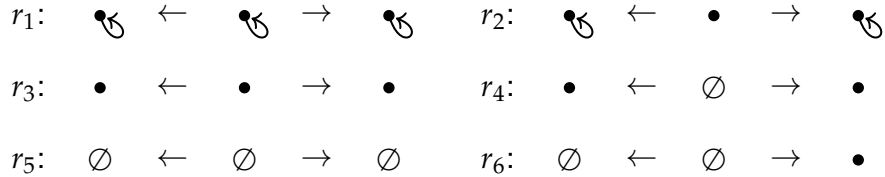
\begin{figure}[H]
\centering
\scalebox{.96}{\begin{tikzpicture}[every node/.style={align=center}]
    \node (a) at (0.0,-0.05) {$r_1$:};
    \node (b) at (1.0,0.0)   [draw, circle, thick, fill=black, scale=0.3] {\,};
    \node (c) at (2.0,0.0)   {$\leftarrow$};
    \node (d) at (3.0,0.0)   [draw, circle, thick, fill=black, scale=0.3] {\,};
    \node (e) at (4.0,0.0)   {$\rightarrow$};
    \node (f) at (5.0,0.0)   [draw, circle, thick, fill=black, scale=0.3] {\,};

    \node (g) at (6.5,-0.05) {$r_2$:};
    \node (h) at (7.5,0.0)   [draw, circle, thick, fill=black, scale=0.3] {\,};
    \node (i) at (8.5,0.0)   {$\leftarrow$};
    \node (j) at (9.5,0.0)   [draw, circle, thick, fill=black, scale=0.3] {\,};
    \node (k) at (10.5,0.0)  {$\rightarrow$};
    \node (l) at (11.5,0.0)  [draw, circle, thick, fill=black, scale=0.3] {\,};

    \draw (b) edge[->,in=-20,out=-70,loop,thick] (b)
          (d) edge[->,in=-20,out=-70,loop,thick] (d)
          (f) edge[->,in=-20,out=-70,loop,thick] (f);

    \draw (h) edge[->,in=-20,out=-70,loop,thick] (h)
          (l) edge[->,in=-20,out=-70,loop,thick] (l);

    \node (a) at (0.0,-1.05) {$r_3$:};
    \node (b) at (1.0,-1.0)  [draw, circle, thick, fill=black, scale=0.3] {\,};
    \node (c) at (2.0,-1.0)  {$\leftarrow$};
    \node (d) at (3.0,-1.0)  [draw, circle, thick, fill=black, scale=0.3] {\,};
    \node (e) at (4.0,-1.0)  {$\rightarrow$};
    \node (f) at (5.0,-1.0)  [draw, circle, thick, fill=black, scale=0.3] {\,};

    \node (g) at (6.5,-1.05) {$r_4$:};
    \node (h) at (7.5,-1.0)  [draw, circle, thick, fill=black, scale=0.3] {\,};
    \node (i) at (8.5,-1.0)  {$\leftarrow$};
    \node (j) at (9.5,-1.0)  {$\emptyset$};
    \node (k) at (10.5,-1.0) {$\rightarrow$};
    \node (l) at (11.5,-1.0) [draw, circle, thick, fill=black, scale=0.3] {\,};

    \node (a) at (0.0,-2.05) {$r_5$:};
    \node (b) at (1.0,-2.0)  {$\emptyset$};
    \node (c) at (2.0,-2.0)  {$\leftarrow$};
    \node (d) at (3.0,-2.0)  {$\emptyset$};
    \node (e) at (4.0,-2.0)  {$\rightarrow$};
    \node (f) at (5.0,-2.0)  {$\emptyset$};

    \node (g) at (6.5,-2.05) {$r_6$:};
    \node (h) at (7.5,-2.0)  {$\emptyset$};
    \node (i) at (8.5,-2.0)  {$\leftarrow$};
    \node (j) at (9.5,-2.0)  {$\emptyset$};
    \node (k) at (10.5,-2.0) {$\rightarrow$};
    \node (l) at (11.5,-2.0) [draw, circle, thick, fill=black, scale=0.3] {\,};
\end{tikzpicture}}
\vspace{-0.2em}
\caption{Example Rules Demonstrating Non-Equivalence}
\label{fig:gtequiv}
\end{figure}

\vspace{0.4em}
\begin{proof}
Let \(T_1\), \(T_2\) be GT systems over some \(\mathcal{L}\), with rule sets \(\mathcal{R}_1\), \(\mathcal{R}_2\). Within this proof, rules \(r_1, r_2, r_3, r_4, r_5, r_6\) can be found in Figure \ref{fig:gtequiv}. Suppose \(T_1 \cong T_2\). Then the \(\cong\)-classes of \(\mathcal{R}_1\) correspond to those of \(\mathcal{R}_2\). Clearly, if we find the normal form of each class, then the correspondence between these classes of normal forms is preserved. So \(T_1 \simeq_N T_2\). To see the inclusion is strict, consider the two systems \((\mathcal{L}, \{r_1\})\), \((\mathcal{L}, \{r_2\})\). They are non-isomorphic, but are normalisation equivalent.

Next suppose \(T_1 \simeq_N T_2\). Then by Theorem \ref{thm:wellbehavedderivations}, the choice of representative element from each class is irrelevant, that is, the derivations possible are identical. Now, since the \(\simeq_N\)-classes of \(\mathcal{R}_1\) and \(\mathcal{R}_2\) are identical, combining all possible derivations from the classes leaves us with identical possible derivations for each. Thus, it is immediate that the induced ARS is identical. To see the inclusion is strict, consider the two systems \((\mathcal{L}, \{r_3\})\), \((\mathcal{L}, \{r_3, r_4\})\). They are not normalisation equivalent, but are step-wise.

Finally, suppose \(T_1 \simeq_S T_2\). Then the induced ARS relations \(\rightarrow_{\mathcal{R}_1}, \rightarrow_{\mathcal{R}_2}\) are equal, so clearly \(f_{T_1} = f_{T_2}\). To see the inclusion is strict, consider the two systems \((\mathcal{L}, \{r_5\})\), \((\mathcal{L}, \{r_6\})\) are not step-wise equivalent since \(r_5\) is always applicable with no effect, but \(r_6\) is also always applicable, adding a new node. They are, however, semantically equivalent since their semantic functions both evaluate to \(\{\bot\}\) on all inputs.
\end{proof}
\vspace{-0.2em}

\vspace{-0.4em}
\vspace{0.2em}
\vspace{-0.2em}

In general, we might be interested in more than proving just equivalence. That is, when does one GT system \enquote{refine} the other. The (stepwise) refinement of programs was originally proposed by Dijkstra \parencite{Dijkstra68} \parencite{Dijkstra72} and Wirth \parencite{Wirth71}. Thinking in terms of GT systems, one may want to consider compatibility of the semantic function. Development of a refinement calculus that behaves properly with rooted GT systems remains open research.


\section{Complexity Theorems} \label{section:complexitythms}

The Graph Matching Problem (Definition \ref{dfn:gmp}) and Rule Application Problem (Definition \ref{dfn:rap}) can be considered in this our setting. When we say \enquote{bounded degree}, we mean the degree of each node has a constant upper bound. We will see that if we have an input graph with bounded degree and a bounded number of root nodes, and a finite set of \enquote{fast} rules, then we can perform matching in constant time.

\newpage

\begin{definition}
We call a rule \(r = \langle L \leftarrow K \rightarrow R \rangle\) \textbf{fast} iff every connected component (Definition \ref{dfn:connectedcomponent}) of \(L\) contains a root node.
\end{definition}

\vspace{-0.6em}
Just like in Lemma \ref{prop:gmptime}, we need to set up some assumptions about the complexity of various problems. We will again be using the assumptions from Figure \ref{fig:perfass}, assuming that the rootedness of any node can be accessed in constant time and that we can access the set of root nodes in a graph in \(O(\abs{X})\) time, given that there are \(\abs{X}\) root nodes.

\begin{lemma}
Given a \textbf{TLRG} \(G\) of \textbf{bounded degree} containing a \textbf{bounded} number of root nodes, and a \textbf{fast} rule \(r\), then the GMP (Definition \ref{dfn:gmp}) requires \(O(\abs{r})\) time and produces \(O(\abs{r})\) matches.
\end{lemma}

\vspace{0.4em}
\begin{proof}
Under the same assumption as in Dodds' Thesis \parencite[p. 39]{Dodds08}, this is easy to see, since there are only a constant number of subgraphs to consider. The full proof is a minor modification of Dodds' proof, with the major difference being the bounded number of root nodes in \(G\), allowing us to conclude \(O(\abs{r})\) time rather than \(O(\abs{V_G})\) time.
\end{proof}
\vspace{-0.2em}

\begin{lemma}
Given a \textbf{TLRG} \(G\) of \textbf{bounded degree}, a rule \(r\), and an \textbf{injective match} \(g\), then RAP (Definition \ref{dfn:rap}) requires \(O(\abs{r})\) time.
\end{lemma}

\vspace{0.4em}
\begin{proof}
Obvious modifications of the proofs in Dodds' Thesis.
\end{proof}
\vspace{-0.2em}

\begin{definition}
We say that a rule \(r = \langle L \leftarrow K \rightarrow R \rangle\) is \textbf{root non-increasing} iff \(\abs{p_L^{-1}(\{1\})} \geq \abs{p_R^{-1}(\{1\})}\).
\end{definition}

\begin{definition}
A rule \(r = \langle L \leftarrow K \rightarrow R \rangle\) is \textbf{degree non-increasing} iff \(\forall v \in (V_R \setminus V_K), \operatorname{deg}_R(v) \leq N\) and \(\forall v \in V_K, \operatorname{deg}_L(v) \geq \operatorname{deg}_R(v)\), where \(N\) is our upper bound on the degree of nodes.
\end{definition}

\begin{theorem}[Fast Derivations] \label{thm:fastderivations}
Given a \textbf{TLRG} \(G\) of \textbf{bounded degree} containing a \textbf{bounded} number of root nodes, and a GT system \(T = (\mathcal{L}, \mathcal{R})\) where each rule is \textbf{fast}, then one can decide in \textbf{constant time} the \textbf{direct successors} (Definition \ref{dfn:ars2}) of \(G\), up to isomorphism.
\end{theorem}

\vspace{0.4em}
\begin{proof}
Combine the above lemmas. There is a constant number of rules to apply. For each rule, a bounded number of matches are produced in constant time, and then the RAP takes constant time for each match.
\end{proof}
\vspace{-0.2em}

\begin{corollary} \label{corollary:lineargt}
Given \(G\), \(T\) as above, where each rule is additionally \textbf{root non-increasing} and \textbf{degree non-increasing}, and \(T\) terminating with maximum derivation length \(N \in \mathbb{N}\), then one can find a \textbf{normal form} (Definition \ref{dfn:ars2}) of \(G\) in \(O(N)\) time, up to isomorphism.
\end{corollary}

\vspace{0.4em}
\begin{proof}
By induction, the application of a rule satisfying the stated conditions will preserve the bound on the number of root nodes and the bound on the degree of the nodes. Thus, we have the result.
\end{proof}
\vspace{-0.2em}

Thus, we have shown that if we have a set of rules as per Corollary \ref{corollary:lineargt}, we need only consider the maximum length of derivations when reasoning about time complexity, as mentioned at the end of Section \ref{section:rooted}.

\chapter{Recognising Trees} \label{chapter:treerecognision}

The language of all unlabelled trees is well-known to be expressible using classical graph transformation systems, using a single rule. The question of recognising trees efficiently is less understood. We present a GT system that can test if a graph is a tree in linear time, given the input is of \enquote{bounded degree}: a new result for graph transformation systems.

We have submitted a version of this chapter for publication as part of a co-authored paper \parencite{Campbell-Courtehoute-Plump19} looking at linear time algorithms in GP\,2.


\section{Generating Trees}

Writing a graph grammar that generates all unlabelled trees (Definition \ref{dfn:tree}) is straightforward. Simply start with the trivial tree (a single node), and arbitrarily add edges pointing to a new node, away from this start node.

\begin{example}[Tree Grammar]
\vspace{0.2em}
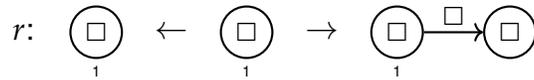
\begin{figure}[H]
\centering
\noindent
\begin{tikzpicture}[every node/.style={inner sep=0pt, text width=6.5mm, align=center}]
    \node (a) at (0.0,0) {$r$:};

    \node (b) at (1.0,0) [draw, circle, thick] {\(\Square\)};

    \node (c) at (2.0,0) {$\leftarrow$};

    \node (d) at (3.0,0) [draw, circle, thick] {\(\Square\)};

    \node (e) at (4.0,0) {$\rightarrow$};

    \node (f) at (5.0,0) [draw, circle, thick] {\(\Square\)};
    \node (g) at (6.5,0) [draw, circle, thick] {\(\Square\)};

    \node (B) at (1.0,-.52) {\tiny{1}};
    \node (D) at (3.0,-.52) {\tiny{1}};
    \node (F) at (5.0,-.52) {\tiny{1}};

    \draw (f) edge[->,thick] node[above, yshift=2.5pt] {\(\Square\)} (g);
\end{tikzpicture}
\vspace{-0.3em}
\caption{Tree Grammar Rules}
\label{fig:tree1}
\end{figure}
\vspace{-0.2em}

Let \(\pmb{TREE} = (\mathcal{L}, \mathcal{N}, S, \mathcal{R})\) where:
\begin{enumerate}[itemsep=-0.6ex,topsep=-0.6ex]
\item \(\mathcal{L} = (\{\Square\}, \{\Square\})\) where \(\Square\) denotes the empty label;
\item \(\mathcal{N} = (\emptyset, \emptyset)\);
\item \(S\) be the graph with a single node labelled with \(\Square\);
\item \(\mathcal{R} = \{r\}\).
\end{enumerate}

To see that this grammar generates the set of all trees, we must show that every graph in the language is a tree, and then that every tree is in the language. This is easy to see by induction.
\end{example}

Notice how the above construction has given us a decision procedure for testing if \([G] \in \pmb{L}(\pmb{TREE})\) (together with Proposition \ref{prop:inversegram}):

\begin{proposition}
\([G] \in \pmb{L}(\pmb{TREE})\) iff \(G \Rightarrow_{{r}^{-1}} S\). Moreover, this procedure always terminates, since the system is acyclic and globally finite.
\end{proposition}

It is easy to see via critical pair analysis (Section \ref{section:critpairs}) that this system is confluent, since it has no \enquote{critical pairs}. Unfortunately, it is not \enquote{fast} due to the fact that in each derivation, we must consider the entire host graph when finding a match. In the next session, we will see that rooted graph transformation rules can actually recognise trees in linear time.


\section{Linear Time Recognition} \label{sec:linearrec}

\vspace{-0.5em}
\vspace{0.2em}
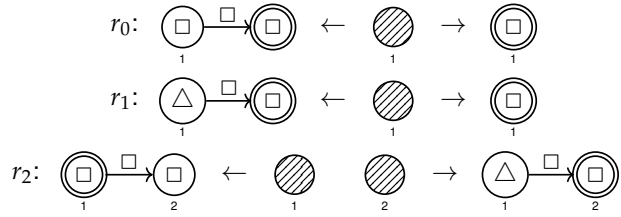
\begin{figure}[H]
\noindent
\centering
\scalebox{.8}{\begin{tikzpicture}[every node/.style={inner sep=0pt, text width=6.5mm, align=center}]
    \node (a) at (0.0,0) {$r_0$:};

    \node (b) at (1.0,0) [draw, circle, thick] {\(\Square\)};
    \node (c) at (2.5,0) [draw, circle, thick, double, double distance=0.4mm] {\(\Square\)};

    \node (d) at (3.5,0) {$\leftarrow$};

    \node (e) at (4.5,0) [draw, circle, thick, pattern=north east lines] {\,};

    \node (f) at (5.5,0) {$\rightarrow$};

    \node (g) at (6.5,0) [draw, circle, thick, double, double distance=0.4mm] {\(\Square\)};

    \node (B) at (1.0,-.52) {\tiny{1}};
    \node (E) at (4.5,-.52) {\tiny{1}};
    \node (G) at (6.5,-.52) {\tiny{1}};

    \draw (b) edge[->,thick] node[above, yshift=2.5pt] {\(\Square\)} (c);
\end{tikzpicture}}

\vspace{0.4em}

\scalebox{.8}{\begin{tikzpicture}[every node/.style={inner sep=0pt, text width=6.5mm, align=center}]
    \node (a) at (0.0,0) {$r_1$:};

    \node (b) at (1.0,0) [draw, circle, thick] {\(\triangle\)};
    \node (c) at (2.5,0) [draw, circle, thick, double, double distance=0.4mm] {\(\Square\)};

    \node (d) at (3.5,0) {$\leftarrow$};

    \node (e) at (4.5,0) [draw, circle, thick, pattern=north east lines] {\,};

    \node (f) at (5.5,0) {$\rightarrow$};

    \node (g) at (6.5,0) [draw, circle, thick, double, double distance=0.4mm] {\(\Square\)};

    \node (B) at (1.0,-.52) {\tiny{1}};
    \node (E) at (4.5,-.52) {\tiny{1}};
    \node (G) at (6.5,-.52) {\tiny{1}};

    \draw (b) edge[->,thick] node[above, yshift=2.5pt] {\(\Square\)} (c);
\end{tikzpicture}}

\vspace{0.4em}

\scalebox{.8}{\begin{tikzpicture}[every node/.style={inner sep=0pt, text width=6.5mm, align=center}]
    \node (a) at (0.0,0) {$r_2$:};

    \node (b) at (1.0,0) [draw, circle, thick, double, double distance=0.4mm] {\(\Square\)};
    \node (c) at (2.5,0) [draw, circle, thick] {\(\Square\)};

    \node (d) at (3.5,0) {$\leftarrow$};

    \node (e) at (4.5,0) [draw, circle, thick, pattern=north east lines] {\,};
    \node (f) at (6.0,0) [draw, circle, thick, pattern=north east lines] {\,};

    \node (g) at (7.0,0) {$\rightarrow$};

    \node (h) at (8.0,0) [draw, circle, thick] {\(\triangle\)};
    \node (i) at (9.5,0) [draw, circle, thick, double, double distance=0.4mm] {\(\Square\)};

    \node (B) at (1.0,-.52) {\tiny{1}};
    \node (C) at (2.5,-.52) {\tiny{2}};
    \node (E) at (4.5,-.52) {\tiny{1}};
    \node (F) at (6.0,-.52) {\tiny{2}};
    \node (H) at (8.0,-.52) {\tiny{1}};
    \node (I) at (9.5,-.52) {\tiny{2}};

    \draw (b) edge[->,thick] node[above, yshift=2.5pt] {\(\Square\)} (c)
          (h) edge[->,thick] node[above, yshift=2.5pt] {\(\Square\)} (i);
\end{tikzpicture}}
\vspace{-0.3em}
\caption{Tree Recognition Rules}
\label{fig:tree2}
\end{figure}
\vspace{-0.2em}

Let \(\mathcal{L} = (\{\Square, \triangle\}, \{\Square\})\), and \(\mathcal{R} = \{r_0, r_1, r_2\}\). We are going to show that \(\mathcal{R}\) induces a linear time algorithm for testing if a graph is a tree. Intuitively, this works by pushing a special node (a \enquote{root} node) to the bottom of a branch, and then pruning. If we start with a tree and run this until we cannot do it anymore, we must be left with a single node. The triangle labels are necessary so that, in the case that the input graph is not a tree, we could \enquote{get stuck} in a directed cycle.

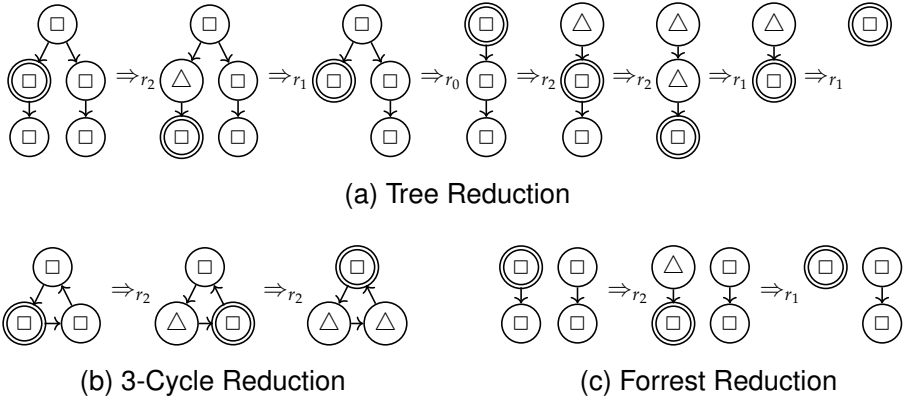
\begin{figure}[H]
\begin{subfigure}{1\textwidth}
    \centering
    \scalebox{.75}{\begin{tikzpicture}[every node/.style={inner sep=0pt, text width=6.5mm, align=center}]

\node (a) at (0,0)     [draw,circle,thick] {\(\Square\)};
\node (b) at (-0.5,-1) [draw,circle,thick, double, double distance=0.4mm] {\(\Square\)};
\node (c) at (0.5,-1)  [draw,circle,thick] {\(\Square\)};
\node (d) at (-0.5,-2) [draw,circle,thick] {\(\Square\)};
\node (e) at (0.5,-2)  [draw,circle,thick] {\(\Square\)};

\draw (a) edge[->,thick] (b)
      (a) edge[->,thick] (c)
      (b) edge[->,thick] (d)
      (c) edge[->,thick] (e);

\node (t) at (1.35,-1) {$\Rightarrow_{r_2}$};

\node (a) at (2.7,0)  [draw,circle,thick] {\(\Square\)};
\node (b) at (2.2,-1) [draw,circle,thick] {\(\triangle\)};
\node (c) at (3.2,-1) [draw,circle,thick] {\(\Square\)};
\node (d) at (2.2,-2) [draw,circle,thick, double, double distance=0.4mm] {\(\Square\)};
\node (e) at (3.2,-2) [draw,circle,thick] {\(\Square\)};

\draw (a) edge[->,thick] (b)
      (a) edge[->,thick] (c)
      (b) edge[->,thick] (d)
      (c) edge[->,thick] (e);

\node (t) at (4.05,-1) {$\Rightarrow_{r_1}$};

\node (a) at (5.4,0)  [draw,circle,thick] {\(\Square\)};
\node (b) at (4.9,-1) [draw,circle,thick, double, double distance=0.4mm] {\(\Square\)};
\node (c) at (5.9,-1) [draw,circle,thick] {\(\Square\)};
\node (e) at (5.9,-2) [draw,circle,thick] {\(\Square\)};

\draw (a) edge[->,thick] (b)
      (a) edge[->,thick] (c)
      (c) edge[->,thick] (e);

\node (t) at (6.75,-1) {$\Rightarrow_{r_0}$};

\node (a) at (7.6,0)    [draw,circle,thick, double, double distance=0.4mm] {\(\Square\)};
\node (c) at (7.6,-1) [draw,circle,thick] {\(\Square\)};
\node (e) at (7.6,-2) [draw,circle,thick] {\(\Square\)};

\draw (a) edge[->,thick] (c)
      (c) edge[->,thick] (e);

\node (t) at (8.45,-1) {$\Rightarrow_{r_2}$};

\node (a) at (9.3,0)  [draw,circle,thick] {\(\triangle\)};
\node (c) at (9.3,-1) [draw,circle,thick, double, double distance=0.4mm] {\(\Square\)};
\node (e) at (9.3,-2) [draw,circle,thick] {\(\Square\)};

\draw (a) edge[->,thick] (c)
      (c) edge[->,thick] (e);

\node (t) at (10.15,-1) {$\Rightarrow_{r_2}$};

\node (a) at (11.0,0)  [draw,circle,thick] {\(\triangle\)};
\node (c) at (11.0,-1) [draw,circle,thick] {\(\triangle\)};
\node (e) at (11.0,-2) [draw,circle,thick, double, double distance=0.4mm] {\(\Square\)};

\draw (a) edge[->,thick] (c)
      (c) edge[->,thick] (e);

\node (t) at (11.85,-1) {$\Rightarrow_{r_1}$};

\node (a) at (12.7,0)  [draw,circle,thick] {\(\triangle\)};
\node (c) at (12.7,-1) [draw,circle,thick, double, double distance=0.4mm] {\(\Square\)};

\draw (a) edge[->,thick] (c);

\node (t) at (13.55,-1) {$\Rightarrow_{r_1}$};

\node (a) at (14.4,0)  [draw,circle,thick, double, double distance=0.4mm] {\(\Square\)};

\end{tikzpicture}}
    \vspace{0.1em}
    \caption{Tree Reduction}
\end{subfigure}
\begin{subfigure}{.5\textwidth}
    \centering
    \vspace{2.0em}
    \scalebox{.75}{\begin{tikzpicture}[every node/.style={inner sep=0pt, text width=6.5mm, align=center}]

\node (a) at (0,0)     [draw,circle,thick] {\(\Square\)};
\node (b) at (-0.5,-1) [draw,circle,thick, double, double distance=0.4mm] {\(\Square\)};
\node (c) at (0.5,-1)  [draw,circle,thick] {\(\Square\)};

\draw (a) edge[->,thick] (b)
      (b) edge[->,thick] (c)
      (c) edge[->,thick] (a);

\node (t) at (1.35,-0.5) {$\Rightarrow_{r_2}$};

\node (a) at (2.7,0)  [draw,circle,thick] {\(\Square\)};
\node (b) at (2.2,-1) [draw,circle,thick] {\(\triangle\)};
\node (c) at (3.2,-1) [draw,circle,thick, double, double distance=0.4mm] {\(\Square\)};

\draw (a) edge[->,thick] (b)
      (b) edge[->,thick] (c)
      (c) edge[->,thick] (a);

\node (t) at (4.05,-0.5) {$\Rightarrow_{r_2}$};

\node (a) at (5.4,0)  [draw,circle,thick, double, double distance=0.4mm] {\(\Square\)};
\node (b) at (4.9,-1) [draw,circle,thick] {\(\triangle\)};
\node (c) at (5.9,-1) [draw,circle,thick] {\(\triangle\)};

\draw (a) edge[->,thick] (b)
      (b) edge[->,thick] (c)
      (c) edge[->,thick] (a);

\end{tikzpicture}}
    \vspace{0.1em}
    \caption{3-Cycle Reduction}
\end{subfigure}
\begin{subfigure}{.5\textwidth}
    \centering
    \vspace{2.0em}
    \scalebox{.75}{\begin{tikzpicture}[every node/.style={inner sep=0pt, text width=6.5mm, align=center}]

\node (a) at (-0.5,0)  [draw,circle,thick, double, double distance=0.4mm] {\(\Square\)};
\node (b) at (0.5,0)   [draw,circle,thick] {\(\Square\)};
\node (c) at (-0.5,-1) [draw,circle,thick] {\(\Square\)};
\node (d) at (0.5,-1)  [draw,circle,thick] {\(\Square\)};

\draw (a) edge[->,thick] (c)
      (b) edge[->,thick] (d);

\node (t) at (1.35,-0.5) {$\Rightarrow_{r_2}$};

\node (a) at (2.2,0)  [draw,circle,thick] {\(\triangle\)};
\node (b) at (3.2,0)  [draw,circle,thick] {\(\Square\)};
\node (c) at (2.2,-1) [draw,circle,thick, double, double distance=0.4mm] {\(\Square\)};
\node (d) at (3.2,-1) [draw,circle,thick] {\(\Square\)};

\draw (a) edge[->,thick] (c)
      (b) edge[->,thick] (d);

\node (t) at (4.05,-0.5) {$\Rightarrow_{r_1}$};

\node (a) at (4.9,0)  [draw,circle,thick, double, double distance=0.4mm] {\(\Square\)};
\node (b) at (5.9,0)  [draw,circle,thick] {\(\Square\)};
\node (d) at (5.9,-1) [draw,circle,thick] {\(\Square\)};

\draw (b) edge[->,thick] (d);

\end{tikzpicture}}
    \vspace{0.1em}
    \caption{Forrest Reduction}
\end{subfigure}
\vspace{0.6em}
\caption{Example Reductions}
\label{fig:tree-red}
\end{figure}

\begin{example}
Figure \ref{fig:tree-red} shows a reduction of a tree and non-trees.
\end{example}

\vspace{-1.5em}
\noindent
\vspace{0.2em}
\vspace{-0.2em}

\begin{definition}
Given a graph \(G\), we define \(G^{\ominus}\) to be exactly \(G\), but with every node unrooted, and everything labelled by \(\Square\). That is, \(G^{\ominus} = (V_G, E_G, s_G, t_G, V_G \times \{\Square\}, E_G \times \{\Square\}, V_G \times \{0\})\).
\end{definition}

\begin{definition}
By \enquote{input graph}, we mean any TLRG containing exactly one \enquote{root} node, with edges and vertices all labelled \(\Square\). By \enquote{input tree}, we mean an \enquote{input graph} that is also a tree (Definition \ref{dfn:tree}).
\end{definition}

\begin{lemma} \label{lem:treederlen}
The system \((\mathcal{L}, \mathcal{R})\) is terminating. Moreover, derivations have length at most \(2 \abs{V_G}\).
\end{lemma}

\vspace{0.4em}
\begin{proof}
Let \(\# G = \abs{V_G}\), \(\Square G = \abs{\{v \in V_G \mid l_G(v) = \Square\}}\), for any TLRG \(G\). If \(G \Rightarrow_{r_0} H\) or \(G \Rightarrow_{r_1} H\), then \(\# G > \# H\) and \(\Square G > \Square H\). If \(G \Rightarrow_{r_2} H\) then \(\# G = \# H\) and \(\Square G > \Square H\). Suppose there were an infinite sequence of derivations \(G_0 \Rightarrow_{\mathcal{R}} G_1 \Rightarrow_{\mathcal{R}} G_2 \Rightarrow_{\mathcal{R}} \cdots\), then there would be an infinite descending chain of natural numbers \(\# G_0 + \Square G_0 > \# G_1 + \Square G_1 > \# G_2 + \Square G_2 > \cdots\), which contradicts the well-ordering of \(\mathbb{N}\). To see the last part, notice that \(\Square G \leq \# G\) for all TLRGs \(G\), so the result is immediate since there are only \(2 \# G\) natural numbers less than \(2 \# G\).
\end{proof}
\vspace{-0.2em}

\begin{lemma} \label{lem:treegarbagesep}
If \(G\) is a tree and \(G \Rightarrow_{\mathcal{R}} H\), then \(H\) is a tree. If \(G\) is not a tree and \(G \Rightarrow_{\mathcal{R}} H\), then \(H\) is not a tree.
\end{lemma}

\vspace{0.4em}
\begin{proof}
Clearly, the application of \(r_2\) preserves structure. Suppose \(G\) is a tree. \(r_0\) or \(r_1\) are applicable iff node 2 is matched against a leaf node due to the dangling condition. Upon application, the leaf node and its incoming edge is removed. Clearly the result graph is still a tree.

If \(G\) is not a tree and one of \(r_0\) or \(r_1\) is applicable, then we can see the properties of not being a tree are preserved. That is, if \(G\) is not connected, \(H\) is certainly not connected. If \(G\) had parallel edges, due to the dangling condition, they must exist in \(G \setminus g(L)\), so \(H\) has parallel edges. Similarly, cycles are preserved. Finally, if \(G\) had a node with incoming degree greater than one, then \(H\) must too, since the node in \(G\) that is deleted in \(H\) had incoming degree one, and the degree of all other nodes is preserved.
\end{proof}
\vspace{-0.2em}

\begin{corollary} \label{corollary:treepreserving}
If \(G\) is an input graph and \(G \Rightarrow_{\mathcal{R}}^* H\), then \(G\) is a tree iff \(H\) is a tree.
\end{corollary}

\vspace{0.4em}
\begin{proof}
Induction.
\end{proof}
\vspace{-0.2em}

\begin{lemma} \label{lemma:oneroot}
If \(G\) is an input graph and \(G \Rightarrow_{\mathcal{R}}^{*} H\), then \(H\) has exactly one root node. Moreover, there is no derivation sequence that derives the empty graph.
\end{lemma}

\vspace{0.4em}
\begin{proof}
In each application of \(r_0, r_1, r_2\), the number of root nodes is invariant (Corollary \ref{cor:invariantroots}), and so the result holds by induction. To see that the empty graph cannot be derived, notice that each derivation reduces \(\# G\) by at most one, and no rules are applicable when \(\# G = 1\).
\end{proof}
\vspace{-0.2em}

\begin{remark}
In Bak's system (and hence GP\,2), Lemma \ref{lemma:oneroot} is still true, however a more direct proof is needed. Since the root node in the LHS of each rule must be matched against a root node in the host graph, so the other non-roots can only be matched against non-roots.
\end{remark}

\begin{lemma} \label{lem:trianglechains}
If \(G\) is an input graph and \(G \Rightarrow_{\mathcal{R}}^{*} H\). Then, every \(\triangle\)-node in \(H\) either has a child \(\triangle\)-node or a root-node child.
\end{lemma}

\vspace{0.4em}
\begin{proof}
Clearly \(G\) satisfies this, as there are no \(\triangle\)-nodes. We now proceed by induction. Suppose \(G \Rightarrow_{\mathcal{R}}^* H \Rightarrow_{\mathcal{R}} H'\) where \(H\) satisfies the condition. If \(r_0\) or \(r_1\) is applicable, we introduce no new \(\triangle\)-nodes. Additionally, in the case of \(r_1\), any \(\triangle\)-node parents of the image 1 are preserved. So \(H'\) satisfies the condition. Finally, if \(r_2\) is applied, then the new \(\triangle\)-node has a root-node child, and the \(\triangle\)-nodes in \(H' \setminus h(R)\) have the same children, so \(H'\) satisfies the condition.
\end{proof}
\vspace{-0.2em}

\begin{corollary} \label{lem:trianglechild}
Let \(G\) be an input tree and \(G \Rightarrow_{\mathcal{R}}^{*} H\). Then the root-node in \(H\) has no \(\triangle\)-node children.
\end{corollary}

\vspace{0.4em}
\begin{proof}
By Lemma \ref{lemma:oneroot}, \(H\) has exactly one root node, and by Lemma \ref{lem:trianglechains}, all chains of \(\triangle\)-nodes terminate with a root-node. If said root-node were to have a \(\triangle\)-node child, then we would have a cycle, which contradicts that \(H\) is a tree (Corollary \ref{corollary:treepreserving}).
\end{proof}
\vspace{-0.2em}

\begin{lemma} \label{lemma:treereduce}
Let \(G\) be an input tree and \(G \Rightarrow_{\mathcal{R}}^{*} H\). Then, either \(\abs{V_H} = 1\) or \(H\) is not in normal form.
\end{lemma}

\vspace{0.4em}
\begin{proof}
By Lemma \ref{lemma:oneroot}, \(\abs{V_H} \geq 1\). If \(\abs{V_G} = 1\), then \(G\) is in normal form. Otherwise, either the root node has no children, or it has at least one \(\Square\)-child. In the first case, \(r_0\) must be applicable, and in the second, \(r_2\).

Suppose \(G \Rightarrow_{\mathcal{R}}^{*} H\). If \(\abs{V_H} = 1\), then \(H\) is in normal form by the proof to Lemma \ref{lemma:oneroot}. Otherwise, by Corollary \ref{corollary:treepreserving} \(H\) is a tree and \(\abs{V_H} > 1\). Now, the root-node in \(H\) (Lemma \ref{lemma:oneroot}) must have a non-empty neighbourhood. If it has no children, then \(r_0\) or \(r_1\) must be applicable. Otherwise, \(r_2\) must be applicable, since by Corollary \ref{lem:trianglechild}, there must be a \(\Square\)-node child. So \(H\) is not in normal form.
\end{proof}
\vspace{-0.2em}

\vspace{-0.3em}
We now present the main result of this chapter:

\begin{theorem}[Tree Recognition] \label{thm:treerecthm}
Given an input graph \(G\), one may use the system \((\mathcal{L}, \mathcal{R})\) from \(G\) to find a normal form for \(G\), say \(H\). \(H\) is the single root-node graph labelled by \(\Square\) iff \([G^{\ominus}] \in \pmb{L}(\pmb{TREE})\). Moreover, for input graphs of bounded degree, we terminate in linear time.
\end{theorem}

\vspace{0.4em}
\begin{proof}
By Lemma \ref{lem:treederlen}, our system is terminating and derivations have maximum length \(2 \# G\). By Corollary \ref{corollary:treepreserving} and Lemma \ref{lemma:treereduce}, \(G\) is a tree iff we can derive the singleton tree without backtracking. Finally, by Corollary \ref{corollary:lineargt}, the algorithm terminates in linear time, since our ruleset satisfies the necessary conditions.
\end{proof}
\vspace{-0.2em}


\section{GP\,2 Implementation}

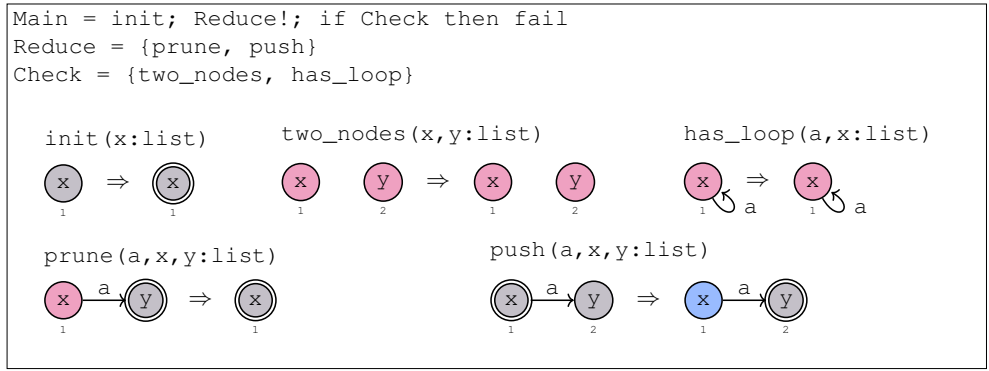
\begin{figure}[H]
\centering
\noindent
\scalebox{.73}{\fbox{\begin{minipage}{17.5cm}
\begin{allintypewriter}
Main = init; Reduce!; if Check then fail

Reduce = \{prune, push\}

Check = \{two\_nodes, has\_loop\}

\medskip
\setlength{\tabcolsep}{16pt}
\vspace{2.5mm}
\begin{tabular}{  p{3.2cm}  p{6.2cm}  p{4.6cm}  }
    
    \vspace{-1mm} init(x:list) & \vspace{-2mm} two\_nodes(x,y:list) & \vspace{-2mm} has\_loop(a,x:list) \\

    \vspace{-2mm}
    \adjustbox{valign=t}{\begin{tikzpicture}[every node/.style={inner sep=0pt, text width=6.5mm, align=center}]
        \node (a) at (0.0,0) [draw,circle,thick,fill=gp2grey] {x};

        \node (b) at (1.0,0) {$\Rightarrow$};
        
        \node (c) at (2.0,0) [draw,circle,thick,fill=gp2grey,double,double distance=0.4mm] {x};
        
        \node (A) at (0.0,-.52) {\tiny{1}};
        \node (C) at (2.0,-.52) {\tiny{1}};
        
    \end{tikzpicture}}
    &
    
    \vspace{-2mm}
    \adjustbox{valign=t}{\begin{tikzpicture}[every node/.style={inner sep=0pt, text width=6.5mm, align=center}]
        \node (a) at (0.0,0) [draw, circle, fill=gp2pink, thick] {x};
        \node (b) at (1.5,0) [draw, circle, fill=gp2pink, thick] {y};

        \node (c) at (2.5,0) {$\Rightarrow$};

        \node (d) at (3.5,0) [draw, circle, fill=gp2pink, thick] {x};
        \node (e) at (5.0,0) [draw, circle, fill=gp2pink, thick] {y};

        \node (A) at (0.0,-.52) {\tiny{1}};
        \node (B) at (1.5,-.52) {\tiny{2}};
        \node (D) at (3.5,-.52) {\tiny{1}};
        \node (E) at (5.0,-.52) {\tiny{2}};
    \end{tikzpicture}}
    &
    
    \vspace{-2mm}
    \adjustbox{valign=t}{\begin{tikzpicture}[every node/.style={inner sep=0pt, text width=6.5mm, align=center}]
        \node (a) at (0.0,0) [draw,circle,thick,fill=gp2pink] {x};
        
        \node (b) at (1.0,0) {$\Rightarrow$};
        
        \node (c) at (2.0,0) [draw,circle,thick,fill=gp2pink] {x};
        
        \node (A) at (0.0,-.52) {\tiny{1}};
        \node (C) at (2.0,-.52) {\tiny{1}};
        
        \draw (a) edge[->,in=-30,out=-60,loop,thick] node[right, yshift=1.5pt] {a} (a)
              (c) edge[->,in=-30,out=-60,loop,thick] node[right, yshift=1.5pt] {a} (c);
    \end{tikzpicture}}
    \\
\end{tabular}
\begin{tabular}{  p{7cm}  p{7cm}  }

    \vspace{-1mm} prune(a,x,y:list) & \vspace{-2mm} push(a,x,y:list) \\
    
    \vspace{-2mm}
    \adjustbox{valign=t}{\begin{tikzpicture}[every node/.style={inner sep=0pt, text width=6.5mm, align=center}]
        \node (a) at (0.0,0) [draw,circle,fill=gp2pink,thick] {x};
        \node (b) at (1.5,0) [draw,circle,fill=gp2grey,thick, double, double distance=0.4mm] {y};
        
        \node (c) at (2.5,0) {$\Rightarrow$};
        
        \node (d) at (3.5,0) [draw,circle,fill=gp2grey,thick, double, double distance=0.4mm] {x};
        
        \node (A) at (0.0,-.52) {\tiny{1}};
        \node (D) at (3.5,-.52) {\tiny{1}};
        
        \draw (a) edge[->,thick] node[above, yshift=2.5pt] {a} (b);
    \end{tikzpicture}}
    
    & 

    \vspace{-2mm}
    \adjustbox{valign=t}{\begin{tikzpicture}[every node/.style={inner sep=0pt, text width=6.5mm, align=center}]
    \node (a) at (0.0,0)     [draw, circle, fill=gp2grey, thick, double, double distance=0.4mm] {x};
    \node (b) at (1.5,0)     [draw, circle, fill=gp2grey, thick] {y};
    
    \node (c) at (2.5,0)     {$\Rightarrow$};
    
    \node (d) at (3.5,0)     [draw, circle, fill=gp2blue, thick] {x};
    \node (e) at (5,0)       [draw, circle, fill=gp2grey, thick, double, double distance=0.4mm] {y};
    
    \node (A) at (0.0,-.52) {\tiny{1}};
    \node (B) at (1.5,-.52) {\tiny{2}};
    \node (D) at (3.5,-.52) {\tiny{1}};
    \node (E) at (5.0,-.52) {\tiny{2}};
    
    \draw (a) edge[->,thick] node[above, yshift=2.5pt] {a} (b)
          (d) edge[->,thick] node[above, yshift=2.5pt] {a} (e);
    \end{tikzpicture}}
    \\
\end{tabular}
\end{allintypewriter}
\end{minipage}
}}
\caption{GP\,2 Implementation}
\label{fig:gp2-impl}
\end{figure}

Our algorithm can be implemented in GP\,2. The program (Figure \ref{fig:gp2-impl}) expects an arbitrary labelled input graph with every node coloured grey, no \enquote{root} nodes, and no additional \enquote{marks}. It will fail iff the input is not a tree. Given an input graph of bounded degree, it will always terminate in linear time with respect to (w.r.t.) the number of nodes in the input graph.

To see that the program is correct follows mostly from our existing proofs. Grey nodes encode the \(\Square\) label, and blue nodes, \(\triangle\). The \enquote{init} rule will fail if the input graph is empty, otherwise, it will make exactly one node rooted, in at most linear time. The \enquote{Reduce!} step is then exactly our previous GT system, which we have shown to be correct, and terminates in linear time. Finally, the \enquote{Check} step checks for garbage in linear time. There is no need to check the host graph is not equal to the empty graph (Lemma \ref{lemma:oneroot}).

\vspace{0.4em}
\vspace{-0.2em}

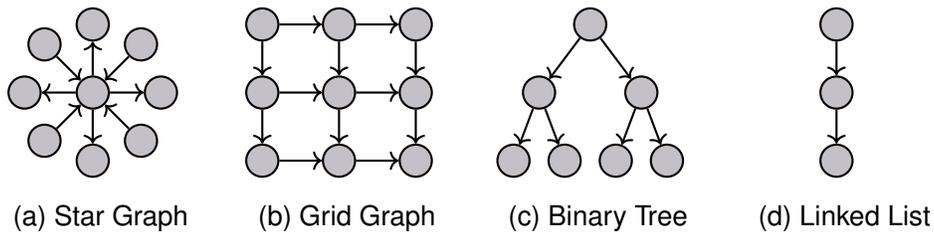
\begin{figure}[H]
\begin{subfigure}{.24\textwidth}
    \centering
    \begin{tikzpicture}[scale=0.68]
    \node (a) at (0.000,0.000)   [draw,circle,thick,fill=gp2grey] {\,};
    \node (b) at (0.000,1.333)   [draw,circle,thick,fill=gp2grey] {\,};
    \node (c) at (0.943,0.943)   [draw,circle,thick,fill=gp2grey] {\,};
    \node (d) at (1.333,0.000)   [draw,circle,thick,fill=gp2grey] {\,};
    \node (e) at (0.943,-0.943)  [draw,circle,thick,fill=gp2grey] {\,};
    \node (f) at (0.000,-1.333)  [draw,circle,thick,fill=gp2grey] {\,};
    \node (g) at (-0.943,-0.943) [draw,circle,thick,fill=gp2grey] {\,};
    \node (h) at (-1.333,0.000)  [draw,circle,thick,fill=gp2grey] {\,};
    \node (i) at (-0.943,0.943)  [draw,circle,thick,fill=gp2grey] {\,};
    
    \draw (a) edge[->, thick] (b)
          (c) edge[->, thick] (a)
          (a) edge[->, thick] (d)
          (e) edge[->, thick] (a)
          (a) edge[->, thick] (f)
          (g) edge[->, thick] (a)
          (a) edge[->, thick] (h)
          (i) edge[->, thick] (a);
\end{tikzpicture}
    \vspace{0.2em}
    \caption{Star Graph}
\end{subfigure}
\begin{subfigure}{.25\textwidth}
    \centering
    \begin{tikzpicture}[scale=0.68]
    \node (a) at (-1.500,1.333)  [draw,circle,thick,fill=gp2grey] {\,};
    \node (b) at (0.000,1.333)   [draw,circle,thick,fill=gp2grey] {\,};
    \node (c) at (1.500,1.333)   [draw,circle,thick,fill=gp2grey] {\,};
    \node (d) at (-1.500,0.000)  [draw,circle,thick,fill=gp2grey] {\,};
    \node (e) at (0.000,0.000)   [draw,circle,thick,fill=gp2grey] {\,};
    \node (f) at (1.500,0.000)   [draw,circle,thick,fill=gp2grey] {\,};
    \node (g) at (-1.500,-1.333) [draw,circle,thick,fill=gp2grey] {\,};
    \node (h) at (0.000,-1.333)  [draw,circle,thick,fill=gp2grey] {\,};
    \node (i) at (1.500,-1.333)  [draw,circle,thick,fill=gp2grey] {\,};
    
    \draw (a) edge[->, thick] (b)
          (a) edge[->, thick] (d)
          (b) edge[->, thick] (c)
          (b) edge[->, thick] (e)
          (c) edge[->, thick] (f)
          (d) edge[->, thick] (e)
          (d) edge[->, thick] (g)
          (e) edge[->, thick] (f)
          (e) edge[->, thick] (h)
          (f) edge[->, thick] (i)
          (g) edge[->, thick] (h)
          (h) edge[->, thick] (i);
\end{tikzpicture}
    \vspace{0.2em}
    \caption{Grid Graph}
\end{subfigure}
\begin{subfigure}{.25\textwidth}
    \centering
    \begin{tikzpicture}[scale=0.68]
    \node (a) at (0.000,1.333)   [draw,circle,thick,fill=gp2grey] {\,};
    \node (b) at (1.000,0.000)   [draw,circle,thick,fill=gp2grey] {\,};
    \node (c) at (-1.000,0.000)  [draw,circle,thick,fill=gp2grey] {\,};
    \node (d) at (1.500,-1.333)  [draw,circle,thick,fill=gp2grey] {\,};
    \node (e) at (0.500,-1.333)  [draw,circle,thick,fill=gp2grey] {\,};
    \node (f) at (-0.500,-1.333) [draw,circle,thick,fill=gp2grey] {\,};
    \node (g) at (-1.500,-1.333) [draw,circle,thick,fill=gp2grey] {\,};
    
    \draw (a) edge[->, thick] (b)
          (a) edge[->, thick] (c)
          (b) edge[->, thick] (d)
          (b) edge[->, thick] (e)
          (c) edge[->, thick] (f)
          (c) edge[->, thick] (g);
\end{tikzpicture}
    \vspace{0.2em}
    \caption{Binary Tree}
\end{subfigure}
\begin{subfigure}{.24\textwidth}
    \centering
    \begin{tikzpicture}[scale=0.68]
    \node (a) at (0.000,1.333)  [draw,circle,thick,fill=gp2grey] {\,};
    \node (b) at (0.000,0.000)  [draw,circle,thick,fill=gp2grey] {\,};
    \node (c) at (0.000,-1.333) [draw,circle,thick,fill=gp2grey] {\,};
    
    \draw (a) edge[->, thick] (b)
          (b) edge[->, thick] (c);
\end{tikzpicture}
    \vspace{0.2em}
    \caption{Linked List}
\end{subfigure}
\vspace{0.8em}
\caption{Graph Classes}
\label{fig:graph-types}
\end{figure}

We have performed empirical benchmarking to verify the complexity of the program, testing it with linked lists, binary trees, grid graphs, and star graphs (Figure \ref{fig:graph-types}). Formal definitions of each of these graph classes can be found in Section \ref{sec:graphclasses}. We have exclusively used \enquote{perfect} binary trees, and \enquote{square} grid graphs in our testing.

\noindent
\vspace{-1.2em}
\vspace{-0.2em}

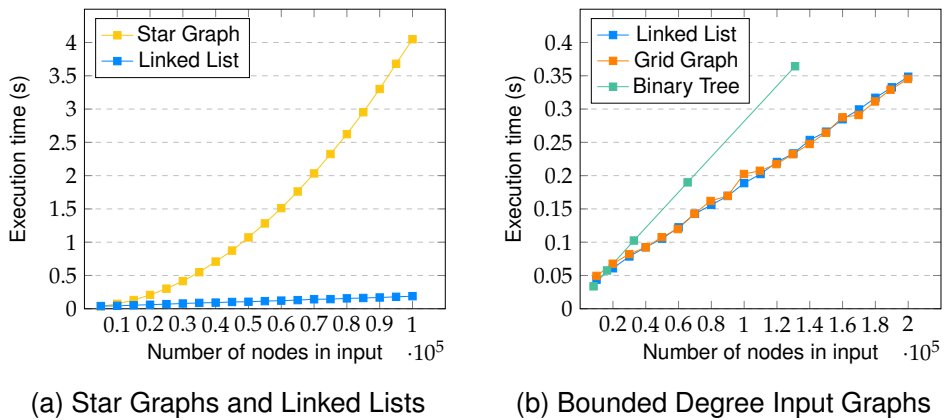
\begin{figure}[H]
\begin{subfigure}{.5\textwidth}
    \centering
    \begin{tikzpicture}[scale=0.7]
\begin{axis}[
xlabel={Number of nodes in input},
ylabel={Execution time (s)},
xmin=0, xmax=110000,
ymin=0, ymax=4.5,
xtick={10000,20000,30000,40000,50000,60000,70000,80000,90000,100000},
ytick={0.0,0.5,1.0,1.5,2.0,2.5,3.0,3.5,4.0},
legend pos=north west,
ymajorgrids=true,
grid style=dashed,
yticklabel style={/pgf/number format/fixed},
]
\addplot[color=performanceYellow, mark=square*] 
coordinates {
    (5000,0.03734)
    (10000,0.07405)
    (15000,0.13061)
    (20000,0.20765)
    (25000,0.30175)
    (30000,0.41571)
    (35000,0.54940)
    (40000,0.70865)
    (45000,0.87360)
    (50000,1.07186)
    (55000,1.28239)
    (60000,1.51184)
    (65000,1.76158)
    (70000,2.03321)
    (75000,2.32459)
    (80000,2.62288)
    (85000,2.95250)
    (90000,3.30216)
    (95000,3.67980)
    (100000,4.04914)
};
\addplot[color=performanceBlue, mark=square*] 
coordinates {
    (5000,0.04032)
    (10000,0.04375)
    (15000,0.05350)
    (20000,0.06085)
    (25000,0.06980)
    (30000,0.07863)
    (35000,0.08824)
    (40000,0.09247)
    (45000,0.10207)
    (50000,0.10537)
    (55000,0.11482)
    (60000,0.12228)
    (65000,0.13004)
    (70000,0.14266)
    (75000,0.14482)
    (80000,0.15604)
    (85000,0.16180)
    (90000,0.16944)
    (95000,0.17961)
    (100000,0.18871)
};
\addlegendentry{Star Graph}
\addlegendentry{Linked List}
\end{axis}
\end{tikzpicture}
    \caption{Star Graphs and Linked Lists}
\end{subfigure}
\begin{subfigure}{.5\textwidth}
    \centering
    \begin{tikzpicture}[scale=0.7]
\begin{axis}[
xlabel={Number of nodes in input},
ylabel={Execution time (s)},
xmin=0, xmax=220000,
ymin=0, ymax=0.45,
xtick={20000,40000,60000,80000,100000,120000,140000,160000,180000,200000},
ytick={0.00,0.05,0.10,0.15,0.20,0.25,0.30,0.35,0.40},
legend pos=north west,
ymajorgrids=true,
grid style=dashed,
yticklabel style={/pgf/number format/fixed},
]
\addplot[color=performanceBlue, mark=square*] 
coordinates {
    (10000,0.04375)
    (20000,0.06085)
    (30000,0.07863)
    (40000,0.09247)
    (50000,0.10537)
    (60000,0.12228)
    (70000,0.14266)
    (80000,0.15604)
    (90000,0.16944)
    (100000,0.18871)
    (110000,0.20241)
    (120000,0.22049)
    (130000,0.23322)
    (140000,0.25353)
    (150000,0.26601)
    (160000,0.28470)
    (170000,0.29935)
    (180000,0.31694)
    (190000,0.33262)
    (200000,0.34859)
};
\addplot[color=orange, mark=square*] 
coordinates {
    (10000,0.04936)
    (19881,0.06766)
    (29929,0.08206)
    (40000,0.09235)
    (49729,0.10750)
    (59536,0.11979)
    (69696,0.14284)
    (79524,0.16192)
    (90000,0.16971)
    (99856,0.20246)
    (109561,0.20725)
    (119716,0.21733)
    (129600,0.23215)
    (139876,0.24756)
    (149769,0.26442)
    (160000,0.28757)
    (169744,0.29114)
    (179776,0.31145)
    (189225,0.32908)
    (199809,0.34526)
};
\addplot[color=gp2green, mark=square*] 
coordinates {
    (8191,0.03370)
    (16383,0.05749)
    (32767,0.10249)
    (65535,0.18989)
    (131071,0.36431)
};
\addlegendentry{Linked List}
\addlegendentry{Grid Graph}
\addlegendentry{Binary Tree}
\end{axis}
\end{tikzpicture}
    \caption{Bounded Degree Input Graphs}
\end{subfigure}
\vspace{0.8em}
\caption{Measured Performance}
\label{fig:tree-bench}
\end{figure}

\vspace{-0.2em}
Star Graphs are not of bounded degree, so we saw quadratic time complexity as expected. The other graphs are of bounded degree, thus we observed linear time complexity (Figure \ref{fig:tree-bench}).

\vspace{0.4em}
\vspace{0.2em}
\vspace{-0.2em}

\chapter{Confluence Analysis} \label{chapter:confluenceanalysis}

Efficient testing of language membership is an important problem in graph transformation \parencite{ArnborgCourcelleBrunoSeese93} \parencite{BodlaenderFluiter01} \parencite{Plump10}. Our GT system for testing if a graph is a tree is actually not confluent, but if the input is a tree, then it has exactly one normal form, so it was in some sense confluent. We can formalise this with the new notion of \enquote{confluence modulo garbage}. The name is attributed to Plump, however it appears in no published work.


\section{Confluence Modulo Garbage}

In this section, we shall be working with standard GT systems, as defined in Appendix \ref{appendix:transformation}, but without relabelling. That is, all graphs are totally labelled, including interface graphs. All the results in this section will actually generalise to systems with relabelling, or the systems defined in Chapter \ref{chapter:newtheory}.

\begin{definition}
Let \(T = (\mathcal{L}, \mathcal{R})\) be a GT system, and \(D \subseteq \mathcal{G}(\mathcal{L})\) be a set of abstract graphs. Then, a graph \(G\) is called \textbf{garbage} iff \([G] \not\in D\).
\end{definition}

\begin{definition}
Let \(T = (\mathcal{L}, \mathcal{R})\), and \(D \subseteq \mathcal{G}(\mathcal{L})\). \(T\) is \textbf{weakly garbage separating} with respect to \(D\) iff for all \(G\), \(H\) such that \(G \Rightarrow_\mathcal{R} H\), if \([G] \in D\) then \([H] \in D\). \(T\) is \textbf{garbage separating} iff we have \([G] \in D\) iff \([H] \in D\).
\end{definition}

\vspace{-0.4em}
This set of abstract graphs \(D\) represents the \enquote{good input}, and the \enquote{garbage} is the graphs that are not in this set. \(D\) need not be explicitly generated by a graph grammar. For example, it could be defined by some (monadic second-order \parencite{Courcelle88}) logical formula.

There are a couple of immediately obvious results:

\begin{proposition} \label{lem:gsimpliesweak}
\textbf{Garbage separation} \(\Rightarrow\) \textbf{weak garbage separation}.
\end{proposition}

\begin{proposition} \label{lem:wgsclosure}
Given \(T = (\mathcal{L}, \mathcal{R})\) \textbf{weakly garbage separating} with respect to \(D \subseteq \mathcal{G}(\mathcal{L})\), then for all graphs \(G\), \(H\) such that \(G \Rightarrow_\mathcal{R}^* H\), if \([G] \in D\), then \([H] \in D\).
\end{proposition}

\begin{figure}[H]
\centering
\noindent
\scalebox{.8}{\begin{tikzpicture}[every node/.style={align=center}]
    \node (a) at (0.0,-0.05) {$r_1$:};
    \node (b) at (1.0,0.0)   [draw, circle, thick, fill=black, scale=0.3] {\,};
    \node (c) at (2.0,0.0)   [draw, circle, thick, fill=black, scale=0.3] {\,};
    \node (d) at (3.0,0.0)   {$\leftarrow$};
    \node (e) at (4.0,0.0)   [draw, circle, thick, fill=black, scale=0.3] {\,};
    \node (f) at (5.0,0.0)   [draw, circle, thick, fill=black, scale=0.3] {\,};
    \node (g) at (6.0,0.0)   {$\rightarrow$};
    \node (h) at (7.0,0.0)   [draw, circle, thick, fill=black, scale=0.3] {\,};
    \node (i) at (8.0,0.0)   [draw, circle, thick, fill=black, scale=0.3] {\,};

    \node (j) at (9.5,-0.05) {$r_2$:};
    \node (k) at (10.5,0.0)  [draw, circle, thick, fill=black, scale=0.3] {\,};
    \node (l) at (11.5,0.0)  [draw, circle, thick, fill=black, scale=0.3] {\,};
    \node (m) at (12.5,0.0)  {$\leftarrow$};
    \node (n) at (13.5,0.0)  [draw, circle, thick, fill=black, scale=0.3] {\,};
    \node (o) at (14.5,0.0)  {$\rightarrow$};
    \node (p) at (15.5,0.0)  [draw, circle, thick, fill=black, scale=0.3] {\,};

    \node (B) at (1.0,-.18)  {\tiny{1}};
    \node (C) at (2.0,-.18)  {\tiny{2}};
    \node (E) at (4.0,-.18)  {\tiny{1}};
    \node (F) at (5.0,-.18)  {\tiny{2}};
    \node (H) at (7.0,-.18)  {\tiny{1}};
    \node (I) at (8.0,-.18)  {\tiny{2}};
    \node (K) at (10.5,-.18) {\tiny{1}};
    \node (N) at (13.5,-.18) {\tiny{1}};
    \node (P) at (15.5,-.18) {\tiny{1}};

    \draw (b) edge[->,thick] (c)
          (k) edge[->,thick] (l);
\end{tikzpicture}}
\vspace{-1.3em}
\caption{Example Reduction Rules}
\label{fig:eg-reduct-rules}
\end{figure}
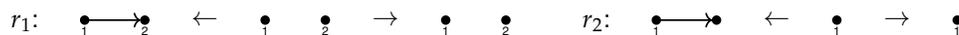

\begin{example}
Consider the reduction rules in Figure \ref{fig:eg-reduct-rules}. The GT system \(((\{\Square\}, \{\Square\}), \{r_1\})\) is \textbf{weakly garbage separating} w.r.t. the language of acyclic graphs, and \(((\{\Square\}, \{\Square\}), \{r_2\})\) \textbf{garbage separating} w.r.t. the language of trees or the language of forests.
\end{example}

\vspace{-0.1em}

We can now define (local) confluence modulo garbage, allowing us to say that, ignoring the garbage graphs, a system is (locally) confluent.

\newpage

\begin{definition}
Let \(T = (\mathcal{L}, \mathcal{R})\), \(D \subseteq \mathcal{G}(\mathcal{L})\). If for all graphs \(G\), \(H_1\), \(H_2\), such that \([G] \in D\), if \(H_1 \Leftarrow_{\mathcal{R}} G \Rightarrow_{\mathcal{R}} H_2\) implies that \(H_1\), \(H_2\) are \textbf{joinable}, then \(T\) is \textbf{locally confluent modulo garbage} with respect to \(D\).
\end{definition}

\begin{definition}
Let \(T = (\mathcal{L}, \mathcal{R})\), \(D \subseteq \mathcal{G}(\mathcal{L})\). If for all graphs \(G\), \(H_1\), \(H_2\), such that \([G] \in D\), if \(H_1 \Leftarrow_{\mathcal{R}}^* G \Rightarrow_{\mathcal{R}}^* H_2\) implies that \(H_1\), \(H_2\) are \textbf{joinable}, then \(T\) is \textbf{confluent modulo garbage} with respect to \(D\).
\end{definition}

\begin{definition}
Let \(T = (\mathcal{L}, \mathcal{R})\), \(D \subseteq \mathcal{G}(\mathcal{L})\). If there is no infinite derivation sequence \(G_0 \Rightarrow_{\mathcal{R}} G_1 \Rightarrow_{\mathcal{R}} G_2 \Rightarrow_{\mathcal{R}} \cdots\) such that \([G_0] \in D\), then \(T\) is \textbf{terminating modulo garbage} with respect to \(D\).
\end{definition}

\begin{lemma}
Let \(T = (\mathcal{L}, \mathcal{R})\), \(D \subseteq \mathcal{G}(\mathcal{L})\), \(E \subseteq D\). Then (\textbf{local}) \textbf{confluence} (\textbf{termination}) \textbf{modulo garbage} with respect to \(D\) implies (\textbf{local}) \textbf{confluence} (\textbf{termination}) \textbf{modulo garbage} with respect to \(E\).
\end{lemma}

\vspace{0.4em}
\begin{proof}
Immediate consequence of set inclusion!
\end{proof}
\vspace{-0.2em}

\begin{corollary} \label{cor:garbageimplications}
Let \(T = (\mathcal{L}, \mathcal{R})\), \(D \subseteq \mathcal{G}(\mathcal{L})\). Then (\textbf{local}) \textbf{confluence} (\textbf{termination}) implies (\textbf{local}) \textbf{confluence} (\textbf{termination}) \textbf{modulo garbage}.
\end{corollary}

\vspace{0.4em}
\begin{proof}
Local confluence (confluence, termination) is exactly local confluence (confluence, termination) modulo garbage with respect to \(\mathcal{G}(\mathcal{L})\).
\end{proof}
\vspace{-0.2em}

\begin{example}
Looking again at \(r_1\) and \(r_2\) from our first example, it is easy to see that \(r_1\) is in fact \textbf{terminating} and \textbf{confluent modulo garbage} w.r.t. the language of acyclic graphs. Similarly, \(r_2\) is \textbf{terminating} and \textbf{confluent modulo garbage} w.r.t. the language of trees.
\end{example}

\begin{lemma} \label{lem:garbagears}
Let \(T = (\mathcal{L}, \mathcal{R})\), \(D \subseteq \mathcal{G}(\mathcal{L})\). Then, if \(T\) is \textbf{weakly garbage separating}, the \textbf{induced ARS} \((D, \rightarrow)\) where \([G] \rightarrow [H]\) iff \(G \Rightarrow_{\mathcal{R}} H\) is closed and well-defined. Moreover, it is (\textbf{locally}) \textbf{confluent} (\textbf{terminating}) whenever \(T\) is, \textbf{modulo garbage} with respect to \(D\).
\end{lemma}

\vspace{0.4em}
\begin{proof}
Since \(T\) is weakly garbage separating, by Proposition \ref{lem:wgsclosure}, the induced ARS \((D, \rightarrow)\) where \([G] \rightarrow [H]\) iff \(G \Rightarrow_{\mathcal{R}} H\) is closed, and clearly it is well-defined due to the uniqueness of derivations up to isomorphism. Clearly this induced ARS is (locally) confluent (terminating) if \(T\) is (locally) confluent (terminating) modulo garbage with respect to \(D\).
\end{proof}
\vspace{-0.2em}

We can now show an analogy to \textbf{Newman's Lemma} (Theorem \ref{thm:newmanlem}).

\begin{theorem}[Newman-Garbage Lemma] \label{thm:newmangarbage}
Let \(T = (\mathcal{L}, \mathcal{R})\), \(D \subseteq \mathcal{G}(\mathcal{L})\). If \(T\) is \textbf{terminating mod. garbage} and \textbf{weakly garbage separating}, then it is \textbf{confluent mod. garbage} iff it is \textbf{locally confluent mod. garbage}.
\end{theorem}

\vspace{0.4em}
\begin{proof}
By Lemma \ref{lem:garbagears}, the induced ARS \((D, \rightarrow)\) is well-defined, closed, and terminating. Thus, by Theorem \ref{thm:newmanlem} it is confluent iff it is locally confluent, as required.
\end{proof}


\section{Non-Garbage Critical Pairs}

In 1970, Knuth and Bendix showed that confluence checking of terminating term rewriting systems is decidable \parencite{Knuth-Bendix70}. Moreover, it suffices to compute all \enquote{critical pairs} and check their joinability \parencite{Huet80} \parencite{Baader98} \parencite{Terese03}. Unfortunately, for (terminating) graph transformation systems, confluence is not decidable (Theorem \ref{thm:undecidablegtsrealdeal}), and joinability of critical pairs does not imply local confluence. In 1993, Plump showed that \enquote{strong joinability} of all critical pairs is sufficient but not necessary to show local confluence \parencite{Plump93} \parencite{Plump05}. We have summarised these results in Section \ref{section:critpairs}.

We would like to generalise Theorem \ref{thm:critpairlem} to allow us to determine when we have local confluence modulo garbage. For this, we need to define a notion of subgraph closure and non-garbage critical pairs. In this section, we shall be working with standard GT systems, as defined in Appendix \ref{appendix:transformation}, but without relabelling. That is, all graphs are totally labelled, including interface graphs.

\begin{definition}
Let \(D \subseteq \mathcal{G}(\mathcal{L})\) be a set of abstract graphs. Then \(D\) is \textbf{subgraph closed} iff for all graphs \(G\), \(H\), such that \(H \subseteq G\), if \([G] \in D\), then \([H] \in D\). The \textbf{subgraph closure} of \(D\), denoted \(\overbar{D}\), is the smallest set containing \(D\) that is \textbf{subgraph closed}.
\end{definition}

\begin{lemma}
Given \(D \subseteq \mathcal{G}(\mathcal{L})\), \(\overbar{D}\) always \textbf{exists}, and is \textbf{unique}. Moreover, \(D = \overbar{D}\) iff \(D\) is \textbf{subgraph closed}.
\end{lemma}

\vspace{0.4em}
\begin{proof}
The key observations are that the subgraph relation is transitive, and each graph has only finitely many subgraphs. Clearly, the smallest possible set containing \(D\) is just the union of all subgraphs of the elements of \(D\), up to isomorphism. This is the unique subgraph closure of \(D\).
\end{proof}
\vspace{-0.8em}

\begin{remark}
\(\overbar{D}\) always exists, however it need not be decidable, even when \(D\) is! It is not obvious what conditions on \(D\) ensure that \(\overbar{D}\) is decidable. Interestingly, the classes of regular and context-free string languages are actually closed under substring closure \parencite{Berstel79}.
\end{remark}

\begin{example}
\(\emptyset\) and \(\mathcal{G}(\mathcal{L})\) are subgraph closed.
\end{example}

\begin{example}
The language of discrete graphs is subgraph closed.
\end{example}

\begin{example}
The subgraph closure of the language of trees is the language of forests. The subgraph closure of the language connected graphs is the language of all graphs.
\end{example}

\begin{definition}
Let \(T = (\mathcal{L}, \mathcal{R})\), \(D \subseteq \mathcal{G}(\mathcal{L})\). A \textbf{critical pair} (Definition \ref{dfn:critpair}) \(H_1 \Leftarrow G \Rightarrow H_2\) is \textbf{non-garbage} iff \([G] \in \overbar{D}\).
\end{definition}

\begin{lemma} \label{lem:fincritpairs}
Let \(T = (\mathcal{L}, \mathcal{R})\), \(D \subseteq \mathcal{G}(\mathcal{L})\). Then there are only finitely many \textbf{non-garbage critical pairs} up to isomorphism. If \(\overbar{D}\) is \textbf{decidable}, then one can find all the \textbf{non-garbage critical pairs} in \textbf{finite time}.
\end{lemma}

\vspace{0.4em}
\begin{proof}
By Theorem \ref{thm:critpairlem} and Remark \ref{remark:finitecritpairs}, there are only finitely many critical pairs for \(T\), up to isomorphism, and there exists a terminating procedure for generating them. Thus, there are only finitely many non-garbage critical pairs up to isomorphism. It remains to show that we can decide if a critical pair is garbage.
Since \(\overbar{D}\) has a computable membership function, we can test if the start graph in each pair is garbage in finite time.
\end{proof}
\vspace{-0.2em}

\begin{corollary}
Let \(T = (\mathcal{L}, \mathcal{R})\), \(D \subseteq \mathcal{G}(\mathcal{L})\) be such that \(T\) is \textbf{terminating modulo garbage} and \(\overbar{D}\) is \textbf{decidable}. Then, one can \textbf{decide} if all the \textbf{non-garbage critical pairs} are \textbf{strongly joinable} (Definition \ref{dfn:strongjoin}).
\end{corollary}

\vspace{0.4em}
\begin{proof}
By Lemma \ref{lem:fincritpairs}, we can find the finitely many pairs in finite time, and since \(T\) is terminating modulo garbage and finitely branching (Lemma \ref{lem:gtprops}), both sides of each pair have only finitely many successors (Lemma \ref{lem:arsbranching}), thus we can test for strong joinability in finite time.
\end{proof}
\vspace{-0.4em}

\begin{lemma}
Let \(T = (\mathcal{L}, \mathcal{R})\), \(D \subseteq \mathcal{G}(\mathcal{L})\). Then, the \textbf{non-garbage critical pairs} are \textbf{complete}. That is, for each pair of \textbf{parallelly independent} (Definition \ref{dfn:parindep}) direct derivations, \(H_1 \Leftarrow_{r_1,g_1} G \Rightarrow_{r_2,g_2} H_2\) such that \([G] \in D\), there is a \textbf{critical pair} \(P_1 \Leftarrow_{r_1,o_1} K \Rightarrow_{r_2,o_2} P_2\) with extension diagrams (1), (2), and an inclusion morphism \(m: K \to G\).

\vspace{-0.2em}
\begin{figure}[H]
\centering
\noindent
\scalebox{.85}{\begin{tikzpicture}[every node/.style={inner sep=0pt, text width=6.5mm, align=center}]
    \node (a) at (0.0,0.0) {$P_1$};
    \node (b) at (1.0,0.0) {$\Longleftarrow$};
    \node (c) at (2.0,0.0) {$K$};
    \node (d) at (3.0,0.0) {$\Longrightarrow$};
    \node (e) at (4.0,0.0) {$P_2$};

    \node (f) at (0.0,-0.8) {$\big\downarrow$};
    \node (g) at (1.0,-0.8) {(1)};
    \node (h) at (2.0,-0.8) {$\big\downarrow$};
    \node (i) at (3.0,-0.8) {(2)};
    \node (j) at (4.0,-0.8) {$\big\downarrow$};

    \node (k) at (0.0,-1.6) {$H_1$};
    \node (l) at (1.0,-1.6) {$\Longleftarrow$};
    \node (m) at (2.0,-1.6) {$G$};
    \node (n) at (3.0,-1.6) {$\Longrightarrow$};
    \node (o) at (4.0,-1.6) {$H_2$};
\end{tikzpicture}}
\vspace{-0.3em}
\caption{Pair Factorisation Diagram}
\end{figure}
\vspace{0em}
\end{lemma}

\vspace{0.4em}
\begin{proof}
By Lemma 6.22 in \parencite{Ehrig06}, critical pairs are complete when \(D = \mathcal{G}(\mathcal{L})\). Now if we only consider derivations from start graphs \(G\) such that \([G] \in D\) where \(D \subseteq \mathcal{G}(\mathcal{L})\), clearly all factorings with critical pairs are such that \(K\) can be embedded into \(G\), so \([K] \in \overbar{D}\). Thus, the non-garbage critical pairs are complete.
\end{proof}
\vspace{-0.4em}

\begin{samepage}
\begin{theorem}[Non-Garbage Critical Pair Lemma] \label{thm:ngcritpairlem}
Let \(T = (\mathcal{L}, \mathcal{R})\), \(D \subseteq \mathcal{G}(\mathcal{L})\). If all its \textbf{non-garbage critical pairs} are \textbf{strongly joinable}, then \(T\) is \textbf{locally confluent modulo garbage} with respect to \(D\).
\end{theorem}

\vspace{0.4em}
\begin{proof}
By the proof of Theorem 6.28 in \parencite{Ehrig06}, strong joinability of critical pairs implies local confluence due to completeness. But, the non-garbage critical pairs are complete with respect to \(D\), so we have the result.
\end{proof}
\vspace{-0.2em}

\begin{corollary}
Let \(T = (\mathcal{L}, \mathcal{R})\), \(D \subseteq \mathcal{G}(\mathcal{L})\). If \(T\) is \textbf{terminating modulo garbage}, \textbf{weakly garbage separating}, and all its \textbf{non-garbage critical pairs} are \textbf{strongly joinable} then \(T\) is \textbf{confluent modulo garbage}.
\end{corollary}

\vspace{0.4em}
\begin{proof}
By the above theorem, \(T\) is \textbf{locally confluent modulo garbage}, so by the Newman-Garbage Lemma (Theorem \ref{thm:newmangarbage}), \(T\) is \textbf{confluent modulo garbage} as required.
\end{proof}
\vspace{-0.2em}
\end{samepage}


\section{Extended Flow Diagrams}

In 1976, Farrow, Kennedy and Zucconi presented \enquote{semi-structured flow graphs}, defining a grammar with confluent reduction rules \parencite{Farrow-Kennedy-Zucconi76}. Plump has considered a restricted version of this language: \enquote{extended flow diagrams} \parencite{Plump05}. The reduction rules for \enquote{extended flow diagrams} are not confluent, however we will see that they are confluent modulo garbage and terminating. Thus we have an efficient mechanism for testing for language membership, since we need not \enquote{backtrack}, just like in Theorem \ref{thm:treerecthm}.

\begin{definition}
The language of \textbf{extended flow diagrams} is generated by \(\pmb{EFD} = (\mathcal{L}, \mathcal{N}, \mathcal{R}, S)\) where \(\mathcal{L}_V = \{\bullet, \Square, \Diamond\}\), \(\mathcal{L}_E = \{t, f, \square\}\), \(\mathcal{N}_V = \mathcal{N}_E = \emptyset\), \(\mathcal{R} = \{seq, while, ddec, dec1, dec2\}\), and \(S = \) \tikz[baseline]{ \tikzstyle{ann} = [draw=none,fill=none,right] \node (a) at (0.0,0.1) [draw, circle, thick, fill=black, scale=0.3] {}; \node[rectangle] (b) at (0.5,0.1) [minimum height=0.2cm,minimum width=0.2cm,draw] {}; \node (c) at (1.0,0.1) [draw, circle, thick, fill=black, scale=0.3] {}; \draw (a) edge[->,thick] (b) (b) edge[->,thick] (c);}.
\end{definition}

\vspace{0.4em}
\begin{figure}[H]
\centering
\vspace{-0.9em}
\includegraphics[totalheight=6.8cm]{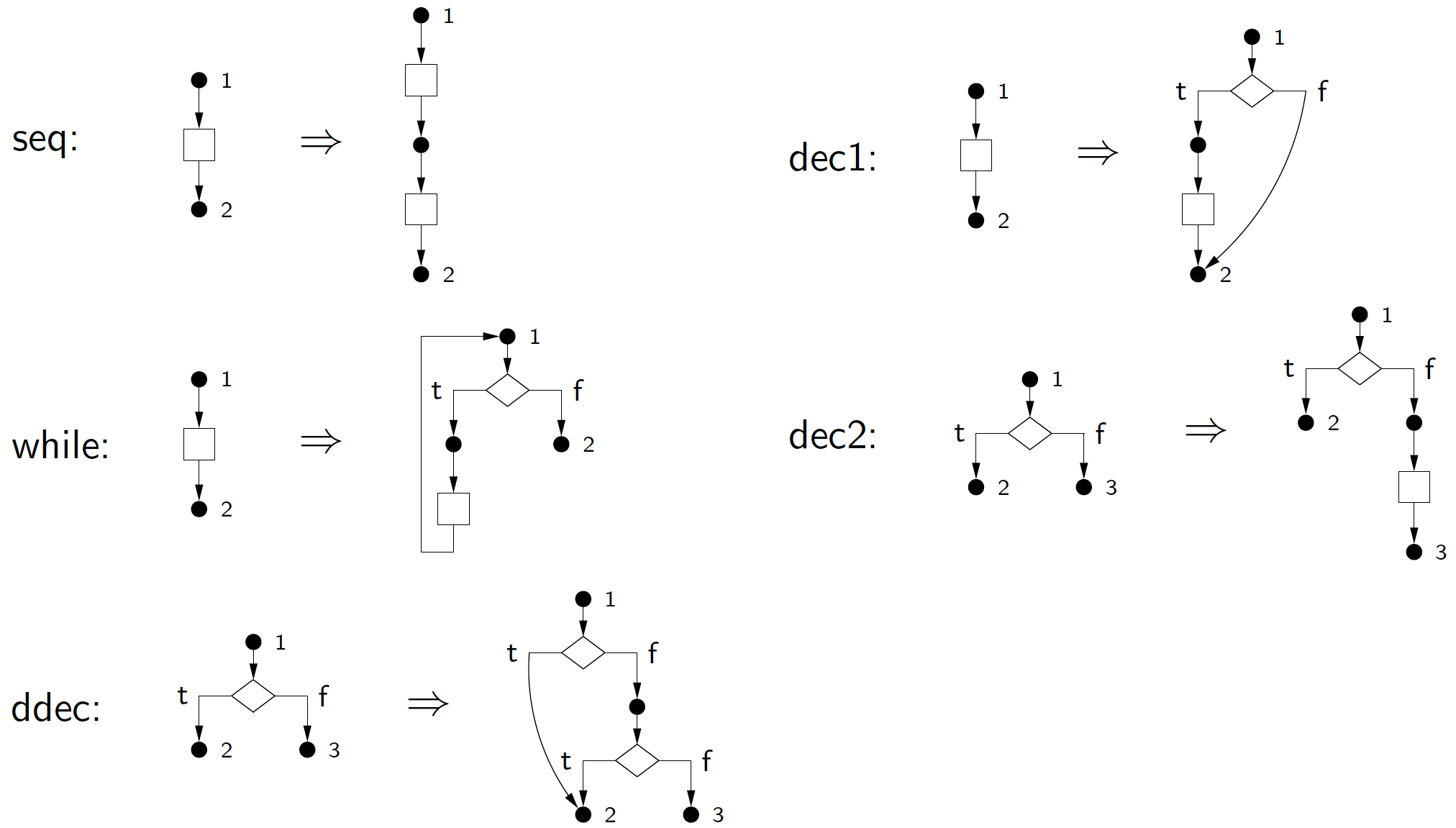}
\caption{EFD Grammar Rules}
\vspace{0.4em}
\end{figure}
\vspace{-0.25em}

\begin{lemma} \label{lem:efd1}
\(EFD^{-1} = (\mathcal{L}, \mathcal{R}^{-1})\) is \textbf{terminating}. Moreover, it is \textbf{garbage separating} w.r.t. \(\pmb{L}(\pmb{EFD})\).
\end{lemma}

\vspace{0.4em}
\begin{proof}
Termination is clear since the rules are size reducing. Weak garbage separation can be seen by induction.
\end{proof}
\vspace{-0.2em}

\begin{lemma} \label{lem:efd2}
Every \textbf{directed cycle} in a graph in the \textbf{subgraph closure} of \(\pmb{L}(\pmb{EFD})\) contains a \(t\)-labelled edge.
\end{lemma}

\vspace{0.4em}
\begin{proof}
Induction.
\end{proof}
\vspace{-0.2em}

Now that we have all the intermediate results we need, we are ready to see that \(EFD^{-1}\) is not confluent, but is confluent modulo garbage. Moreover, that non-garbage critical pair analysis is sufficient to prove this!

\begin{theorem}[EFD Recognition]
\(EFD^{-1} = (\mathcal{L}, \mathcal{R}^{-1})\) is \textbf{confluent modulo garbage} w.r.t. \(\pmb{L}(\pmb{EFD})\), but not \textbf{confluent}.
\end{theorem}

\begin{figure}[H]
\centering
\vspace{-0.9em}
\includegraphics[totalheight=2.4cm]{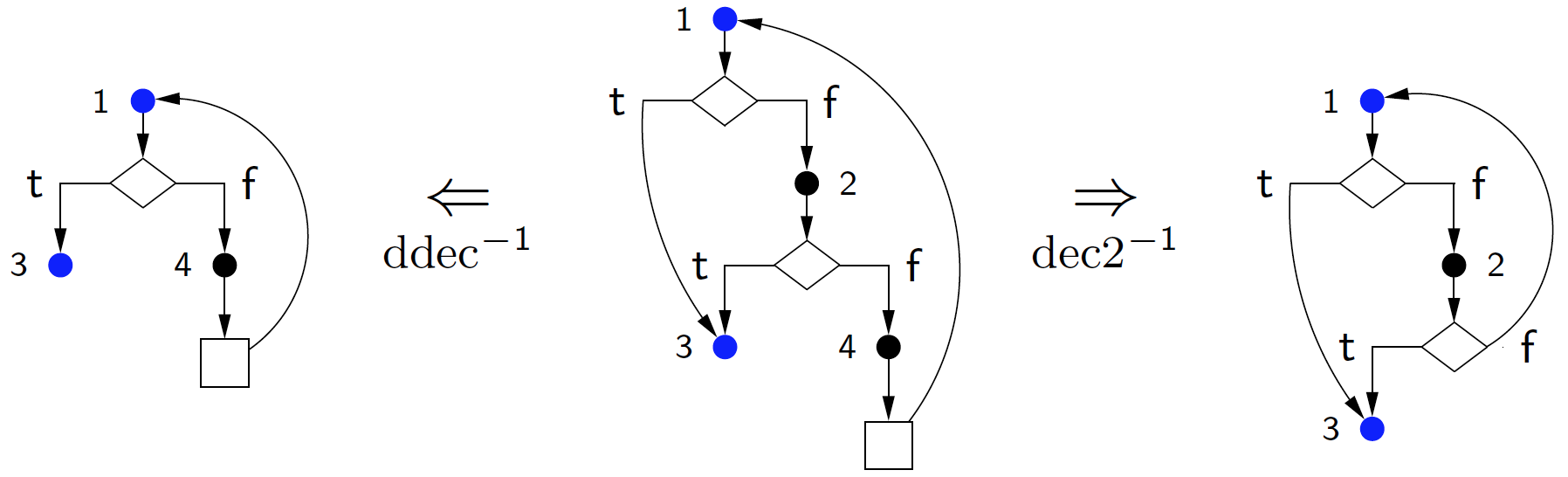}
\vspace{-0.4em}
\caption{Non-Joinable Critical Pair}
\label{fig:efd-pair}
\end{figure}

\vspace{0.4em}
\begin{proof}
By Lemma \ref{lem:efd1} and the Newman-Garbage Lemma (Theorem \ref{thm:newmangarbage}), it suffices to show local confluent modulo garbage. \(EFD^{-1}\) has ten critical pairs \parencite{Plump19}, all but one of which are strongly joinable. Thus, we do not have confluence, however by Lemma \ref{lem:efd2}, the non-joinable critical pair (Figure \ref{fig:efd-pair}) is garbage, so by the Non-Garbage Critical Pair Lemma (Theorem \ref{thm:ngcritpairlem}), we have local confluence modulo garbage, as required.
\end{proof}
\vspace{-0.2em}

\vspace{0.2em}
\vspace{-0.2em}

\vspace{-0.1em}
\begin{remark}
This special case of \textbf{weak garbage separation} with respect to the language we are recognising has actually been considered before by Bakewell \parencite{Bakewell03}. He called this property \textbf{closedness}.
\end{remark}


\section{Encoding Partial Labelling} \label{encodinglabelling}

We now turn our attention to encoding partially labelled graphs and morphisms as totally labelled graphs and morphisms. The reason for doing this is that if we can show that our encoded rules are confluent modulo garbage, this must mean our original rules with relabelling were. Thus, we can attempt to determine local confluence of a system with relabelling by performing non-garbage critical pair analysis of the encoded rules!

Let \(\mathcal{L} = (\mathcal{L}_V, \mathcal{L}_E)\) be an arbitrary label alphabet, and suppose without loss of generality (w.l.o.g.) that \(\mathcal{L}_V \cap \mathcal{L}_E = \emptyset\) and \(\{\Square\} \not\in \mathcal{L}_V \cup \mathcal{L}_E\). We will start by showing that partially labelled graphs (Definition \ref{dfn:plgraph}), morphisms, and rules can be encoded by totally labelled systems.

\begin{definition}
Let \(G\) be a \textbf{partially labelled graph} over \(\mathcal{L}\), and w.l.o.g., suppose \(V_G \cap E_G = \emptyset\). Define \(e(G) = (V, E, s, t, l, m)\) where:
\begin{multicols}{2}
\begin{enumerate}[itemsep=-0.6ex,topsep=-0.6ex]
\item \(V = V_G\)
\item \(E = E_G \cup \,l^{-1}(\mathcal{L}_V)\)
\item\(
s(e) =
\begin{cases}
s_G(e) \text{\,\,if } e \in E_G\\
e \text{\,\,\,\,\,\,\,\,\,\,\,\,\,\,otherwise}
\end{cases}
\)\item\(
t(e) =
\begin{cases}
t_G(e) \text{\,\,if } e \in E_G\\
e \text{\,\,\,\,\,\,\,\,\,\,\,\,\,\,otherwise}
\end{cases}
\)
\item \(l(v) = \Square\)
\item\(
m(e) =
\begin{cases}
m_G(e) \text{\,\,if } e \in E_G\\
l_G(e) \text{\,\,\,\,\,otherwise}
\end{cases}
\)
\end{enumerate}
\end{multicols}
\end{definition}

\vspace{-0.6em}
\begin{proposition} \label{prop:objencode}
\(e(G)\) is a \textbf{totally labelled graph} over the \textbf{encoded label alphabet} \(e(\mathcal{L}) = (\{\Square\}, \mathcal{L}_V \cup \mathcal{L}_E)\).
\end{proposition}

\begin{figure}[H]
\centering
\noindent
\scalebox{.8}{\begin{tikzpicture}[every node/.style={inner sep=0pt, text width=6.5mm, align=center}]
    \node (a) at (0.0,-0.05) {$G$:};
    \node (b) at (1.0,0.0)   [draw, circle, thick] {x};
    \node (c) at (2.5,0.0)   [draw, circle, thick] {x};
    \node (d) at (4.0,0.0)   [draw, circle, thick] {\,};

    \node (e) at (6.15,-0.05) {$e(G)$:};
    \node (f) at (7.5,0.0)   [draw, circle, thick] {$\Square$};
    \node (g) at (9.0,0.0)   [draw, circle, thick] {$\Square$};
    \node (h) at (10.5,0.0)  [draw, circle, thick] {$\Square$};

    \draw (b) edge[->] node[above, yshift=2.5pt] {y} (c)
          (c) edge[->,in=-25,out=-60,loop,thick] node [right, yshift=1pt] {z} (c);

    \node (B) at (1.0,-0.52)  {\tiny{1}};
    \node (C) at (2.5,-0.52)  {\tiny{2}};
    \node (D) at (4.0,-0.52)  {\tiny{3}};
    \node (F) at (7.5,-0.52)  {\tiny{1}};
    \node (G) at (9.0,-0.52)  {\tiny{2}};
    \node (H) at (10.5,-0.52) {\tiny{3}};

    \draw (f) edge[->] node[above, yshift=2.5pt] {y} (g)
          (f) edge[->,in=-25,out=-60,loop,thick] node [right, yshift=1pt] {x} (f)
          (g) edge[->,in=-25,out=-60,loop,thick] node [right, yshift=1pt] {x} (g)
          (g) edge[->,in=60,out=25,loop,thick] node [right, yshift=1pt] {z} (g);
\end{tikzpicture}}
\vspace{-0.6em}
\caption{Example Encoded Partially Labelled Graph}
\label{fig:encoding}
\end{figure}
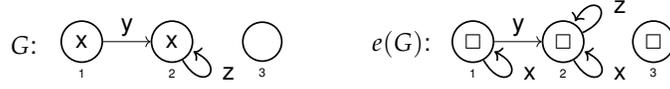

\begin{example}
Let \(\mathcal{L} = (\{x\}, \{y, z\})\). Then Figure \ref{fig:encoding} shows an example partially labelled graph and its encoding as a totally labelled graph.
\end{example}

\newpage

\vspace{-1.4em}
\vspace{-0.4em}

\begin{definition}
Given two \textbf{partially labelled graphs} \(G\), \(H\), and a \textbf{morphism} \(g: G \to H\), define \(e(g) = (g_V', g_E')\) where:
\begin{multicols}{2}
\begin{enumerate}[itemsep=-0.6ex,topsep=-0.6ex]
\item \(g_V'(v) = g_V(v)\)
\item\(
g_E'(e) =
\begin{cases}
g_E(e) \text{\,\,\,if } e \in E_G\\
g_V(e) \text{\,\,otherwise}
\end{cases}
\)
\end{enumerate}
\end{multicols}
\end{definition}

\vspace{-0.8em}
\begin{proposition} \label{prop:morencode}
Clearly, \(g\) is source/target/label preserving, and thus it is a \textbf{morphism} between the \textbf{totally labelled graphs} \(e(G)\) and \(e(H)\).
\end{proposition}

\begin{lemma}
\(e\) is a \textbf{fully faithful covariant functor} from the category of \textbf{partially labelled graphs} to the category of \textbf{totally labelled graphs}.
\end{lemma}

\vspace{0.4em}
\begin{proof}
Clearly each graph and morphism has a distinct encoding, which is well-defined by Propositions \ref{prop:objencode} and \ref{prop:morencode}. Clearly composition of morphisms behaves properly too.
\end{proof}
\vspace{-0.2em}

\begin{theorem}[Partial Labelling Simulation]
Given a \textbf{rule} \(r = \langle L \leftarrow K \rightarrow R\rangle\) where \(L\) and \(R\) are totally labelled graphs, and \(K\) partially labelled, then for all \textbf{totally labelled graphs} \(G\), \(H\), \(G \Rightarrow_r H\) iff \(e(G) \Rightarrow_{e(r)} e(H)\).
\end{theorem}

\vspace{0.4em}
\begin{proof}
Firstly, given a fixed graph \(G\), every injective morphism \(g: L \to G\) satisfying the dangling condition can be encoded. Its encoding must also be injective (since encoding is an injective functor), and it is easy to check it must also satisfy the dangling condition. To see the other inclusion, suppose there was an injective morphism satisfying the dangling condition in the encoded system \(g': e(L) \to e(G)\). Then, we must be able to decode the morphism, to give an injective morphism. Again, it is easy to check the decoded morphism satisfies the dangling condition.

Finally, it is routine to check that the encoding of result graph for each match is exactly the same as the encoded result graph, derived using the encoded system, by using the explicit definition of rule application.
\end{proof}
\vspace{-0.2em}

\begin{corollary}
Given a \textbf{GT system} \((\mathcal{L}, \mathcal{R})\), \((e(\mathcal{L}), e(\mathcal{R}))\) is \textbf{weakly garbage separating} with respect to \(e(\mathcal{G}(\mathcal{L}))\).
\end{corollary}

\vspace{0.4em}
\begin{proof}
By the theorem, the encoded system can only derive encoded totally labelled graphs from encoded totally labelled graphs.
\end{proof}
\vspace{-0.2em}

\begin{corollary} \label{cor:simcorol}
The \textbf{GT system} \((\mathcal{L}, \mathcal{R})\) is (\textbf{locally}) \textbf{confluent} (\textbf{terminating}) iff \((e(\mathcal{L}), e(\mathcal{R}))\) is (\textbf{locally}) \textbf{confluent} (\textbf{terminating}) \textbf{modulo garbage} with respect to \(e(\mathcal{G}(\mathcal{L}))\).
\end{corollary}

\vspace{0.4em}
\begin{proof}
By the theorem, we have a correspondence between derivations and derivations in the encoded system, so it is immediate that these notions line up with the notions in the encoded system.
\end{proof}
\vspace{-0.2em}


\section{Tree Recognition Revisited}

It is possible to rephrase the results from Section \ref{sec:linearrec} in terms of our new notion of garbage:

\begin{proposition} \label{prop:treegarb}
Let \(\mathcal{L} = (\{\Square, \triangle\}, \{\Square\})\), \(\mathcal{R} = \{r_0, r_1, r_2\}\), where the rules are as in Figure \ref{fig:tree2}. Then, \(T = (\mathcal{L}, \mathcal{R})\) is \textbf{garbage separating} w.r.t. to \(D = \{[G] \in \mathcal{G}^{\varoplus}(\mathcal{L}) \mid [G^{\varominus}] \in \pmb{L}(\pmb{TREE}), \abs{p_G^{-1}(\{1\})} = 1\}\) and \textbf{confluent modulo garbage} w.r.t. \(E = \{[G] \in D \mid l_G(V_G) = \{\Square\}\}\).
\end{proposition}

\vspace{0.4em}
\begin{proof}
Garbage separation is due to Lemma \ref{lem:treegarbagesep} and confluence modulo garbage due to Theorem \ref{thm:treerecthm}.
\end{proof}
\vspace{-0.2em}

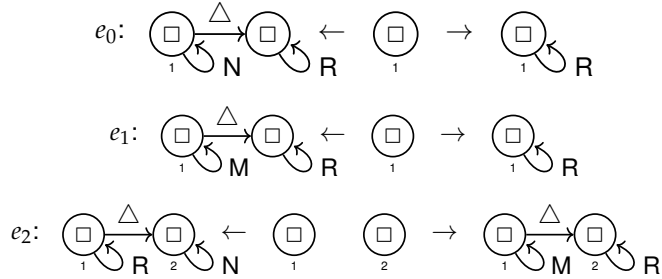
\begin{figure}[H]
\centering
\noindent
\scalebox{.85}{\begin{tikzpicture}[every node/.style={inner sep=0pt, text width=6.5mm, align=center}]
    \node (a) at (0.0,0) {$e_0$:};

    \node (b) at (1.0,0) [draw, circle, thick] {\(\Square\)};
    \node (c) at (2.5,0) [draw, circle, thick] {\(\Square\)};

    \node (d) at (3.5,0) {$\leftarrow$};

    \node (e) at (4.5,0) [draw, circle, thick] {\(\Square\)};

    \node (f) at (5.5,0) {$\rightarrow$};

    \node (g) at (6.5,0) [draw, circle, thick] {\(\Square\)};

    \node (B) at (1.0,-.52) {\tiny{1}};
    \node (E) at (4.5,-.52) {\tiny{1}};
    \node (G) at (6.5,-.52) {\tiny{1}};

    \draw (b) edge[->,thick] node[above, yshift=2.5pt] {\(\triangle\)} (c)
          (b) edge[->,in=-25,out=-60,loop,thick] node [right, yshift=1pt] {N} (b)
          (c) edge[->,in=-25,out=-60,loop,thick] node [right, yshift=1pt] {R} (c)
          (g) edge[->,in=-25,out=-60,loop,thick] node [right, yshift=1pt] {R} (g);
\end{tikzpicture}}

\vspace{0.4em}

\scalebox{.8}{\begin{tikzpicture}[every node/.style={inner sep=0pt, text width=6.5mm, align=center}]
    \node (a) at (0.0,0) {$e_1$:};

    \node (b) at (1.0,0) [draw, circle, thick] {\(\Square\)};
    \node (c) at (2.5,0) [draw, circle, thick] {\(\Square\)};

    \node (d) at (3.5,0) {$\leftarrow$};

    \node (e) at (4.5,0) [draw, circle, thick] {\(\Square\)};

    \node (f) at (5.5,0) {$\rightarrow$};

    \node (g) at (6.5,0) [draw, circle, thick] {\(\Square\)};

    \node (B) at (1.0,-.52) {\tiny{1}};
    \node (E) at (4.5,-.52) {\tiny{1}};
    \node (G) at (6.5,-.52) {\tiny{1}};

    \draw (b) edge[->,thick] node[above, yshift=2.5pt] {\(\triangle\)} (c)
          (b) edge[->,in=-25,out=-60,loop,thick] node [right, yshift=1pt] {M} (b)
          (c) edge[->,in=-25,out=-60,loop,thick] node [right, yshift=1pt] {R} (c)
          (g) edge[->,in=-25,out=-60,loop,thick] node [right, yshift=1pt] {R} (g);
\end{tikzpicture}}

\vspace{0.4em}

\scalebox{.8}{\begin{tikzpicture}[every node/.style={inner sep=0pt, text width=6.5mm, align=center}]
    \node (a) at (0.0,0) {$e_2$:};

    \node (b) at (1.0,0) [draw, circle, thick] {\(\Square\)};
    \node (c) at (2.5,0) [draw, circle, thick] {\(\Square\)};

    \node (d) at (3.5,0) {$\leftarrow$};

    \node (e) at (4.5,0) [draw, circle, thick] {\(\Square\)};
    \node (f) at (6.0,0) [draw, circle, thick] {\(\Square\)};

    \node (g) at (7.0,0) {$\rightarrow$};

    \node (h) at (8.0,0) [draw, circle, thick] {\(\Square\)};
    \node (i) at (9.5,0) [draw, circle, thick] {\(\Square\)};

    \node (B) at (1.0,-.52) {\tiny{1}};
    \node (C) at (2.5,-.52) {\tiny{2}};
    \node (E) at (4.5,-.52) {\tiny{1}};
    \node (F) at (6.0,-.52) {\tiny{2}};
    \node (H) at (8.0,-.52) {\tiny{1}};
    \node (I) at (9.5,-.52) {\tiny{2}};

    \draw (b) edge[->,thick] node[above, yshift=2.5pt] {\(\triangle\)} (c)
          (h) edge[->,thick] node[above, yshift=2.5pt] {\(\triangle\)} (i)
          (b) edge[->,in=-25,out=-60,loop,thick] node [right, yshift=1pt] {R} (b)
          (c) edge[->,in=-25,out=-60,loop,thick] node [right, yshift=1pt] {N} (c)
          (h) edge[->,in=-25,out=-60,loop,thick] node [right, yshift=1pt] {M} (h)
          (i) edge[->,in=-25,out=-60,loop,thick] node [right, yshift=1pt] {R} (i);
\end{tikzpicture}}
\vspace{-0.6em}
\caption{Encoded Tree Recognition Rules}
\label{fig:tree3}
\end{figure}

Finally, in Section \ref{encodinglabelling} we have seen that we can encode a GT system with relabelling as a standard GT system (where interface graphs are totally labelled). One can pull a similar trick to encode rootedness of nodes, using looped edges with special labels. We give an encoding of the tree recognition rules from Figure \ref{fig:tree2}: \(T' = ((\{\Square\}, \{R, N, M, \triangle\}), \{e_0, e_1, e_2\})\), where the rules are defined in Figure \ref{fig:tree3}.

\vspace{-0.2em}
\vspace{0.2em}
\vspace{-0.2em}

\begin{figure}[H]
\centering
\noindent
\scalebox{.75}{\begin{tikzpicture}[every node/.style={inner sep=0pt, text width=6.5mm, align=center}]
    \node (a) at (2.0,0.0)   [draw, circle, thick] {\(\Square\)};
    \node (b) at (1.0,-1.5)  [draw, circle, thick] {\(\Square\)};
    \node (c) at (3.0,-1.5)  [draw, circle, thick] {\(\Square\)};

    \node (d) at (4.5,-0.8)  {$\Leftarrow_{e_2}$};

    \node (e) at (7.0,0.0)   [draw, circle, thick] {\(\Square\)};
    \node (f) at (6.0,-1.5)  [draw, circle, thick] {\(\Square\)};
    \node (g) at (8.0,-1.5)  [draw, circle, thick] {\(\Square\)};

    \node (h) at (9.5,-0.8)  {$\Rightarrow_{e_2}$};

    \node (i) at (12.0,0.0)  [draw, circle, thick] {\(\Square\)};
    \node (j) at (11.0,-1.5) [draw, circle, thick] {\(\Square\)};
    \node (k) at (13.0,-1.5) [draw, circle, thick] {\(\Square\)};

    \node (A) at (2.0,-0.52)  {\tiny{1}};
    \node (B) at (1.0,-2.02)  {\tiny{2}};
    \node (C) at (3.0,-2.02)  {\tiny{3}};

    \node (E) at (7.0,-0.52)  {\tiny{1}};
    \node (F) at (6.0,-2.02)  {\tiny{2}};
    \node (G) at (8.0,-2.02)  {\tiny{3}};

    \node (I) at (12.0,-0.52) {\tiny{1}};
    \node (J) at (11.0,-2.02) {\tiny{2}};
    \node (K) at (13.0,-2.02) {\tiny{3}};

    \draw (a) edge[->,thick] node[above, yshift=-3pt, xshift=-9pt] {\(\triangle\)} (b)
          (a) edge[->,thick] node[above, yshift=-3pt, xshift=9pt] {\(\triangle\)} (c)
          (e) edge[->,thick] node[above, yshift=-3pt, xshift=-9pt] {\(\triangle\)} (f)
          (e) edge[->,thick] node[above, yshift=-3pt, xshift=9pt] {\(\triangle\)} (g)
          (i) edge[->,thick] node[above, yshift=-3pt, xshift=-9pt] {\(\triangle\)} (j)
          (i) edge[->,thick] node[above, yshift=-3pt, xshift=9pt] {\(\triangle\)} (k)
          (a) edge[->,in=20,out=-20,loop,thick] node [right, yshift=1pt] {M} (a)
          (b) edge[->,in=20,out=-20,loop,thick] node [right, yshift=1pt] {R} (b)
          (c) edge[->,in=20,out=-20,loop,thick] node [right, yshift=1pt] {N} (c)
          (e) edge[->,in=20,out=-20,loop,thick] node [right, yshift=1pt] {R} (e)
          (f) edge[->,in=20,out=-20,loop,thick] node [right, yshift=1pt] {N} (f)
          (g) edge[->,in=20,out=-20,loop,thick] node [right, yshift=1pt] {N} (g)
          (i) edge[->,in=20,out=-20,loop,thick] node [right, yshift=1pt] {M} (i)
          (j) edge[->,in=20,out=-20,loop,thick] node [right, yshift=1pt] {N} (i)
          (k) edge[->,in=20,out=-20,loop,thick] node [right, yshift=1pt] {R} (k);
\end{tikzpicture}}
\vspace{-0.4em}
\caption{Non-Strongly Joinable Encoded Critical Pair}
\label{fig:tree4}
\end{figure}
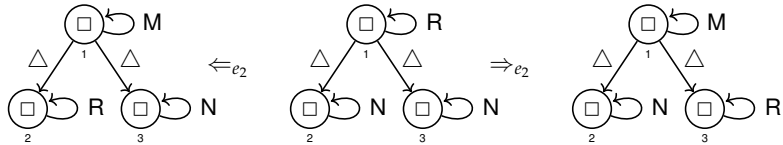

\vspace{-0.5em}
One would hope that we could then perform non-garbage critical pair analysis on the encoding of \(D\) (where \(D\) is as in Proposition \ref{prop:treegarb}) in order to demonstrate local confluence modulo garbage of the original system. Every non-garbage critical pair is joinable, but unfortunately, one of them is not strongly joinable (Figure \ref{fig:tree4}), so we are unable to make any conclusion about local confluence modulo garbage using the Non-Garbage Critical Pair Lemma (Theorem \ref{thm:ngcritpairlem}).

\vspace{-0.2em}
\vspace{0.2em}
\vspace{-0.2em}

\vspace{-0.5em}
Thus, just like Plump's Critical Pair Lemma, strong joinability of non-garbage critical pairs is sufficient, but not necessary to imply local confluence modulo garbage. As discussed in the next chapter, it remains future work to develop stronger (non-garbage) critical pair analysis theorems.

\chapter{Conclusion}


We have reviewed the current state of graph transformation, with a particular focus on the \enquote{injective DPO} approach with relabelling and graph programming languages, establishing issues with the current approach to rooted graph transformation. We developed a new type of graph transformation system that supports relabelling and root nodes, but where derivations are invertible, and looked at a case study, showing that rooted graph transformation systems can recognise trees in linear time. This work on tree recognition has been submitted for publication as part of \parencite{Campbell-Courtehoute-Plump19}. We have also defined some notions of equivalence for our new type of graph transformation system, and briefly discussed a possible theory of refinement.

Finally, in Chapter \ref{chapter:confluenceanalysis}, we have introduced the new notion of confluence modulo garbage for graph transformation systems, that allows us to have confluence, except in the cases we do not care about. Moreover, we have shown that it is sufficient to only analyse the non-garbage critical pairs to establish confluence modulo garbage. We have applied this to see that Extended Flow Diagrams (EFDs) can be recognised by a system that is confluent modulo garbage, and that we can rephrase the question of confluence of less well understood systems in terms of confluence modulo garbage of an encoded standard graph transformation system.


\section{Evaluation}

We regard this project as a success, having achieved our four original goals as detailed in the Executive Summary. Our first goal was to review rooted DPO graph transformation with relabelling. We have done this in Chapter \ref{chapter:theoryintro}, looking at labelled GT systems with the DPO approach with injective matching, and how relabelling and root nodes have been implemented, providing further detail in Appendices \ref{appendix:ars} and \ref{appendix:transformation}. We also briefly reviewed DPO-based graph programming languages.

Our second goal was to address the problem that the current theory of rooted graph transformation does not have invertible derivations. We have fixed this problem in Chapter \ref{chapter:newtheory} by defining rootedness using a partial function onto a two-point set rather than pointing graphs with root nodes. We have shown rule application corresponds to NDPOs, how Dodds' complexity theory applies in our system, and briefly discussed the equivalence of and refinement of GT systems.

Our third goal was to show a new example of how rooted graph transformation can be applied. We showed a new result that the graph class of trees can be recognised by a rooted GT system in linear time, given an input graph of bounded degree. Moreover, we have given empirical evidence by implementing the algorithm in GP\,2 and collecting timing results. We have submitted our program and results for publication \parencite{Campbell-Courtehoute-Plump19}.

Our final goal was to develop new confluence analysis theory. We have defined a new notion of confluence modulo garbage and non-garbage critical pairs, and shown that it is sufficient to require strong joinability of only the non-garbage critical pairs to establish confluence modulo garbage. We have applied this theory to EFDs and the encoding of partially labelled (rooted) GT systems as standard GT systems, performing non-garbage critical pair analysis on the encoded system. We look to publish this work.


\section{Future Work}

Developing a fully-fledged theory of correctness and refinement for (rooted) GT systems remains future work, extending the work from Section \ref{sec:gtequiv}. Additionally, extending this notion to GP\,2, or other graph transformation based languages, and looking at the automated introduction of root nodes in order to improve time complexity remains open. Overcoming the restriction of host graphs to be of bounded degree in Theorem \ref{thm:fastderivations} remains open too.

Further exploring the relationship between (local) confluence modulo garbage and weak garbage separation remains open work. In fact, confluence analysis of GT systems remains an underexplored area in general. Developing a stronger version of the Non-Garbage Critical Pair Lemma that allows for the detection of persistent nodes that need not be identified in the joined graph would allow conclusions of confluence modulo garbage where it was previously not determined, remains future work.

Additional future work in the foundations of our new theory of rooted graph transformation would be to attempt to establish if the Local Church-Rosser and Parallelism Theorems hold \parencite{EhrigGolasHabelLambersOrejas2014}, which have applications in database systems \parencite{EhrigKreowski80} and algebraic specifications \parencite{ParisiPresicce89}. It has been shown by Habel and Plump that this is the case with only relabelling \parencite{HabelPlump12}. It is likely that our new system with root nodes is \(\mathcal{M},\mathcal{N}\)-adhesive. Moreover, showing an analogy to the Extension Theorem and Critical Pair Lemma \parencite{EhrigGolasHabelLambersOrejas2012} would be excellent. Based on Section \ref{encodinglabelling}, we think that this is possible.

Finally, it remains open research, to explore the overlap between graph transformation systems and the study of \enquote{reversible computation} \parencite{Frank05}. Our new foundations of rooted graph transformation allows for the specification of both efficient and reversible GT systems. Since graph transformation is a uniform way of expressing many problems in computer science, it is only natural that its applications in reversible computation is explored.

\appendix
\addtocontents{toc}{\protect\newpage}

\chapter{Mathematical Prelude} \label{appendix:notation}


\section{Sets I}

There is not time to develop ZF(C) Set Theory and the foundational logic required. For the most part, a naive approach will suffice. We split the \enquote{Sets} section into two halves. This section is derived from \parencite{Sutherland09}.

\begin{definition}
We let \(\emptyset\) denote the \textbf{empty set}. If \(A\) is a \textbf{set}, then we write \(a \in A\) to say that \(a\) \enquote{belongs to} \(A\). We say that \(B\) is a \textbf{subset} of \(A\), \(B \subseteq A\) iff \(\forall x \in B, x \in A\). We say \(A = B\) iff \(A \subseteq B\) and \(B \subseteq A\).
\end{definition}

\begin{definition}
If \(A, B\) are sets, then we define:
\begin{enumerate}[itemsep=-0.6ex,topsep=-0.6ex]
\item \textbf{Set union}: \(A \cup B = \{x \mid x \in A \text{ or } x \in B\}\).
\item \textbf{Set intersection}: \(A \cap B = \{x \mid x \in A \text{ and } x \in B\}\).
\item \textbf{Set difference}: \(A \setminus B = \{x \mid x \in A \text{ and } x \not\in B\}\).
\item \textbf{Cartesian product}: \(A \times B = \{(a, b) \mid a \in A \text{ and } b \in B\}\).
\item \textbf{Power set}: \(\mathcal{P}(A) = \{X \mid X \subseteq A\}\), \(\mathcal{P}_1(A) = \mathcal{P}(A) \setminus \emptyset\).
\end{enumerate}
\end{definition}

\begin{definition}
Let \(\mathbb{N} = \{0, 1, 2, \ldots\}\), and \(\mathbb{Z} = \{\ldots, -1, 0, 1, \ldots\}\).
\end{definition}


\section{Functions}

This section is derived from Chapter 3 of \parencite{Sutherland09} and Chapter 1 of \parencite{Howie95}. We use the conventional order of composition.

\begin{definition}
Let \(A, B\) be sets. A \textbf{function} \(f\) from \(A\) to \(B\) is a rule which assigns to each \(a \in A\) a \textbf{unique} \(b \in B\). We write \(b = f(a)\), \(f: A \to B\), and call \(a\) the \textbf{argument} of \(f\).
Formally, a \textbf{function} from \(A\) to \(B\) is a subset of \(A \times B\) such that for each \(a \in A\) there is \textbf{exactly one} element \((a, b)\) in \(f\).
\end{definition}

\begin{definition}
Let \(A, B, C, D\) be sets. If \(f: A \to B\), \(g: C \to D\) are \textbf{functions}, then \(f\) and \(g\) are \textbf{equal} (\(f = g\)) iff they are equal as sets.
\end{definition}

\begin{definition}
Let \(A, B, C\) be sets. If \(f: A \to B\), \(g: B \to C\) are \textbf{functions}, then we form a new \textbf{function} \((g \circ f): A \to C\) the \textbf{composite} of \(f\) and \(g\) by the rule \((g \circ f)(a) = g(f(a))\).
\end{definition}

\begin{proposition}
\textbf{Composition} of functions is \textbf{associative}. That is, given \(f: A \to B\), \(g: B \to C\), \(h: C \to D\), then \(h \circ (g \circ f) = (h \circ g) \circ f\).
\end{proposition}

\begin{definition}
For any set \(A\), the \textbf{identity function} on \(A\), \(I_A: A \to A\) is defined by \(\forall a \in A, I_A(a) = a\).
\end{definition}

\newpage
\begin{proposition}
If \(f: B \to A\), then \(I_A \circ f = f\). If \(g: A \to C\), \(g \circ I_A= g\).
\end{proposition}

\begin{definition}
Let \(f: A \to B\) be a \textbf{function}. Then a \textbf{function} \(g: B \to A\) is the \textbf{inverse} of \(f\) iff \(g \circ f = I_A\) and \(f \circ g = I_B\)
\end{definition}

\begin{proposition}
Let \(f: A \to B\) be a \textbf{function}. Then, if an \textbf{inverse} exists, it is unique, and is denoted \(f^{-1}: B \to A\).
\end{proposition}

\begin{definition}
Let \(f: A \to B\) be a \textbf{function}. Then \(f\) is \textbf{injective} iff \(\forall a, b \in A, f(a) = f(b)\) implies \(a = b\). \(f\) is \textbf{surjective} iff \(\forall a \in A, \exists b \in B, f(a) = b\). If \(f\) satisfies both properties, then it is \textbf{bijective}.
\end{definition}

\begin{lemma}
A \textbf{function} has an \textbf{inverse} iff it is a \textbf{bijection}.
\end{lemma}

\begin{definition}
Let \(f: A \to B\) be a \textbf{function}, \(X \subseteq A\), and \(Y \subseteq B\). Then the \textbf{image} of \(A\) under \(f\) is \(f(A) = \{f(a) \mid a \in A\} \subseteq B\), and the \textbf{preimage} of \(B\) is \(f^{-1}(B) = \{a \in A \mid f(a) \in B\} \subseteq A\).
\end{definition}

\begin{remark}
This does not imply the existence of an inverse, but if it does exist, then preimage of \(f\) coincides with the image of \(f^{-1}\).
\end{remark}

\begin{definition}
Let \(f: A \to B\) be a \textbf{function}, and \(X \subseteq A\). Then the \textbf{restriction} of \(f\) to \(X\) is \(\restr{f}{X}: X \to B\) is defined by \(\forall x \in X, \restr{f}{X}(x) = f(x)\).
\end{definition}

\begin{definition}
A \textbf{partial function} \(f: A \to B\) is a subset \(f\) of \(A \times B\) such that there is \textbf{at most one} element \((a, b)\) in \(f\).
\end{definition}


\section{Binary Relations}

This section is derived from Chapter 2 of \parencite{Sutherland09}, Chapter 1 of \parencite{Howie95} and Appendix A of \parencite{Baader98}.
\vspace{-0.1em}

\begin{definition} \label{def:binrel}
Let \(A\) be a set. Then a \textbf{binary relation} on \(A\) is a subset \(R\) of \(A \times A\). For any \(a, b \in A\), we write \(a R b\) iff \((a, b) \in R\).
\end{definition}
\vspace{-0.1em}

\begin{definition}
Let \(A\) be a set. Then, the \textbf{identity relation} on \(A\) is \(\iota_A = \{(a, a) \mid a \in A\}\), and the \textbf{universal relation} on \(A\) is \(\omega_A = A \times A\).
\end{definition}
\vspace{-0.1em}

\begin{samepage}
\begin{definition}
Let \(A\) be a set. Then we call a \textbf{binary relation} \(R\) on \(A\) \textbf{functional} iff for any \(a, b, c \in A\), \(a R b\) and \(a R c\) implies \(b = c\).
\end{definition}
\vspace{-0.1em}

\begin{definition} \label{dfn:relcomposition}
Let \(A\) be a set, and \(R, S\) be \textbf{binary relations} on \(A\). Then the \textbf{composition} of \(R\) and \(S\) is \(S \circ R = \{(x, y) \in A \times A \mid \exists z \in A \text{ with } x R z \text{ and } z S y\}\). Define \(R^0 = \iota_A\), and \(\forall n \in \mathbb{N}^+, R^n = R \circ R^{n-1}\).
\end{definition}
\vspace{-0.1em}

\begin{definition}
Let \(A\) be a set. Then the \textbf{inverse} of a \textbf{binary relation} \(R\) on \(A\) is \(R^{-1} = \{(b, a) \in A \times A \mid a R b\}\).
\end{definition}
\vspace{-0.1em}

\begin{proposition}
When considered as \textbf{binary relations}, \textbf{functions} and \textbf{partial functions} are \textbf{functional}. Moreover, the definitions of \textbf{composition} and \textbf{inverses} coincide.
\end{proposition}
\vspace{-0.2em}
\end{samepage}

\begin{definition}
A \textbf{binary relation} \(R\) on \(A\) is:
\begin{enumerate}[itemsep=-0.6ex,topsep=-0.6ex]
\item \textbf{Reflexive} iff \(\iota_A \subseteq R\).
\item \textbf{Irreflexive} iff \(\iota_A \cap R = \emptyset\).
\item \textbf{Symmetric} iff \(R = R^{-1}\).
\item \textbf{Antisymmetric} iff \(R \cap R^{-1} \subseteq \iota_A\).
\item \textbf{Transitive} iff \(R \circ R \subseteq R\).
\item \textbf{Connex} iff \(\omega_A \setminus \, \iota_A \subseteq R \cup R^{-1}\).
\end{enumerate}
\end{definition}

\begin{definition} \label{dfn:preorder}
A \textbf{binary relation} is a \textbf{preorder} iff it is \textbf{reflexive} and \textbf{transitive}. A \textbf{symmetric preorder} is called an \textbf{equivalence}.
\end{definition}

\begin{proposition}
The \textbf{classes} \([a] = \{b \in A \mid a \sim b\}\) of an \textbf{equivalence} \(\sim\) on \(A\) \textbf{partition} \(A\) into a union of pairwise disjoint non-empty subsets.
\end{proposition}

\begin{definition} \label{dfn:quotient}
Given an \textbf{equivalence} \(\sim\), define \(A /\!\!\sim \,= \{[a] \mid a \in A\}\).
\end{definition}


\section{Orders}

This section is derived from Chapter 1 of \parencite{Howie95} and Appendix A and of \parencite{Baader98}.

\begin{definition}
An \textbf{antisymmetric preorder} \(\leq\) on \(X\) is called a \textbf{partial order}, and we call \((X, \leq)\) a \textbf{partially ordered set} (\textbf{poset}).
\end{definition}

\begin{definition}
A \textbf{partial order} satisfying the \textbf{connex} property is called a \textbf{total order}, giving a \textbf{totally ordered set}.
\end{definition}

\begin{definition}
A \textbf{strict order} is an \textbf{irreflexive}, \textbf{transitive} relation.
\end{definition}

\begin{proposition}
Every \textbf{partial order} \(\leq\) induces a \textbf{strict order} \(\leq \setminus \,\, \iota\), and every \textbf{strict order} \(<\) induces a \textbf{partial order} \(< \cup \,\, \iota\).
\end{proposition}

\begin{definition}
Let \((X, \leq)\) be a poset, and \(\emptyset \neq Y \subseteq X\). Then:
\begin{enumerate}[itemsep=-0.6ex,topsep=-0.6ex]
\item \(a \in Y\) is \textbf{minimal} iff \(\forall y \in Y, y \leq a\) implies \(y = a\);
\item \(b \in Y\) is the \textbf{minimum} iff \(\forall y \in b \leq y\);
\item \(c \in X\) is a \textbf{lower bound} for \(Y\) iff \(\forall y \in Y, c \leq y\).
\end{enumerate}
\end{definition}

\begin{proposition}
Let \((X, \leq)\) be a poset, and \(\emptyset \neq Y \subseteq X\). Then every \textbf{minimum} element of \(Y\) is \textbf{minimal}, \(Y\) and has \textbf{at most} one \textbf{minimum}.
\end{proposition}

\begin{definition} \label{dfn:wellfounded}
We say that the poset \((X, \leq)\) satisfies the \textbf{minimal condition} (\textbf{well-founded}) iff every non-empty subset of \(X\) has a \textbf{minimal} element. If \(\leq\) is also a \textbf{total order}, then we say it is \textbf{well-ordered}.
\end{definition}

\begin{definition} \label{dfn:monotone}
Let \((X, \leq_X)\), \((Y, \leq_Y)\) be posets. Then a function \(\varphi: X \to Y\) is called \textbf{monotone} iff \(a \leq_X b\) implies \(\varphi(a) \leq_Y \varphi(b)\).
\end{definition}


\section{Sets II}

This section is derived from Chapter 7 of \parencite{Weiss08}, Part I of \parencite{Munkres18}, Chapter 8 of \parencite{Martin11}, and Chapter 1 of \parencite{Weihrauch13}.

\begin{theorem}[Well-Ordered Sets]
Every set can be \textbf{well-ordered}, and every \textbf{well-ordered} set is \textbf{isomorphic} to an \textbf{ordinal} (see \parencite{Weiss08} for details).
\end{theorem}

\begin{proposition}
We define the \textbf{cardinality} of \(A\) (\(\abs{A}\)), to be the \textbf{least ordinal} \(\kappa\) such that there is some \textbf{bijection} \(f: A \to \kappa\). Every set has \textbf{unique} cardinality. All sets with \textbf{cardinality} \(\leq\) to that of \(\mathbb{N}\) are \textbf{countable}.
\end{proposition}

\begin{theorem}[Countable Sets] \label{thm:setprod}
The \textbf{Cartesian product} of two \textbf{countable} sets is \textbf{countable}, and a \textbf{countable union} of \textbf{countable} sets is \textbf{countable}. If \(A\) is \textbf{finite}, then \(\mathcal{P}(A)\) is \textbf{finite}. The set \(\mathcal{P}(\mathbb{N})\) is \textbf{uncountable}.
\end{theorem}

\begin{definition}
A \textbf{partial function} \(f: \mathbb{N} \to \mathbb{N}\) is \textbf{computable} iff there exists a \textbf{Turing Machine} that computes \(f\) (see \parencite{Martin11} for details).
\end{definition}

\begin{definition} \label{dfn:decidable}
A \textbf{countable} set \(A \subseteq \mathbb{N}\) has \textbf{characteristic function} \(\chi_A: \mathbb{N} \to \{0, 1\}\) defined by \(\forall x \in \mathbb{N}, \chi_A(x) = 1\) iff \(x \in A\). \(A\) is \textbf{decidable} or \textbf{recursive} iff \(\chi_A\) is \textbf{computable}. Otherwise, \(A\) is \textbf{undecidable}.
\end{definition}


\section{Categories}

This section is derived from Chapter 1 of \parencite{MacLane98} and Chapter 1 of \parencite{Awodey10}.

\begin{definition} \label{def:cat}
A \textbf{category} consists of the following data:
\begin{enumerate}[itemsep=-0.6ex,topsep=-0.6ex]
\item \textbf{Objects}: \(A\), \(B\), \(C\), \dots
\item \textbf{Arrows}: \(f\), \(g\), \(h\), \dots
\item For each arrow \(f\), there are given objects \(\operatorname{dom}(f)\), \(\operatorname{cod}(f)\), and we write \(f: A \to B\) to indicate that \(A = \operatorname{dom}(f)\), \(B = \operatorname{cod}(f)\);
\item Given arrows \(f: A \to B\), \(g: B \to C\), \(g \circ f\) is an arrow such that \(A = \operatorname{dom}(g \circ f)\), \(C = \operatorname{cod}(g \circ f)\);
\item For each object \(A\) there is given an arrow \(1_A: A \to A\);
\end{enumerate}
such that for all \(f: A \to B, g: B \to C, h: C \to D\), \(h \circ (g \circ f) = (h \circ g) \circ f\) and \(f \circ 1_A = f = 1_B \circ f\). Moreover, it is \textbf{locally small} iff the collection of arrows between any two objects is a \textbf{set}.
\end{definition}

\begin{definition}
A \textbf{functor} \(F: \pmb{C} \to \pmb{D}\) between categories \(\pmb{C}\), \(\pmb{D}\) is a mapping such that:
\begin{enumerate}[itemsep=-0.6ex,topsep=-0.6ex]
\item \(F(f\!: A \to B) = F(f)\!: F(A) \to F(B)\);
\item \(F(1_A) = 1_{F(A)}\);
\item \(F(g \circ f) = F(g) \circ F(f)\).
\end{enumerate}
\end{definition}

\chapter{Abstract Reduction Systems} \label{appendix:ars}

The definitions and theorems in this appendix are derived from Chapter 2 of \parencite{Baader98}, Section 2.2 of \parencite{Plump99}, and Section 1.1 of \parencite{Book-Otto93}.


\section{Basic Definitions}

\begin{definition} \label{dfn:ars0}
An \textbf{abstract reduction system} (ARS) is a pair \((A, \rightarrow)\) where \(A\) is a \textbf{set} and \(\rightarrow\) a \textbf{binary relation} on \(A\).
\end{definition}

\begin{definition} \label{dfn:ars1}
Let \((A, \rightarrow)\) be an ARS. We define the notation:
\begin{enumerate}[itemsep=-0.7ex,topsep=-0.7ex]
\item \textbf{Composition}: \(\xrightarrow{n} \defeq \rightarrow^n\) (\(n \geq 0\)).
\item \textbf{Transitive closure}: \(\xrightarrow{+} \defeq \bigcup_{n \geq 1} \xrightarrow{n}\).
\vspace{-0.2em}
\item \textbf{Reflexive transitive closure}: \(\xrightarrow{*} \defeq \xrightarrow{+} \cup \xrightarrow{0}\).
\item \textbf{Reflexive closure}: \(\xrightarrow{=} \defeq \rightarrow \cup \xrightarrow{0}\).
\item \textbf{Inverse}: \(\leftarrow \defeq \rightarrow^{-1}\).
\item \textbf{Symmetric closure}: \(\leftrightarrow \defeq \rightarrow \cup \leftarrow\).
\end{enumerate}
\end{definition}

\begin{remark}
It is usual that \(\rightarrow\) is \textbf{decidable}. This \textbf{does not} imply that \(\xrightarrow{+}\) is \textbf{decidable}, only that \(\xrightarrow{n}\) is \textbf{decidable}.
\end{remark}

\begin{definition} \label{dfn:ars2}
Let \((A, \rightarrow)\) be an ARS. We say that:
\begin{enumerate}[itemsep=-0.7ex,topsep=-0.7ex]
\item \(x\) is \textbf{reducible} iff there is a \(y\) s.t. \(x \rightarrow y\).
\item \(x\) is \textbf{in normal form} iff \(x\) is not reducible.
\item \(y\) is \textbf{a normal form} of \(x\) iff \(x \xrightarrow{*} y\) and \(y\) is in normal form. If \(x\) has a \textbf{unique normal form}, it is denoted \(x\!\!\downarrow\).
\vspace{-0.2em}
\item \(y\) is a \textbf{successor} to \(x\) iff \(x \xrightarrow{+} y\), and a \textbf{direct successor} iff \(x \rightarrow y\).
\vspace{-0.2em}
\item \(x\) and \(y\) are \textbf{joinable} iff there is a \(z\) s.t. \(x \xrightarrow{*} z \xleftarrow{*} y\). We write \(x \downarrow y\).
\end{enumerate}
\end{definition}

\begin{definition} \label{dfn:ars3}
Let \((A, \rightarrow)\) be an ARS. Then \(\rightarrow\) is called:
\begin{enumerate}[itemsep=-0.7ex,topsep=-0.7ex]
\item \textbf{Church-Rosser} iff \(x \xleftrightarrow{*} y\) implies \(x \downarrow y\).
\vspace{-0.2em}
\item \textbf{Semi-confluent} iff \(y_1 \leftarrow x \xrightarrow{*} y_2\) implies \(y_1 \downarrow y_2\).
\vspace{-0.2em}
\item \textbf{Confluent} iff \(y_1 \xleftarrow{*} x \xrightarrow{*} y_2\) implies \(y_1 \downarrow y_2\).
\item \textbf{Terminating} iff there is no infinite descending chain \(x_0 \rightarrow x_1 \rightarrow \dots\).
\item \textbf{Normalising} iff every element has a normal form.
\item \textbf{Convergent} iff it is both confluent and terminating.
\end{enumerate}
\end{definition}

\begin{remark}
Other texts call a \textbf{terminating} reduction \textbf{uniformly terminating} or \textbf{Noetherian}, or say it satisfies the \textbf{descending chain condition}.
\end{remark}


\newpage
\section{Noetherian Induction}

The principle of \textbf{Noetherian induction} (\textbf{well-founded induction}) is a generalisation of induction from \((\mathbb{N}, >)\) to any \textbf{terminating} reduction system.

\begin{definition}
Let \((A, \rightarrow)\) be an ARS, and \(P\) is some property of the elements of \(A\). Then the inference rule for \textbf{Noetherian Induction} is:
\noindent
\vspace{-0.2em}
\begin{align*}
\infer{\forall x \in A, P(x)}{\forall x \in A, (\forall y \in A, x \xrightarrow{+} y \Rightarrow P(y)) \Rightarrow P(x)}
\end{align*}
\end{definition}

\begin{theorem}[Noetherian Induction] \label{thm:ninduct}
Let \((A, \rightarrow)\) The following are equivalent for an ARS:
\begin{enumerate}[itemsep=-0.7ex,topsep=-0.7ex]
\item The principle of \textbf{Noetherian induction} holds.
\item \(\rightarrow\) is \textbf{well-founded} (Definition \ref{dfn:wellfounded}).
\item \(\rightarrow\) is \textbf{terminating} (Definition \ref{dfn:ars3}).
\end{enumerate}
\end{theorem}

\begin{definition} \label{dfn:branching}
Let \((A, \rightarrow)\) be an ARS. Then \(\rightarrow\) is called
\begin{enumerate}[itemsep=-0.7ex,topsep=-0.7ex]
\item \textbf{Finitely branching} iff each \(a\) has only finitely many direct successors.
\item \textbf{Globally finite} iff each \(a\) has only finitely many successors.
\vspace{-0.2em}
\item \textbf{Acyclic} iff there is no \(a\) such that \(a \xrightarrow{+} a\).
\end{enumerate}
\end{definition}

\begin{lemma} \label{lem:arsbranching}
Let \((A, \rightarrow)\) be an ARS. Then:
\begin{enumerate}[itemsep=-0.7ex,topsep=-0.7ex]
\item If \(\rightarrow\) is \textbf{finitely branching} and \textbf{terminating}, then it is \textbf{globally finite}.
\item If \(\rightarrow\) is \textbf{acyclic} and \textbf{globally finite}, then it is \textbf{terminating}.
\vspace{-0.2em}
\item \(\rightarrow\) is \textbf{acyclic} iff \(\xrightarrow{+}\) is a \textbf{strict order}.
\end{enumerate}
\end{lemma}


\section{Confluence and Termination}

\begin{theorem}[Church-Rosser]
Let \((A, \rightarrow)\) be an ARS. Then, \(\rightarrow\) has the \textbf{Church-Rosser} property iff it is \textbf{semi-confluent} iff it is \textbf{confluent}.
\end{theorem}

\begin{theorem}[Normal Forms]
Let \((A, \rightarrow)\) be an ARS. If \(\rightarrow\) is \textbf{confluent}, then every element has at most one \textbf{normal form}. Moreover, if \(\rightarrow\) is \textbf{confluent} and \textbf{normalising}, then \(x \xleftrightarrow{*} y\) iff \(x\!\!\downarrow = y\!\!\downarrow\).
\end{theorem}

\begin{definition}
Let \((A, \rightarrow)\) be an ARS. Then \(\rightarrow\) is called \textbf{locally confluent} iff \(y_1 \leftarrow x \rightarrow y_2\) implies \(y_1 \downarrow y_2\).
\end{definition}

\begin{theorem}[Newman's Lemma] \label{thm:newmanlem}
A \textbf{terminating} relation is \textbf{confluent} iff it is \textbf{locally confluent}.
\end{theorem}

\begin{lemma} \label{lem:monoembedding}
A \textbf{finitely branching} reduction \textbf{terminates} iff there is a \textbf{monotone} (Definition \ref{dfn:monotone}) embedding into \((\mathbb{N}, >)\).
\end{lemma}

\chapter{Graph Transformation} \label{appendix:transformation}

We give a quick introduction to the theory of algebraic graph transformation derived from my earlier literature review \parencite{Campbell18}, which is in turn derived from \parencite{Ehrig06}. We generalise to \textbf{partially labelled graphs} using \parencite{Habel-Plump02} and \parencite{Plump19}, in that we allow \textbf{relabelling} of \textbf{totally labelled graphs}.

The additional section on pushouts and pullbacks is derived from \parencite{Ehrig06} and \parencite{Plump19}, the section on critical pair analysis is derived from \parencite{Plump19}, and the section on rooted graphs is derived from \parencite{Bak-Plump12}. The definition of an unlabelled graph can be found in Section \ref{sec:maingraphs}. The proof of Theorem \ref{theorem:uniquederivations} is given by \parencite{Habel-Plump02}, and of Theorem \ref{thm:undecidablegts} is given by the proof of Theorem \ref{thm:undecidablegtsrealdeal}.


\section{Partially Labelled Graphs} \label{sec:graphpl}

\begin{definition} \label{dfn:labelalphabet}
A label alphabet \(\mathcal{L} = (\mathcal{L}_V, \mathcal{L}_E)\) consists of \textbf{finite} sets of \textbf{node labels} \(\mathcal{L}_V\) and \textbf{edge labels} \(\mathcal{L}_E\).
\end{definition}

\begin{definition} \label{dfn:plgraph}
A \textbf{concrete partially labelled graph} over a label alphabet \(\mathcal{L}\) is a \textbf{concrete graph} equipped with two \textbf{partial} label maps \(l: V \to \mathcal{L}_V\), \(m: E \to \mathcal{L}_E\): \(G = (V, E, s, t, l, m)\).
\end{definition}

\vspace{-1.4em}
\vspace{0.2em}
\begin{figure}[H]
\centering
\noindent
\begin{equation*}
\begin{tikzcd}
  \mathcal{L}_E
    & E \arrow[l, "m"] \arrow[r, shift left=1ex, "s"] \arrow[r, shift right=1ex, "t"]
    & V \arrow[r, "l"]
    & \mathcal{L}_V
\end{tikzcd}
\end{equation*}
\vspace{-1.4em}
\caption{Partially Labelled Graph Diagram}
\end{figure}
\vspace{-0.2em}

\begin{remark}
By this definition, we \textbf{do not} work with the free monoid on the alphabet, as in string rewriting systems. Nodes and edges are labelled exactly with the elements from the respective alphabets.
\end{remark}

\begin{definition}
We say that a \textbf{partially labelled graph} \(G\) is \textbf{totally labelled} iff \(l_G\) is total.
\end{definition}

\begin{definition}
Given a common \(\mathcal{L}\), a \textbf{partially labelled graph morphism} \(g: G \to H\) is a graph morphism on the underlying concrete graphs, with the extra constraint that labels must be preserved, if defined. That is:
\begin{enumerate}[itemsep=-0.6ex,topsep=-0.6ex]
\item \(\forall e \in E_G, \, g_V(s_G(e))= s_H(g_E(e))\);                 \hspace{2.185cm} [Sources]
\item \(\forall e \in E_G, \, g_V(t_G(e)) = t_H(g_E(e))\);                \hspace{2.230cm} [Targets]
\item \(\forall e \in E_G, \, m_G(e) = m_H(g_E(e))\);                     \hspace{2.709cm} [Edge Labels]
\item \(\forall v \in l_G^{-1}(\mathcal{L}_V), \, l_G(v) = l_H(g_V(v))\). \hspace{1.905cm} [Node Labels]
\end{enumerate}
\end{definition}

\begin{definition}
Given a common \(\mathcal{L}\), a partially labelled graph morphism \(g: G \to H\) is \textbf{injective}/\textbf{surjective} iff the underlying graph morphism is injective/surjective.
\end{definition}

\begin{definition}
Given a common \(\mathcal{L}\), we say \(H\) is a \textbf{subgraph} of \(G\) iff there exists an \textbf{inclusion morphism} \(H \hookrightarrow G\). This happens iff \(V_H \subseteq V_G\), \(E_H \subseteq E_G\), \(s_H = \restr{s_G}{E_H}\), \(t_H = \restr{t_G}{E_H}\), \(m_H = \restr{m_G}{E_G}\), \(l_H \subseteq l_G\).
\end{definition}

\begin{remark}
Given a \textbf{totally labelled graph} \(G\), and \(H\) \textbf{partially labelled}. If there exists a \textbf{surjective morphism} \(G \to H\), then \(H\) is \textbf{totally labelled}.
\end{remark}

\begin{definition}
We say that graphs \(G, H\) are \textbf{isomorphic} iff there exists an \textbf{injective}, \textbf{surjective} graph morphism \(g: G \to H\) such that \(g^{-1}: H \to G\) is a graph morphism. We write \(G \cong H\), and call \(g\) an \textbf{isomorphism}. This naturally gives rise to \textbf{equivalence classes} \([G]\): the \textbf{countably} many partially labelled abstract graphs over some fixed \(\mathcal{L}\).
\end{definition}


\section{Typed Graphs} \label{section:typed}

\begin{definition}
A \textbf{typed graph} is the tuple \(G_T = (G, type_G)\) where \(G\) is an \textbf{unlabelled graph}, and \(type_G\) is a graph morphism \(G \to TG\) where \(TG\) is an \textbf{unlabelled graph} called a \textbf{type graph}. The vertices and edges of \(TG\) are called the \textbf{node alphabet} and \textbf{edge alphabet}.
\end{definition}

\begin{definition}
Given two \textbf{typed graphs} \(G_T, H_T\), a \textbf{typed graph morphism} is an \textbf{unlabelled graph morphism} \(f: G \to H\) such that \(type_H \circ f = type_G\).
\end{definition}

\begin{theorem}[Typed-Labelled Graph Correspondence]
There is a \textbf{bijective correspondence} between the \textbf{totally labelled graphs} over some fixed \textbf{label alphabet} \(\mathcal{L}\) and the \textbf{typed graphs} over \(\mathcal{L}\).
\end{theorem}


\section{Performance Assumptions}

We will assume that graphs are stored in a format such that the time complexities of various problems are as given in the table \parencite{Dodds08}.

\vspace{-0.8em}
\vspace{0.2em}
\begin{figure}[H]
\noindent
\begin{center}
\begin{tabular}{l|l|l}
Input                   & Output                                             & Time           \\ \hline
label \(l\)             & The set \(X\) of nodes with label \(l\).           & \(O(\abs{X})\) \\
node \(v\)              & Values \(deg(v)\), \(indeg(v)\), \(outdeg(v)\).    & \(O(1)\)       \\
node \(v\), label \(l\) & No. nodes with source \(v\), label \(l\).          & \(O(1)\)       \\
node \(v\), label \(l\) & No. nodes with target \(v\), label \(l\).          & \(O(1)\)       \\
node \(v\), label \(l\) & Set \(X\) of nodes with source \(v\), label \(l\). & \(O(\abs{X})\) \\
node \(v\), label \(l\) & Set \(X\) of nodes with target \(v\), label \(l\). & \(O(\abs{X})\) \\
graph \(G\)             & \(\abs{V_G}\) and \(\abs{E_G}\).                   & \(O(1)\)       \\
\end{tabular}
\end{center}

\vspace{-1.0em}
\caption{Complexity Assumptions Table}
\label{fig:perfass}
\vspace{-1.6em}
\end{figure}
\vspace{-0.2em}


\section{Pushouts and Pullbacks}

Pushouts and pullbacks are limits, in the sense of category theory. Our definitions are for any category (Definition \ref{def:cat}), and the propositions hold in the category of unlabelled concrete graphs, but not necessarily in others.

\vspace{-0.6em}
\vspace{0.2em}
\begin{figure}[H]
\centering
\noindent
\begin{equation*}
\begin{tikzcd}
  A \arrow[r] \arrow[d]
    & B \arrow[d] \arrow[ddr, bend left=20]
    & \text{}
    & A' \arrow[drr, bend left=20] \arrow[ddr, bend right=20] \arrow[dr, dashed]
    & \text{}
    & \text{} \\
  C \arrow[r] \arrow[drr, bend right=20]
    & D \arrow[dr, dashed]
    & \text{}
    & \text{}
    & A \arrow[r] \arrow[d]
    & B \arrow[d] \\
  \text{}
    & \text{}
    & D'
    & \text{}
    & C \arrow[r]
    & D
\end{tikzcd}
\end{equation*}
\vspace{-1.5em}
\caption{Pushout and Pullback}
\vspace{-0.2em}
\end{figure}
\vspace{-0.2em}

\begin{definition} \label{dfn:pushout}
Given graph morphisms \(A \to B\) and \(A \to C\), a graph \(D\) together with graph morphisms \(B \to D\) and \(C \to D\) is a \textbf{pushout} iff:
\begin{enumerate}[itemsep=-0.6ex,topsep=-0.6ex]
\item \textbf{Commutativity}: \(A \to B \to D = A \to C \to D\).
\item \textbf{Universal property}: For all morphisms \(B \to D'\), \(C \to D'\) such that \(A \to B \to D' = A \to C \to D'\), there is a unique morphism \(D \to D'\) such that \(C \to D \to D' = B \to D'\) and \(C \to D \to D' = C \to D'\).
\end{enumerate}
\end{definition}

\begin{proposition}
Every pushout satisfies the following:
\begin{enumerate}[itemsep=-0.6ex,topsep=-0.6ex]
\item \textbf{No junk}: Each item in \(D\) has a preimage in \(B\) or \(C\).
\item \textbf{No confusion}: If \(A \to B\), \(A \to C\) injective, then \(B \to D\), \(C \to D\) injective and an item from \(B\) is merged in \(D\) with an item from \(C\) only if the items have a common preimage in \(A\).
\end{enumerate}
\end{proposition}

\begin{definition} \label{dfn:pullback}
Given graph morphisms \(B \to D\) and \(C \to D\), a graph \(A\) together with graph morphisms \(A \to B\) and \(A \to C\) is a \textbf{pullback} iff:
\begin{enumerate}[itemsep=-0.6ex,topsep=-0.6ex]
\item \textbf{Commutativity}: \(A \to B \to D = A \to C \to D\).
\item \textbf{Universal property}: For all morphisms \(A' \to B\), \(A' \to C\) such that \(A' \to B \to D = A' \to C \to D\), there is a unique morphism \(A' \to A\) such that \(A' \to A \to B = A' \to B\) and \(A' \to A \to C = A' \to C\).
\end{enumerate}
\end{definition}

\begin{definition} \label{dfn:npo}
A \textbf{pushout} that is a \textbf{pullback} is called a \textbf{natural pushout}.
\end{definition}

\begin{proposition}
A pushout is \textbf{natural} if \(A \to B\) is injective.
\end{proposition}

\begin{theorem}[Limit Uniqueness]
In any category, if they exist, in a \textbf{pushout} (\textbf{pullback}), \(D\) (\(A\)) are unique up isomorphism.
\end{theorem}

\begin{definition}
Given graph morphisms \(A \to B\) and \(B \to D\), a (natural) \textbf{pushout complement} is a graph \(C\) together with morphisms \(A \to C\) and \(C \to D\) such that the resulting square is a (natural) \textbf{pushout}.
\end{definition}

\begin{theorem}[Limit Existence]
In the category of unlabelled graphs, \textbf{pushouts}, \textbf{pushout complements}, and \textbf{pullbacks} always exist.
\end{theorem}


\newpage
\section{Rules and Derivations} \label{section:gt1}

Let \(\mathcal{L} = (\mathcal{L}_V, \mathcal{L}_E)\) be the ambient label alphabet, and graphs be concrete.

\begin{definition}
A \textbf{rule} \(r = \langle L \leftarrow K \rightarrow R \rangle\) consists of \textbf{totally labelled graphs} \(L\), \(R\) over \(\mathcal{L}\), the \textbf{partially labelled graph} \(K\) over \(\mathcal{L}\), and \textbf{inclusions} \(K \hookrightarrow L\) and \(K \hookrightarrow R\).
\end{definition}

\begin{definition}
We define the \textbf{inverse rule} to be \(r^{-1} = \langle R \leftarrow K \rightarrow L \rangle\).
\end{definition}

\begin{definition}
If \(r = \langle L \leftarrow K \rightarrow R \rangle\) is a \textbf{rule}, then \(\abs{r} = max \{\abs{L},\abs{R}\}\).
\end{definition}

\begin{definition}
Given a \textbf{rule} \(r = \langle L \leftarrow K \rightarrow R \rangle\) and a \textbf{totally labelled graph} \(G\), we say that an \textbf{injective} morphism \(g: L \hookrightarrow G\) satisfies the \textbf{dangling condition} iff no edge in \(G \setminus g(L)\) is incident to a node in \(g(L \setminus K)\).
\end{definition}

\begin{definition} \label{def:rukeapp}
To \textbf{apply} a rule \(r = \langle L \leftarrow K \rightarrow R \rangle\) to some \textbf{totally labelled graph} \(G\), find an \textbf{injective} graph morphism \(g: L \hookrightarrow G\) satisfying the \textbf{dangling condition}, then:

\begin{enumerate}[itemsep=-0.7ex,topsep=-0.7ex]
\item Delete \(g(L \setminus K)\) from \(G\), and for each unlabelled node \(v\) in \(K\), make \(g_V(v)\) unlabelled, giving the \textbf{intermediate graph} \(D\);
\item Add disjointly \(R \setminus K\) to D, keeping their labels, and for each unlabelled node \(v\) in \(K\), label \(g_V(v)\) with \(l_R(v)\), giving the \textbf{result graph} \(H\).
\end{enumerate}

\noindent
If the \textbf{dangling condition} fails, then the rule is not applicable using the \textbf{match} \(g\). We can exhaustively check all matches to determine applicability.
\end{definition}

\begin{definition}
We write \(G \Rightarrow_{r,g} M\) for a successful application of \(r\) to \(G\) using match \(g\), obtaining result \(M \cong H\). We call Figure \ref {fig:directderivation} a \textbf{direct derivation}, and the injective morphism \(h\) the \textbf{comatch}.
\end{definition}
\vspace{-1.2em}
\vspace{0.2em}
\begin{figure}[H]
\centering
\noindent
\begin{equation*}
\begin{tikzcd}
  L \arrow[d, "g"]
  & K \arrow[l, ""] \arrow[d, "d"] \arrow[r, ""]
  & R \arrow[d, "h"] \\
  G 
    & D \arrow[l, ""] \arrow[r, ""]
    & H
\end{tikzcd}
\end{equation*}
\vspace{-1.4em}
\caption{Direct Derivation}
\label{fig:directderivation}
\end{figure}
\vspace{-0.2em}
\vspace{-0.4em}

\begin{theorem}[Derivation Uniqueness] \label{theorem:uniquederivations}
It turns out that \textbf{deletions} are \textbf{natural pushout complements} and \textbf{gluings} are \textbf{natural pushouts} in the category of partially labelled graphs. Moreover, direct derivations are \textbf{natural double pushouts}, \(D\) and \(H\) are \textbf{unique up to isomorphism}, and \(H\) is \textbf{totally labelled}. Moreover, derivations \(G \Rightarrow_{r,g} H\) are \textbf{invertible}.
\end{theorem}

\begin{definition}
Given a rule set \(\mathcal{R}\), we define \(\mathcal{R}^{-1} = \{r^{-1} \mid r \in \mathcal{R}\}\).
\end{definition}

\begin{definition} \label{def:directderives}
For a given set of rules \(\mathcal{R}\), we write \(G \Rightarrow_{\mathcal{R}} H\) iff \(H\) is \textbf{directly derived} from \(G\) using any of the rules from \(\mathcal{R}\).
\end{definition}

\begin{definition} \label{def:derives}
We write \(G \Rightarrow_{\mathcal{R}}^{+} H\) iff \(H\) is \textbf{derived} from \(G\) in one or more \textbf{direct derivations}, and \(G \Rightarrow_{\mathcal{R}}^{*} H\) iff \(G \cong H\) or \(G \Rightarrow_{\mathcal{R}}^{+} H\).
\end{definition}


\section{Transformation Systems} \label{section:gt2}

\begin{definition}
A \textbf{graph transformation system} \(T = (\mathcal{L}, \mathcal{R})\), consists of a label alphabet \(\mathcal{L} = (\mathcal{L}_V, \mathcal{L}_E)\), and a \textbf{finite} set \(\mathcal{R}\) of rules over \(\mathcal{L}\).
\end{definition}

\begin{proposition}
Given a \textbf{graph transformation system} \(T = (\mathcal{L}, \mathcal{R})\), then one can always decide if \(G \Rightarrow_{\mathcal{R}} H\).
\end{proposition}

\begin{definition}
Given a \textbf{graph transformation system} \(T = (\mathcal{L}, \mathcal{R})\), we define the inverse system \(T^{-1} = (\mathcal{L}, \mathcal{R}^{-1})\).
\end{definition}

\begin{definition}
Given a label alphabet \(\mathcal{L} = (\mathcal{L}_V, \mathcal{L}_E)\), \(\mathcal{P} = (\mathcal{P}_V, \mathcal{P}_E)\) is a \textbf{subalphabet} of \(\mathcal{L}\) iff \(\mathcal{P}_V \subseteq \mathcal{L}_V\) and \(\mathcal{P}_E \subseteq \mathcal{L}_E\).
\end{definition}

\begin{definition}
Given a \textbf{graph transformation system} \(T = (\mathcal{L}, \mathcal{R})\), a subalphabet of \textbf{non-terminals} \(\mathcal{N}\), and a \textbf{start graph} \(S\) over \(\mathcal{L}\), then a \textbf{graph grammar} is the system \(\pmb{G} = (\mathcal{L}, \mathcal{N}, \mathcal{R}, S)\).
\end{definition}

\begin{definition} \label{dfn:graphgrammar}
Given a \textbf{graph grammar} \(\pmb{G}\) as defined above, we say that a graph \(G\) is \textbf{terminally labelled} iff \(l(V) \cap \mathcal{N}_V = \emptyset\) and \(m(E) \cap \mathcal{N}_E = \emptyset\). Thus, we can define the \textbf{graph language} generated by \(\pmb{G}\):
\begin{align*}
\pmb{L}(\pmb{G}) = \{[G] \mid S \Rightarrow_{\mathcal{R}}^{*} G, G \text{ terminally labelled}\}
\end{align*}
\end{definition}

\vspace{-0.4em}
\begin{proposition} \label{prop:inversegram}
Given a \textbf{graph grammar} \(\pmb{G} = (\mathcal{L}, \mathcal{N}, \mathcal{R}, S)\), \(G \Rightarrow_{r} H\) iff \(H \Rightarrow_{r^{-1}} G\), for some \(r \in \mathcal{R}\) (simply use the comatch). Moreover, \([G] \in \pmb{L}(\pmb{G})\) iff \(G \Rightarrow_{\mathcal{R}^{-1}}^* S\) and \(G\) is terminally labelled.
\end{proposition}

\begin{remark}
Graph languages need not be finite. In fact, graph grammars are as powerful as unrestricted string grammars. As such, many questions like if the language is empty, are undecidable in general.
\end{remark}


\section{Confluence and Termination}

\begin{samepage}
Let \(T = (\mathcal{L}, \mathcal{R})\) be a graph transformation system.

\begin{definition}
The graphs \(H_1\), \(H_2\) are \textbf{joinable} iff there is a graph \(M\) such that \(H_1 \Rightarrow_{\mathcal{R}}^{*} M \Leftarrow_{\mathcal{R}}^{*} H_2\).
\end{definition}
\vspace{-0.1em}

\begin{definition}
\(T\) is \textbf{locally confluent} iff for all graphs \(G\), \(H_1\), \(H_2\) such that \(H_1 \Leftarrow_{\mathcal{R}} G \Rightarrow_{\mathcal{R}} H_2\), \(H_1\) and \(H_2\) are \textbf{joinable}.
\end{definition}
\vspace{-0.1em}

\begin{definition}
\(T\) is \textbf{confluent} iff for all graphs \(G\), \(H_1\), \(H_2\) such that \(H_1 \Leftarrow_{\mathcal{R}}^{*} G \Rightarrow_{\mathcal{R}}^{*} H_2\), \(H_1\) and \(H_2\) are \textbf{joinable}.
\end{definition}
\vspace{-0.1em}

\begin{definition}
\(T\) is \textbf{terminating} iff there is no infinite derivation sequence \(G_0 \Rightarrow_{\mathcal{R}} G_1 \Rightarrow_{\mathcal{R}} G_2 \Rightarrow_{\mathcal{R}} G_3 \Rightarrow_{\mathcal{R}} \cdots\).
\end{definition}
\vspace{-0.1em}

\begin{theorem}[Property Undecidability] \label{thm:undecidablegts}
Testing if \(T\) has (\textbf{local}) \textbf{confluence} or is \textbf{terminating} is \textbf{undecidable} in general.
\end{theorem}
\vspace{-0.2em}
\end{samepage}


\section{Critical Pair Analysis} \label{section:critpairs}

Throughout this section, we fix some common label alphabet \(\mathcal{L} = (\mathcal{L}_V, \mathcal{L}_E)\), and also require that the \textbf{interface} in all rules to be \textbf{totally labelled}.

\begin{samepage}
\begin{definition}
The derivations \(G_1 \Rightarrow_{r_1,g_1} H \Rightarrow_{r_2,g_2} G_2\) are \textbf{sequentially independent} iff \((h_1(R_1) \cap g_2(L_2)) \subseteq (h_1(K_1) \cap g_2(K_2))\).
\end{definition}
\vspace{-0.1em}

\begin{lemma}
The derivations \(G_1 \Rightarrow_{r_1,g_1} H \Rightarrow_{r_2,g_2} G_2\) are \textbf{sequentially independent} iff there exist morphisms \(R_1 \to D_2\) and \(L_2 \to D_1\) with \(R_1 \to D_1 \to H = R_1 \to H\) and \(L_1 \to D_2 \to H = L_2 \to H\).
\end{lemma}
\vspace{-0.1em}

\begin{theorem}[Sequential Independence]
If \(G_1 \Rightarrow_{r_1,g_1} H \Rightarrow_{r_2,g_2} G_2\) are \textbf{sequentially independent}, then there exists a graph \(\overbar{H}\) and \textbf{sequentially independent} steps \(G \Rightarrow_{r_2} \overbar{H} \Rightarrow_{r_1} G_2\).
\end{theorem}
\vspace{-0.1em}

\begin{definition} \label{dfn:parindep}
The derivations \(H_1 \Leftarrow_{r_1,g_1} G \Rightarrow_{r_2,g_2} H_2\) are \textbf{parallelly independent} iff \((g_1(L_1) \cap g_2(L_2)) \subseteq (g_1(K_1) \cap g_2(K_2))\).
\end{definition}
\vspace{-0.1em}

\begin{lemma}
The derivations \(H_1 \Leftarrow_{r_1,g_1} G \Rightarrow_{r_2,g_2} H_2\) are \textbf{parallelly independent} iff there exist morphisms \(L_1 \to D_2\) and \(L_2 \to D_1\) with \(L_1 \to D_2 \to G = L_1 \to G\) and \(L_2 \to D_1 \to G = L_2 \to G\).
\end{lemma}
\vspace{-0.1em}

\begin{lemma}
The derivations \(H_1 \Leftarrow_{r_1,g_1} G \Rightarrow_{r_2,g_2} H_2\) are \textbf{parallelly independent} iff \(H_1 \Rightarrow_{r_1^{-1},h_1} G \Rightarrow_{r_2,g_2} H_2\) are \textbf{sequentially independent}.
\end{lemma}
\vspace{-0.1em}

\begin{theorem}[Parallel Independence]
If \(H_1 \Leftarrow_{r_1,g_1} G \Rightarrow_{r_2,g_2} H_2\) are \textbf{parallelly independent}, then there exists a graph \(\overbar{G}\) and \textbf{direct derivations} \(H_1 \Rightarrow_{r_2} \overbar{G} \Leftarrow_{r_1} H_2\) with \(G \Rightarrow_{r_1} H_1 \Rightarrow_{r_2} \overbar{G}\) and \(G \Rightarrow_{r_2} H_2 \Rightarrow_{r_1} \overbar{G}\) \textbf{sequentially independent}.
\end{theorem}
\vspace{-0.1em}

\begin{definition} \label{dfn:critpair}
A pair of \textbf{direct derivations} \(G_1 \Leftarrow_{r_1,g_1} H \Rightarrow_{r_2,g_2} G_2\) is a \textbf{critical pair} iff \(H = g_1(L_1) \cup g_2(L_2)\), the steps are not \textbf{parallelly independent}, and if \(r_1 = r_2\) then \(g_1 \neq g_2\).
\end{definition}
\vspace{-0.1em}

\begin{definition}
Let \(G \Rightarrow H\) be a \textbf{direct derivation}. Then the \textbf{track morphism} is defined to be the partial morphism \(\mathit{tr}_{G \Rightarrow H} = \mathit{in}' \circ \mathit{in}^{-1}\). We define \(\mathit{tr}_{G \Rightarrow^* H}\) inductively as the composition of track morphisms.
\end{definition}
\vspace{-0.1em}

\begin{definition}
The set of \textbf{persistent nodes} of a critical pair \(\Phi : H_1 \Leftarrow G \Rightarrow H_2\) is \(\mathit{Persist}_{\Phi} = \{v \in G_V \mid \mathit{tr}_{G \Rightarrow H_1}(\{v\}), \mathit{tr}_{G \Rightarrow H_2}(\{v\}) \neq \emptyset\}.\)
\end{definition}
\vspace{-0.1em}

\begin{definition} \label{dfn:strongjoin}
A critical pair \(\Phi : H_1 \Leftarrow G \Rightarrow H_2\) is \textbf{strongly joinable} iff there exists a graph \(M\), a derivation \(H_1 \Rightarrow_{\mathcal{R}}^* M \Leftarrow_{\mathcal{R}}^* H_2\) and:
\vspace{-0.4em}
\begin{align*}
\forall v \in \mathit{Persist}_{\Phi}, \mathit{tr}_{G \Rightarrow H_1 \Rightarrow^* \overbar{G}}(\{v\}) = \mathit{tr}_{G \Rightarrow H_2 \Rightarrow^* \overbar{G}}(\{v\}) \neq \emptyset
\end{align*}
\vspace{-2.0em}
\end{definition}
\vspace{-0.2em}

\begin{theorem}[Critical Pair Lemma] \label{thm:critpairlem}
A graph transformation system \(T\) is \textbf{locally confluent} if all its \textbf{critical pairs} are \textbf{strongly joinable}.
\end{theorem}
\vspace{-0.2em}

\begin{remark} \label{remark:finitecritpairs}
Every graph transformation system has, up to isomorphism, only finitely many critical pairs. Thus, the reverse direction of this theorem is false, as this would contradict the undecidability of checking for confluence.
\end{remark}
\vspace{-0.2em}
\end{samepage}


\section{Rooted Graph Transformation} \label{section:rootedgt}

We fix some common label alphabet \(\mathcal{L} = (\mathcal{L}_V, \mathcal{L}_E)\), and allow rules to have a partially labelled interface again.

\begin{definition}
Let \(G\) be a \textbf{partially labelled graph}, and \(P_G \subseteq V_G\) be a set of \textbf{root nodes}. Then a \textbf{rooted partially labelled graph} is the tuple \(\widehat{G} = (G, P_G)\).
\end{definition}

\begin{definition}
Given two \textbf{rooted partially labelled graphs} \(\widehat{G}, \widehat{H}\), a \textbf{partially labelled graph morphism} \(g: G \to H\) is a \textbf{rooted labelled graph morphism} \(\widehat{G} \to \widehat{H}\) iff \(g_V(P_G) \subseteq P_H\). A morphism \(g: \widehat{G} \to \widehat{H}\) is \textbf{injective}/\textbf{surjective} iff the underlying graph morphism is injective/surjective. \textbf{Inclusion morphisms} and \textbf{subgraphs} are defined in the obvious way.
\end{definition}

\begin{definition}
We say that \textbf{rooted partially labelled graphs} \(\widehat{G}, \widehat{H}\) are \textbf{isomorphic} iff there exists an \textbf{injective}, \textbf{surjective} morphism \(g: \widehat{G} \to \widehat{H}\) such that \(g^{-1}: \widehat{H} \to \widehat{G}\) is also a morphism, and we write \(\widehat{G} \cong \widehat{H}\). This naturally gives rise to \textbf{equivalence classes} \([\widehat{G}]\): the \textbf{countably} many rooted partially labelled abstract graphs over some fixed \(\mathcal{L}\).
\end{definition}

\begin{definition}
\textbf{Direct derivations} on \textbf{rooted totally labelled graphs} are defined analogously as for \textbf{totally labelled graphs}, but with the following modifications to the rule application process (Definition \ref{def:rukeapp}):
\begin{enumerate}[itemsep=-0.8ex,topsep=-0.8ex]
  \item The root nodes of the \textbf{intermediate graph} are \(P_G \setminus g_V(P_L \setminus P_K)\).
  \item The root nodes of the \textbf{result graph} are \(P_D \cup h_V(P_R \setminus P_K)\).
\end{enumerate}
We write \(\widehat{G} \Rightarrow_{r,g} \widehat{M}\) for a successful application of \(r\) to \(\widehat{G}\) using match \(g\), obtaining result \(\widehat{H} \cong \widehat{M}\). We call this a \textbf{direct derivation}. Definitions \ref{def:directderives} and \ref{def:derives} are analogous.
\end{definition}

\begin{samepage}
\begin{theorem}[Rooted Derivation Uniqueness] \label{thm:uniquederivations}
The \textbf{result graph} of a \textbf{direct derivation} is \textbf{unique up to isomorphism} and is \textbf{totally labelled}.
\end{theorem}
\vspace{-0.1em}

\begin{definition}
A \textbf{rooted graph transformation system} \(\widehat{T} = (\mathcal{L}, \widehat{\mathcal{R}})\), consists of a label alphabet \(\mathcal{L}\), and a \textbf{finite} set \(\mathcal{\widehat{R}}\) of rules over \(\mathcal{L}\).
\end{definition}
\vspace{-0.1em}

\begin{proposition}
Given a \textbf{rooted graph transformation system} \(\widehat{T} = (\mathcal{L}, \widehat{\mathcal{R}})\), then one can always decide if \(\widehat{G} \Rightarrow_{\widehat{\mathcal{R}}} \widehat{H}\).
\end{proposition}
\vspace{-0.1em}

\begin{definition}
Given a \textbf{rooted graph transformation system} \(\widehat{T} = (\mathcal{L}, \widehat{\mathcal{R}})\), a subalphabet of \textbf{non-terminals} \(\mathcal{N}\), and a \textbf{start graph} \(\widehat{S}\) over \(\mathcal{L}\), then a \textbf{rooted graph grammar} is the system \(\pmb{\widehat{G}} = (\mathcal{L}, \mathcal{N}, \widehat{\mathcal{R}}, \widehat{S})\).
\end{definition}
\vspace{-0.1em}

\begin{definition}
Given a \textbf{rooted graph grammar} \(\pmb{\widehat{G}}\) as defined above, we say that a graph \(\widehat{G}\) is \textbf{terminally labelled} iff \(l(V) \cap \mathcal{N}_V = \emptyset\) and \(m(E) \cap \mathcal{N}_E = \emptyset\). Thus, we can define the \textbf{graph language}:
\begin{align*}
\pmb{L}(\pmb{\widehat{G}}) = \{[\widehat{G}] \mid \widehat{S} \Rightarrow_{\mathcal{\widehat{R}}}^{*} \widehat{G}, \widehat{G} \text{ terminally labelled}\}
\end{align*}
\end{definition}
\vspace{-0.2em}
\end{samepage}

\chapter{Graph Theory} \label{appendix:graphtheory}

In standard literature, \enquote{graph theory} is the mathematical study of \enquote{graphs}, where in this context a graph is a finite set of vertices with (directed) edges between them, without parallel edges. We will present this theory in terms of the more general notion of a (labelled) graph from Appendix \ref{appendix:transformation}. The definitions and theorems in this appendix have been adapted from \parencite{Johnson18}, \parencite{Lozin18}, Chapter 1 of \parencite{Bang-Jensen-Gutin09}, and Chapter 3 of \parencite{Skiena08}.


\section{Basic Definitions}

\begin{definition} \label{def:graphprops}
Given a concrete graph \(G\), \(v \in V_G\), we define the:
\begin{enumerate}[itemsep=-0.5ex,topsep=-0.5ex]
\item \textbf{Incoming degree}: \(\operatorname{indeg}_G(v) = \abs{{t_G}^{-1}(\{v\})}\).
\item \textbf{Outgoing degree}: \(\operatorname{outdeg}_G(v) = \abs{{s_G}^{-1}(\{v\} )}\).
\item \textbf{Degree}: \(\operatorname{deg}_G(v) = \operatorname{indeg}_G(v) + \operatorname{outdeg}_G(v)\).
\item \textbf{Neighbourhood}: \(\operatorname{N}_G(v) = s_G({t_G}^{-1}(\{v\})) \cup t_G({s_G}^{-1}(\{v\}))\).
\item \textbf{Closed neighbourhood}: \(\operatorname{N}_G[v] = \operatorname{N}_G(v) \cup \{v\}\).
\end{enumerate}
\end{definition}

\begin{definition} \label{dfn:leafnode}
Given a concrete graph \(G\), \(v \in V_G\), we:
\begin{enumerate}[itemsep=-0.5ex,topsep=-0.5ex]
\item Say \(v \in V_G\) is a \textbf{leaf node} iff \(\operatorname{outdeg}_G(v) = 0\).
\item Say \(u, v \in V_G\) are \textbf{adjacent} iff \(\{u, v\} \subseteq \operatorname{N}[u] \cap \operatorname{N}[v]\).
\item Say \(e \in E_G\) is \textbf{proper} iff \(s_G(e) \neq t_G(e)\).
\end{enumerate}
\end{definition}

\begin{definition}
We say two proper edges \(e, f \in E_G\) are \textbf{parallel} iff [\(s_G(e) = s_G(f)\) and \(t_G(e) = t_G(f)\)] or [\(s_G(e) = t_G(f)\) and \(s_G(e) = t_G(f)\)].
\end{definition}

\begin{definition} \label{dfn:walksconnected}
Let \(G\) be a concrete graph. Then:
\begin{enumerate}[itemsep=-0.5ex,topsep=-0.5ex]
\item An \textbf{undirected walk} of length \(k\) is a non-empty, finite sequence of alternating vertices and edges in \(G\): \(\langle v_0, e_0, v_1, e_1, \dots, e_{k-1}, v_k \rangle\), such that for each \(e_i\) (\(0 \neq i < k\)), [\(s_G(e_i) = v_i\) and \(t_G(e_i) = v_{i+1}\)] or [\(s_G(e_i) = v_{i+}\) and \(t_G(e_i) = v_{i}\)].
\item A \textbf{walk} is an undirected walk such that for each \(e_i\) (\(0 \leq i < k\)), \(s_G(e_i) = v_i\) and \(t_G(e_i) = v_{i+1}\).
\item We call a (undirected) walk \textbf{closed} iff \(v_0 = v_k\).
\item If the vertices \(v_i\) of a \textbf{walk} are all \textbf{distinct} (except possibly \(v_0 = v_k\)), we call the walk a \textbf{path}.
\item A \textbf{closed walk} is called a \textbf{cycle}; a graph with no cycles is \textbf{acyclic}. Similarly, a \textbf{closed undirected walk} is called an \textbf{undirected cycle}.
\end{enumerate}
\end{definition}

\newpage

\begin{definition} \label{dfn:connectedcomponent}
A graph is called \textbf{connected} iff there is an undirected walk between every pair of distinct vertices. A \textbf{connected component} of a concrete graph \(G\) is a \textbf{maximal} connected subgraph.
\end{definition}

\begin{theorem}[Graph Decomposition]
Every concrete graph \(G\) has a unique decomposition into \textbf{connected components}.
\end{theorem}

\begin{definition}
Given a concrete graph \(G\), \(v \in V_G\) we define the:
\begin{enumerate}[itemsep=-0.6ex,topsep=-0.6ex]
\item \textbf{Children}: \(\operatorname{children}_G(v) = t_G({s_G}^{-1}(\{v\}))\).
\item \textbf{Parents}: \(\operatorname{parents}_G(v) = s_G({t_G}^{-1}(\{v\}))\).
\end{enumerate}
\(u\) is a \textbf{child} of \(v\) iff \(u \in \operatorname{children}_G(v)\), and a \textbf{parent} iff \(u \in \operatorname{parents}_G(v)\).
\end{definition}

\begin{proposition}
Given a concrete graph \(G\), \(v \in V_G\). Then:
\begin{enumerate}[itemsep=-0.6ex,topsep=-0.6ex]
\item \(\operatorname{children}_G(v) \subseteq \operatorname{N}_G(v)\) and \(\operatorname{parents}_G(v) \subseteq \operatorname{N}_G(v)\).
\item \(\abs{\operatorname{children}_G(v)} \leq \operatorname{outdeg}_G(v)\) and \(\abs{\operatorname{parents}_G(v)} \leq \operatorname{indeg}_G(v)\).
\end{enumerate}
\end{proposition}


\section{Classes of Graphs} \label{sec:graphclasses}

\begin{definition} \label{dfn:discretegraph}
A graph is called \textbf{discrete} iff it has no edges.
\end{definition}

\begin{definition} \label{dfn:tree}
A \textbf{tree} is a non-empty connected graph without undirected cycles such that every node has at most one incoming edge. Moreover:
\begin{enumerate}[itemsep=-0.5ex,topsep=-0.5ex]
\item A \textbf{linked list} is a \textbf{tree} such that every node has outgoing degree at most \(1\).
\item A \textbf{binary tree} is a \textbf{tree} such that every node has outgoing degree at most \(2\).
\item A \textbf{perfect binary tree} is a \textbf{binary tree} such that every node has either \(0\) or \(2\) children and every \textbf{maximal path} is the same length.
\item A \textbf{forest} is a graph where each \textbf{connected component} is a \textbf{tree}.
\end{enumerate}
\end{definition}

\begin{definition}
A \(n \times m\)-\textbf{grid graph} is a graph with underlying unlabelled graph isomorphic to \((V, E, s, t)\) where \(V = \mathbb{Z}_n \times \mathbb{Z}_m\), \(E = (\mathbb{Z}_2 \times V) \setminus \{(0, i, m-1), (1, n-1, j) \mid i \in \mathbb{Z}_n, j \in \mathbb{Z}_m\}\), \(s(d, i, j) = (i, j)\), and \(t(d, i, j) = (i+d, j+1-d)\). We call such a graph \textbf{square} iff \(n = m\).
\end{definition}

\begin{definition}
An \(n\)-\textbf{star graph} is a graph with underlying unlabelled graph isomorphic to \((V, E, s, t)\) where \(V = \mathbb{Z}_{n+1}\), \(E = \mathbb{Z}_{n}\), and:

\vspace{-2em}
\begin{multicols}{2}
\[
s(i) =
\begin{cases}
  n & \text{ if } i \equiv 0 \text{ mod } 2 \\
  i & \text{ otherwise}
\end{cases}
\]

\[
t(i) =
\begin{cases}
  n & \text{ if } i \equiv 1 \text{ mod } 2 \\
  i & \text{ otherwise}
\end{cases}
\]
\end{multicols}
\end{definition}

An example linked list, perfect binary tree, square grid graph, and star graph can be found in Figure \ref{fig:graph-types}.

\begingroup
\sloppy
\cleardoublepage
\phantomsection
\addcontentsline{toc}{chapter}{Bibliography}
\printbibliography
\endgroup

\end{document}